%% file: main.tex
\documentclass{vldb}
\usepackage{graphicx}
\usepackage{balance}  
\usepackage{enumitem}
\usepackage{algpseudocode}
\usepackage{graphicx} 
\usepackage{epstopdf}
\usepackage{subfigure}
\usepackage{amsmath,bm}
\usepackage{multirow}
\usepackage{color}
\usepackage{balance}
\usepackage{url}
\usepackage{hyperref}
\hypersetup{hidelinks}
\usepackage{caption}
\usepackage[ruled,linesnumbered]{algorithm2e}

\newtheorem{problem}{Problem}

\newtheorem{definition}{Definition}

\vldbTitle{Efficient and Effective Similar Subtrajectory Search with Deep Reinforcement Learning}
\vldbAuthors{Zheng Wang, Cheng Long, Gao Cong, Yiding Liu}
\vldbDOI{https://doi.org/10.14778/xxxxxxx.xxxxxxx}
\vldbVolume{12}
\vldbNumber{xxx}
\vldbYear{2020}

\begin{document}
\sloppy

\title{Efficient and Effective Similar Subtrajectory Search with Deep Reinforcement Learning}



%
%
%
%

\numberofauthors{1} 
\author{
\alignauthor
   Zheng Wang, Cheng Long, Gao Cong, Yiding Liu\\
   \affaddr{School of Computer Science and Engineering, Nanyang Technological University, Singapore}
   \email{\{wang\_zheng, c.long, gaocong, ydliu\}@ntu.edu.sg}
}

\maketitle

\pagestyle{empty}

\begin{abstract}

\input{abstract}
\end{abstract}

\renewcommand{\thesection}{\arabic{section}}
\setcounter{section}{0}

\input{introduction}

\input{related_v2}

\input{problem}

\input{non-learning}

\input{rl_v2}

\input{effectiveness}
\input{conclusion}

%
\newpage
\bibliographystyle{abbrv}
\balance
\bibliography{ref}
\newpage
\appendix
\input{appendix}
\end{document}

%% file: abstract.tex
Similar trajectory search is a fundamental problem and has been well studied over the past two decades. However, the similar subtrajectory search (SimSub) problem, aiming to return a portion of a trajectory (i.e., a subtrajectory), which is the most similar to a query trajectory, has been mostly disregarded despite that it could capture trajectory similarity in a finer-grained way and many applications take subtrajectories as basic units for analysis. In this paper, we study the SimSub problem and develop a suite of algorithms including both exact and approximate ones.
Among those approximate algorithms, two that are based on deep reinforcement learning stand out and outperform those non-learning based algorithms in terms of effectiveness and efficiency.
We conduct experiments on real-world trajectory datasets, which verify the effectiveness and efficiency of the proposed algorithms.

%% file: introduction.tex
\section{INTRODUCTION}
\label{intro}

Trajectory data, which corresponds to a type of data for capturing the traces of moving objects, is ubiquitous. It has been used for various types of analysis such as clustering~\cite{agarwal2018subtrajectory,lee2007trajectory,buchin2011detecting} and similarity search~\cite{chen2004marriage,chen2005robust,xie2014eds,ranu2015indexing,li2018deep,yi1998efficient}. The majority of existing studies take \emph{a trajectory as a whole} for analysis~\cite{chen2004marriage,chen2005robust,xie2014eds,ranu2015indexing,li2018deep,yi1998efficient}. Motivated by the phenomenon that two trajectories could be dissimilar to each other if each is considered a whole but similar if only some portion of each is considered, there have been a few studies, which take \emph{a portion of a trajectory} as a basic entity for analysis~\cite{agarwal2018subtrajectory,lee2007trajectory,buchin2011detecting,tampakis2019distributed,tampakis2019scalable}. Some examples include subtrajectory clustering~\cite{agarwal2018subtrajectory,lee2007trajectory,buchin2011detecting} and subtrajectory join~\cite{tampakis2019distributed,tampakis2019scalable}. For example, the subtrajectory clustering method in~\cite{lee2007trajectory} first partitions raw trajectories into different subtrajectories using some principle and then groups those subtrajectories that are similar to one another into clusters.

In this paper, we study a query with its goal to search for a portion of a trajectory from a database storing many trajectories called \emph{data trajectories}, which is the most similar to a given trajectory called \emph{query trajectory}. In this query, a portion of a trajectory, called \emph{subtrajectory}, is considered as a basic entity and a query trajectory is taken as a whole for analysis.
Therefore, it captures trajectory similarity in a finer-grained way than conventional similar trajectory search.
For instance, consider a data trajectory and a query trajectory.
When considered as a whole, the data trajectory is not similar to the query trajectory based on some trajectory similarity measurement,
but some portion of it is very similar to the query trajectory.
With the conventional similar trajectory search query,
this data trajectory would be ruled out,
though a portion of it is very similar to the query trajectory, which is interesting.

Moreover, in quite a few real-life applications, subtrajectories are naturally considered as basic units for analysis,
e.g., subtrajectory search~\cite{sha2016chalkboarding}, subtrajectory join~\cite{tampakis2019distributed}, subtrajectory clustering~\cite{buchin2011detecting}, etc.
One application is the subtrajectory search query on sports play data.
In sports such as soccer and basketball, a common practice nowadays is to track the movements of players and/or the ball using some special-purpose camera and/or GPS devices~\cite{wang2019effective}.
The resulting trajectory data is used to capture the semantics of the plays and for different types of data analyses.
One typical task on such sports play data is to search for a portion/segment of play from a database of plays, with its trajectories of players and/or its trajectory of the ball similar to those and/or that of a given query play~\cite{sha2016chalkboarding}. This task is essentially one of searching for similar subtrajectories.
Another potential application is detour route detection.
It first collect those routes that have been reported by passengers to be detour routes and then searches for those subtrajectories of taxis' routes, which are similar to a detour route.
The found subtrajectories are probably detour routes as well. 

A key problem that is involved in answering the query mentioned above is to find a subtrajectory of a data trajectory, which is the most similar to a given query trajectory. We call this problem the \emph{similar subtrajectory search} (SimSub) problem. 
While there are many existing studies on the similar trajectory search problem with each trajectory considered a whole, there are very few studies on the SimSub problem.
Let $T$ be a data trajectory involving $n$ points and $T_q$ be a query trajectory involving $m$ points. 
We design an exact algorithm, which enumerates all possible subtrajectories of $T$, computes the similarity between each subtrajectory and the query trajectory, and returns the one with the greatest similarity. 
We further adopt an \emph{incremental} strategy for computing the similarities that are involved in the exact algorithm, which helps to improve the time complexity by $O(n)$.
%
%
%
We also follow some existing studies on subsequence matching~\cite{kim2013efficient,zhang2010bed} and design an algorithm, which considers only those subtrajectories with their sizes similar to that of the query trajectory and controlled by a user parameter. This would provide a controllable trade-off between efficiency and effectiveness.

To push the efficiency further up, we propose several algorithms, which share the idea of splitting a data trajectory into some subtrajectories to be candidate solutions to the problem and differ in using different methods for splitting the data trajectory. Specifically, the process is to scan the points of a data trajectory one by one sequentially and for each one, it decides whether to split the data trajectory at the point. Some of them use pre-defined heuristics, e.g., a greedy one. Others model the process as a \emph{markov decision process} (MDP)~\cite{puterman2014markov} and use deep reinforcement learning to learn an optimal policy for the MDP, which is then used for splitting the data trajectory. 
These splitting-based algorithms have time complexities much lower than the exact algorithm in general, e.g., for measurements such as t2vec, each splitting-based algorithm runs in $O(n)$ time while the exact algorithm runs in $O(nm)$ time.

The major contributions of this paper are as follows.
\begin{itemize}
	\item We propose the SimSub problem, and this to the best of our knowledge, corresponds to the first systematic study on searching subtrajectories that are similar to a query trajectory.
	The SimSub problem relies on a trajectory similarity measurement, and in this paper, we assume an abstract one, which could be instantiated with any existing measurement.
	
	\item We develop a suite of algorithms for the SimSub problem: (1) one exact algorithm, (2) one approximate algorithm, which provides a controllable trade-off between efficiency and effectiveness, and (3) several splitting-based algorithms including both heuristics-based ones and deep reinforcement learning based ones. These algorithms should cover a wide spectrum of application scenarios in terms of efficiency and effectiveness requirements.
	
	\item Third, we conducted extensive experiments, which verified that splitting-based algorithms in general have good efficiency and among them, the algorithms based on deep reinforcement learning achieve the best effectiveness and efficiency.
\end{itemize}

\noindent
\textbf{Organization}. We review the related work in Section~\ref{related} and provide the problem definition and
some preliminaries in Section~\ref{problem_and_preliminary}.
Section~\ref{sec:non-learning} presents all non-learning based algorithms and
Section~\ref{section:RLS} presents the deep reinforcement learning based algorithms.
We report our experimental results in Section~\ref{experiment} and conclude this paper and discuss some future work in Section~\ref{conclusion}.

%% file: related_v2.tex
\section{RELATED WORK}
\label{related}

\noindent\textbf{(1) Trajectory Similarity Measurements.}
Measuring the similarity between trajectories is a fundamental problem and has been studied extensively. Some classical solutions focus on indexing trajectories and performing similarity computation by the alignment of matching sample points. For example, DTW~\cite{yi1998efficient} is the first attempt at solving the local time shift issue for computing trajectory similarity. Frechet distance~\cite{alt1995computing} is a classical similarity measure that treats each trajectory as a spatial curve and takes into account the location and order of the sampling points. Further, ERP~\cite{chen2004marriage} and EDR~\cite{chen2005robust} are proposed to improve the ability to capture the spatial semantics in trajectories. However, these point-matching methods are inherently sensitive to noise and suffer from quadratic time complexity. EDS~\cite{xie2014eds} and EDwP~\cite{ranu2015indexing} are two segment-matching methods, which operate on segments for matching two trajectories.
In recent years, some learning-based algorithms were proposed to speed up the similarity computation. Li et al.~\cite{li2018deep} propose to learn representations of trajectories in the form of vectors and then measure the similarity between two trajectories as the Euclidean distance between their corresponding vectors.
Some other studies~\cite{wang2018torch,yuan2019distributed,wang2019fast} define similarity measurements on trajectories based on road segments, to which the trajectories are matched.
Yao et al.~\cite{yao2019computing} employ deep metric learning to approximate and accelerate trajectory similarity computation.
In addition, Ma et al.~\cite{ma2012ksq} propose a similarity measurement called p-distance for uncertain trajectories and study the problem of searching for top-k similar trajectories to a given query trajectory based on p-distance.
Different specialized index techniques are developed for these similarity measures, such as DTW distance~\cite{yi1998efficient,keogh2005exact}, LCSS~\cite{vlachos2002discovering}, ERP~\cite{chen2004marriage}, EDR~\cite{chen2005robust}, and EDwP~\cite{ranu2015indexing}. However, these index techniques do not generalize to other similarity measures or subtrajectory similarity search.
%
%
In this paper, we assume an abstract trajectory similarity measurement, which could be instantiated with any of these existing similarity measurements and our techniques still apply.
%

\smallskip
\noindent
\textbf{(2) Subtrajectory Similarity Related Problems.}
Measuring subtrajectory similarity is also a fundamental functionality in many tasks such as clustering~\cite{agarwal2018subtrajectory,lee2007trajectory,buchin2011detecting} and similarity join~\cite{tampakis2019distributed}. Lee et al~\cite{lee2007trajectory} propose a general partition and group framework for subtrajectory clustering. Further, Buchin et al.~\cite{buchin2011detecting} show the hardness of subtrajectory clustering based on Frechet distance, and Agarwal et al.~\cite{agarwal2018subtrajectory} apply the trajectory simplification technique to approximate discrete Frechet to reduce the time cost of subtrajectory clustering. Recently, Tampakis et al.~\cite{tampakis2019distributed,tampakis2019scalable} proposed a distributed solution for subtrajectory join and clustering by utilizing the MapReduce programming model. Although these algorithms need to consider subtrajectory similarity, similarity computation is not their focus and they usually first segment a trajectory into subtrajectories and employ an existing measure, such as Freechet distance.

\smallskip
\noindent
\textbf{(3) Subsequence (Substring) Matching.} Subsequence matching is a related but different problem. It aims to find a subsequence that has the same length as the query in a given candidate sequence, which usually contains millions or even trillions~\cite{rakthanmanon2012searching,rakthanmanon2013addressing} of elements. Efficient pruning algorithms~\cite{gong2019fast,rakthanmanon2013addressing,rakthanmanon2012searching,athitsos2008approximate,han2007ranked,moon2001duality,faloutsos1994fast} have been proposed for the matching, and these pruning algorithms are generally designed for a specific similarity measure, such as DTW~\cite{gong2019fast,rakthanmanon2013addressing,rakthanmanon2012searching,athitsos2008approximate,han2007ranked,park2000efficient,kim2001index} and Euclidean distance~\cite{faloutsos1994fast,moon2001duality}, and cannot generalize to other measures.
On the other hand, substring matching~\cite{kim2013efficient,zhang2010bed} often focuses on approximate matching based on the Edit distance. It aims to find a substring in a string to best match the query. Our problem differs from the substring matching problem mainly in two aspects. First, characters in a string have exact match (0 or 1) in the alphabet; however, the points of a trajectory are different. Second, substring matching techniques are usually designed based on the characteristics of strings. e.g.,
grammar structure patterns, or word concurrence patterns; however, a trajectory does not have such patterns.

\smallskip
\noindent
\textbf{(4) Reinforcement Learning.} The goal of reinforcement learning is to guide agents on how to take actions to maximize a cumulative reward~\cite{sutton2018reinforcement} in an environment, and the environment is generally modeled as a Markov decision process (MDP)~\cite{puterman2014markov}. Recently, RL models have been utilized successfully to solve some database related problems. For example, Zhang et al.~\cite{zhang2019end} and Li et al.~\cite{li2019qtune} use RL model for automatic DBMS tuning. Trummer et al.~\cite{trummer2018skinnerdb} use RL to learn optimal join orders in the SkinnerDB system. Wang et al.~\cite{wang2019adaptive} design an effective RL-based algorithm for bipartite graph matching. Overall, there are two types of popular reinforcement learning methods: (1) model-based methods~\cite{brafman2002r,kearns2002near} that require to understand the environment and learn the parameters of the MDP in advance, and (2) model-free methods~\cite{watkins1992q,mnih2013playing} that make no efforts to learn a model and get feedback from the environment step by step. In this paper, we follow the model-free methods because they are more efficient. Specifically, we make use of a popular reinforcement learning method, namely Deep $Q$ Network (DQN)~\cite{mnih2013playing}, for splitting a trajectory into subtrajectories to be candidate solutions for similar subtrajectory search.

%% file: problem.tex
\section{Problem Definition and Preliminaries}
\label{problem_and_preliminary}


The trace of a moving object such as a vehicle and a mobile user is usually captured by a trajectory. Specifically, a trajectory $T$ has its form as a sequence of time-stamped locations (called points), i.e., $T = <p_1, p_2, ..., p_n>$, where point $p_i = (x_i, y_i, t_i)$ means that the location is $(x_i, y_i)$ at time $t_i$. The size of trajectory $T$, denoted by $|T|$, corresponds to the number of points of $T$.

Given a trajectory $T=<p_1, p_2, ..., p_n>$ and $1\le i\le j\le n$, we denote by $T[i, j]$ the portion of $T$ that starts from the $i^{th}$ point and ends at the $j^{th}$ point, i.e., $T[i, j] = <p_i, p_{i+1}, ..., p_j>$. Besides, we say that $T[i, j]$ for any $1\le i \le j \le n$ is \emph{subtrajectory} of $T$. There are in total $\frac{n(n+1)}{2}$ subtrajectories of $T$. Note that any subtrajectory of a trajectory $T$ belongs to a trajectory itself.

\subsection{Problem Definition}
Suppose we have a database of many trajectories, which we call \emph{data trajectories}. As discussed in Section~\ref{intro}, one common application scenario would be that a user has a trajectory at hand, which we call a \emph{query trajectory} and would like to check what is the portion of the data trajectories that is the most similar to the one at his/her hand. Note that in some cases, by looking each data trajectory as whole, none is similar enough to the query trajectory, e.g., all data trajectories are relatively long while the query trajectory is relatively short.

We note that a more general query is to find the \emph{top-$k$} similar subtrajectories to a query trajectory, which reduces to the user's query as described above when $k = 1$. In this paper, we stick to the setting of $k=1$ since extending the techniques for the setting of $k=1$ to general settings of $k$ is straightforward. Specifically, the techniques for the setting $k=1$ in this paper are all based on a search process, which maintains the most similar subtrajectory found so far and updates it when a more similar subtrajectory is found during the process. These techniques could be adapted to general settings of $k$ by simply maintaining the $k$ most similar subtrajectories and updating them when a subtrajectory that is more similar than the $k^{th}$ most similar subtrajectory.

An intuitive solution to answer the user's query is to scan the data trajectories, and for each one, compute its subtrajectory that is the most similar to the query one based on some similarity measurement and update the most similar subtrajectory found so far if necessary. This solution could be further enhanced by employing indexing techniques such as the R-tree based index and the inverted-file based index for pruning~\cite{yao2019computing,wang2018torch}, e.g., the data trajectories that do not have any overlap with the query trajectory could usually be pruned. The key component of this solution (no matter whether indexing structures are used or not) is to compute for a given data trajectory, its subtrajectory that is the most similar to a query trajectory. We formally define the problem corresponding to this procedure as follows.
\begin{problem}[Similar Subtrajectory Search]
	Given a data trajectory $T = <p_1, p_2, ..., p_n>$ and a query trajectory $T_q=<q_1, q_2, ..., q_m>$,
	the \textbf{similar subtrajectory search (SimSub)} problem is to find a subtrajectory of $T$,
	denoted by $T[i^*, j^*]$ ($1\le i^* \le j^* \le n$), which is the most similar to $T_q$ according to a trajectory similarity measurement $\Theta(\cdot, \cdot)$,
	i.e., $[i^*, j^*] = \arg\max_{1\le i\le j\le n} \Theta(T[i, j], T_q)$.
	\label{subtrajectory-search}
\end{problem}

The SimSub problem relies on a similarity measurement $\Theta(T, T')$, which captures the extent to which two trajectories $T$ and $T'$ are similar to each other. The larger the similarity $\Theta(T, T')$ is, the more similar $T$ and $T'$ are. In the literature, several ``dissimilarity measurements'' have been proposed for $\Theta(\cdot, \cdot)$ such as DTW~\cite{yi1998efficient}, Frechet~\cite{alt1995computing}, LCSS~\cite{vlachos2002discovering}, ERP~\cite{chen2004marriage}, EDR~\cite{chen2005robust}, EDS~\cite{xie2014eds}, EDwP~\cite{ranu2015indexing}, and t2vec~\cite{li2018deep}. Different measurements have different merits and suit for different application scenarios. In this paper, we assume an abstract similarity measurement $\Theta(\cdot, \cdot)$, which could be instantiated with any of these existing measurements by applying some \emph{inverse operation} such as taking the ratio between 1 and a distance.

\subsection{Trajectory Similarity Measurements}

The SimSub problem assumes an abstract similarity measurement and the techniques developed could be applied to any existing measurements. Since the time complexity analysis of the algorithms proposed in this paper relies on the time complexities of computing a specific measurement in several different cases, in this part, we review three existing measurements, namely t2vec~\cite{li2018deep}, DTW~\cite{yi1998efficient}, and Frechet~\cite{alt1995computing}, and discuss their time complexities in different cases as background knowledge. The first two are the most widely used measurements and the last one is the most recently proposed one, which is a data-driven measurement.
%
%

%
We denote by $\Phi$ the time complexity of computing the similarity between a general subtrajectory of $T$ and $T_q$ \emph{from scratch}, $\Phi_{inc}$ be the time complexity of computing $\Theta(T[i, j], T_q)$ ($1\le i < j \le n$) \emph{incrementally} assuming that $\Theta(T[i, j-1], T_q)$ has been computed already, and $\Phi_{ini}$ the time complexity of computing $\Theta(T[i, i], T_q)$ ($1\le i\le n$) from scratch since it cannot be computed incrementally. As will be discussed later, $\Phi_{inc}$ and $\Phi_{ini}$ are usually much smaller than $\Phi$ across different similarity measurements.

\smallskip\noindent\textbf{t2vec~\cite{li2018deep}.}
t2vec is a data-driven similarity measure based on deep representation learning. It adapts a sequence-to-sequence framework based on RNN~\cite{cho2014learning} and takes the final hidden vector of the encoder~\cite{rumelhart1988learning} to represent a trajectory. It computes the similarity between two trajectories based on the Euclidean distance between their representations as vectors.

Given $T$ and $T_q$, it takes $O(n)$ and $O(m)$ time to compute their hidden vectors, respectively and $O(1)$ to compute the Euclidean distance between two vectors~\cite{li2018deep}. Therefore, we know $\Phi = O(n + m + 1) = O(n + m)$. Since in the context studied in this paper, we need to compute the similarities between many subtrajectories and a query trajectory $T_q$, we assume that the representation of $T_q$ under t2vec is computed once and re-used many times, i.e., the cost of computing the representation of $T_q$, which is $O(m)$, could be amortized among all computations of similarity and then that for each one could be neglected. Because of the sequence-to-sequence nature of t2vec, given the representation of $T[i, j-1]$, it would take $O(1)$ to compute that of $T[i, j]$ ($1\le i < j\le n$). Therefore, we know $\Phi_{inc} = O(1)$. Besides, we know $\Phi_{ini} = O(1)$ since the subtrajectory involved in the computation of similarity, i.e., $T[i,i]$ ($1\le i\le n$), has its size equal to 1.

\smallskip\noindent\textbf{DTW~\cite{yi1998efficient}.} Given a data trajectory $T = <p_1, p_2,...,p_n>$ and a query trajectory $T_q = <q_1, q_2,...,q_m>$, the DTW distance is defined as below
\begin{equation}
\label{dtw_sm}
D_{i,j}=\left\{
\begin{aligned}
& \sum\nolimits_{h=1}^{i}d(p_h,q_1)&\text{if } j = 1\\
& \sum\nolimits_{k=1}^{j}d(p_1,q_k)&\text{if } i = 1\\
& d(p_i,q_j) + \\
& \min(D_{i-1,j-1}, D_{i-1,j},D_{i,j-1})&\text{otherwise}
\end{aligned}
\right.
\end{equation}
where $D_{i,j}$ denotes the DTW distance between $T[1,i]$ and $T_q[1, j]$ and $d(p_i,q_j)$ is the distance between $p_i$ and $q_j$ (typically the Euclidean distance, which could be computed in $O(1)$). Consider $\Phi$. It is clear that $\Phi = O(n\cdot m)$ since it needs to compute all pairwise distances between a point in a subtrajectory of $T$ and a point in $T_q$ and in general, the subtrajectory has its size of $O(n)$ and $T_q$ has its size of $m$.
Consider $\Phi_{inc}$. This should be the same as the time complexity of computing $D_{i,m}$ given that $D_{i-1,m}$ has been computed. Since $D_{i-1,m}$ has been computed, we can safely assume that $D_{i-1,1}$, $D_{i-1,2}$, ..., $D_{i-1,m}$ have been computed also according to Equation~\ref{dtw_sm} (note that we can always make this hold by enforcing that we compute $D_{i-1,m}$ or any other DTW distance in this way). Therefore, in order to compute $D_{i,m}$, we compute $D_{i,1}, D_{i,2}, ..., D_{i,m}$ sequentially, each of which would take $O(1)$ time with the information of $D_{i-1,k}$ ($1\le k\le m$) all available. That is, it takes $O(m)$ to compute $D_{i,m}$, and thus we know that $\Phi_{inc} = O(m)$.
Consider $\Phi_{ini}$. We know $\Phi_{ini} = O(m)$ since $T[i, i]$ ($1\le i \le n$) has its size always equal to 1 and $T_q$ has its size of $m$.

\smallskip\noindent\textbf{Frechet~\cite{alt1995computing}.} Given a data trajectory $T = <p_1, p_2,...,p_n>$ and a query trajectory $T_q = <q_1, q_2,...,q_m>$, the Frechet distance is defined as below
\begin{equation}
\label{fre_sm}
F_{i,j}=\left\{
\begin{aligned}
&\max\nolimits_{h=1}^{i}d(p_h,q_1)&\text{if }j = 1\\
&\max\nolimits_{k=1}^{j}d(p_1,q_k)&\text{if }i = 1\\
&\max(d(p_i,q_j),\\
& \min(F_{i-1,j-1},F_{i-1,j},F_{i,j-1}))&\text{otherwise}
\end{aligned}
\right.
\end{equation}
where $F_{i,j}$ denotes the Frechet distance between $T[1,i]$ and $T_q[1, j]$ and $d(p_i,q_j)$ is the distance between $p_i$ and $q_j$ (typically the Euclidean distance, which could be computed in $O(1)$).
When the Frechet distance is used, we have $\Phi = O(n\cdot m)$, $\Phi_{inc} = O(m)$, and $\Phi_{ini} = O(m)$, based on similar analysis as for the DTW distance.

The summary of $\Phi$, $\Phi_{inc}$ and $\Phi_{ini}$ for the similarity measurements corresponding to the distance measurements DTW, Frechet and t2vec is presented in Table~\ref{tab:similarity-complexities}.

\begin{table}[t!]
\centering
	\caption{Time complexities of computing the similarity between a subtrajectory of $T$ and $T_q$ in three cases}
	\begin{tabular}{|l||c|c|c|}
		\hline
		{Time complexities}			&	t2vec 		&	DTW				&	Frechet			\\\hline
		$\Phi$ (general)			& 	$O(n + m)$	&	$O(n\cdot m)$	&	$O(n\cdot m)$	\\\hline
		$\Phi_{inc}$ (incremental)	& 	$O(1)$		&	$O(m)$			&	$O(m)$			\\\hline
		$\Phi_{ini}$ (initial)		& 	$O(1)$		&	$O(m)$			&	$O(m)$			\\\hline
	\end{tabular}
	\label{tab:similarity-complexities}
\end{table}

%% file: non-learning.tex
\section{Non-Learning based Algorithms}
\label{sec:non-learning}

In this part, we introduce three types of algorithms, namely an exact algorithm ExactS, an approximate algorithm SizeS, and splitting-based algorithms including PSS, POS and POS-D. The ExactS algorithm is based on an exhaustive search with some careful implementation and has the highest complexity, the SizeS algorithm is inspired by existing studies on subsequence matching~\cite{kim2013efficient,zhang2010bed} and provides a tunable parameter for controlling the trade-off between efficiency and effectiveness, and the splitting-based algorithms are based on the idea of splitting the data trajectory for constructing subtrajectories as candidates of the solution and run the fastest. A summary of the time complexities of these algorithms is presented in Table~\ref{tab:algorithm-complexities}.

\begin{table*}[t!]
\centering
	\caption{Time complexities of algorithms ($n_1 << n$)}
	\begin{tabular}{|l||c|c|c|c|}
		\hline
		Algorithms		&	abstract similarity	measurement							&	t2vec 					&	DTW								&	Frechet							\\\hline
		ExactS			&	$O(n\cdot ( \Phi_{ini} + n \cdot \Phi_{inc}))$			& 	$O(n^2)$		&	$O(n^2\cdot m)$					&	$O(n^2\cdot m)$					\\\hline
		SizeS			& 	$O(n\cdot( \Phi_{ini} + (m + \xi) \cdot \Phi_{inc}))$	& 	$O((\xi + m)\cdot n)$	&	$O((\xi + m)\cdot n\cdot m)$	&	$O((\xi + m)\cdot n\cdot m)$	\\\hline
		PSS, POS, POS-D	&	$O(n_1\cdot \Phi_{ini} + n\cdot \Phi_{inc})$			& 	$O(n)$					&	$O(n \cdot m)$					&	$O(n \cdot m)$					\\\hline\hline
		\parbox{2.5cm}{RLS, RLS-Skip (learning-based)}				&	$O(n_1\cdot \Phi_{ini} + n\cdot \Phi_{inc})$			& 	$O(n)$					&	$O(n \cdot m)$					&	$O(n \cdot m)$					\\\hline
	\end{tabular}	
	\label{tab:algorithm-complexities}
\end{table*}

\subsection{The ExactS Algorithm}

Let $T$ be a data trajectory and $T_q$ be a query trajectory.
The ExactS algorithm enumerates all possible subtrajectories $T[i, j]$ ($1\le i\le j\le n$) of the data trajectory $T$ and computes the similarity between each $T[i, j]$ and $T_q$, i.e., $\Theta(T[i, j], T_q)$, and then returns the one with the greatest similarity. 
For better efficiency,
ExactS computes the similarities between the subtrajectories and $T_q$ \emph{incrementally} as much as possible as follows. 
It involves $n$ iterations, and in the $i^{th}$ iteration, it computes the similarity between each subtrajectory starting from the $i^{th}$ point and the query trajectory in an ascending order of the ending points, i.e., it computes $\Theta(T[i, i], T_q)$ (from scratch) first and then computes $\Theta(T[i, i+1], T_q)$, ..., $\Theta(T[i, n], T_q)$ sequentially and incrementally. During the process, it maintains the subtrajectory that is the most similar to the query one, among those that have been traversed so far. As could be verified, it would traverse all possible subtrajectories after $n$ iterations. The ExactS algorithm with this implementation is presented in Algorithm~\ref{alg:ExactS}.

Consider the time complexity of ExactS. Since there are $n$ iterations and in each iteration, the time complexity of computing $\Theta(T[i, i], T_q)$ is $\Phi_{ini}$ and the time complexity of computing $\Theta(T[i, i+1], T_q)$, ..., and $\Theta(T[i, n], T_q)$ is $O(n\cdot \Phi_{inc})$, we know that the overall time complexity is $O(n\cdot ( \Phi_{ini} + n \cdot \Phi_{inc}))$. 

We note that for some specific similarity measurement, there may exist algorithms that have better time complexity than ExactS. For example, the Spring algorithm~\cite{sakurai2007stream}, which finds the most similar subsequence of a data time series to a query one, is applicable to the SimSub problem and has the time complexity of $O(nm)$. The major idea of Spring is a dynamic programming process for computing the DTW distance between the data time series and the query one, where the latter is padded with a fictitious point that could be aligned with any point of the data time series with distance equal to 0 (so as to cover all possible suffixes of the data time series). Nevertheless, Spring is designed for the specific similarity DTW while ExactS works for an abstract one that could be instantiated to be any similarity.
\begin{algorithm}[t!]
        \caption{ExactS}\label{alg:ExactS}
		\footnotesize{
		\KwIn{
		A data trajectory $T$ and query trajectory $T_q$;}
		\KwOut{
		A subtrajectory of $T$ that is the most similar to $T_q$;}
		$T_{best} \leftarrow \emptyset$; $\Theta_{best} \leftarrow 0$;\\
		\ForAll{$1 \leq i \leq |T|$}{
		compute $\Theta(T[i, i], T_q)$;\\
		\If{$\Theta(T[i, i], T_q) > \Theta_{best}$}{
		$T_{best} \leftarrow T[i, i]$; $\Theta_{best} \leftarrow \Theta(T[i, i], T_q)$;
		}
		\ForAll{$i+1 \le j\le |T|$}{
		compute $\Theta(T[i, j], T_q)$ based on $\Theta(T[i, j-1], T_q)$;\\
		\If{$\Theta(T[i, j], T_q) > \Theta_{best}$}{
		$T_{best} \leftarrow T[i, j]$; $\Theta_{best} \leftarrow \Theta(T[i, j], T_q)$;
		}
		}
		}
		\textbf{return} $T_{best}$;
	}
\end{algorithm}

\subsection{The SizeS Algorithm}
ExactS explores all possible $\frac{n(n+1)}{2}$ subtrajectories, many of which might be quite dissimilar from the query trajectory and could be ignored. For example, by following some existing studies on subsequence matching~\cite{kim2013efficient,zhang2010bed}, we could restrict our attention to only those subtrajectories, which have similar sizes as the query one for better efficiency. Specifically, we enumerate all subtrajectories that have their sizes within the range $[m - \xi, m + \xi]$, where $\xi\in [0, n-m]$ is a pre-defined parameter that controls the trade-off between the efficiency and effectiveness of the algorithm. Again, we adopt the strategy of incremental computation for the similarities between those subtrajectories starting from the same point and the query trajectory.
We call this algorithm \emph{SizeS} and analyze its time complexity as follows. The time complexity of computing the similarities between all subtrajectories starting from a specific point and having their sizes within the range $[m - \xi, m+\xi]$ is $O(\Phi_{ini} + (m-\xi-1)\cdot \Phi_{inc} + 2\xi \cdot \Phi_{inc})$, where $\Phi_{ini} + (m-\xi-1)\cdot \Phi_{inc}$ is cost of computing $\Theta(T[i, i+m-\xi-1], T_q)$ and $2\xi \cdot \Phi_{inc}$ is the cost of computing $\Theta(T[i, j], T_q)$ for $j\in [i+m-\xi, i+m+\xi-1]$. It could be further reduced to $O(\Phi_{ini} + (m+\xi)\cdot \Phi_{inc})$. Therefore, the overall time complexity of SizeS is $O(n\cdot( \Phi_{ini} + (m + \xi) \cdot \Phi_{inc}))$. For example, when DTW or Frechet is used, it is $O(n\cdot (m + (m + \xi)\cdot m)) = O((\xi + m)\cdot n\cdot m)$ and when t2vec is used, it is $O(n\cdot (1 + (\xi + m)\cdot 1)) = O((\xi + m)\cdot n)$.

In summary, SizeS achieves a better efficiency than ExactS at the cost of its effectiveness. Besides, SizeS still needs to explore $O(\xi \cdot n)$ subtrajectories, which restricts its application on small and moderate datasets only.
Unfortunately, SizeS may return a solution, which is arbitrarily worse than the best one. We illustrate this in the technical report version~\cite{TR} due to the page limit.

\subsection{Splitting-based Algorithms}

\begin{figure*}[htb]
	\centering
\begin{tabular}{cc}
	\centering
  \begin{minipage}{4cm}
    \centering
    \includegraphics[width=4cm]{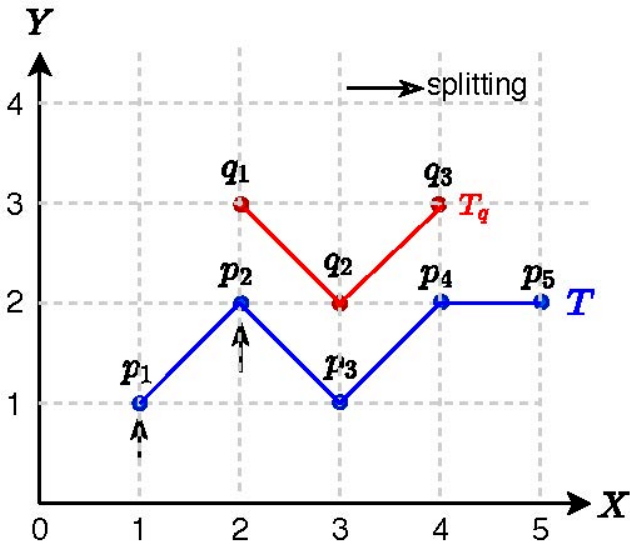}
  \caption{A problem input.}
  \label{fig:input}
  \end{minipage}
  &
  \begin{minipage}{13cm}
    \centering
  	\scriptsize
    \captionof{table}{Illustration of PSS with the DTW distance.}
\begin{tabular}{|c|c|c|c|c|c|c|}
\hline
Initial   & \multicolumn{6}{c|}{$h = 1$, $T_{best} = \emptyset$ and $\Theta_{best} = 0$}                      \\ \hline
Point & Prefix                       & Suffix                           & Split & $h$ & $\Theta_{best}$ & $T_{best}$ \\ \hline
$p_1$         & $\Theta(T[1, 1], T_q)=0.124$ & $\Theta(T[1, 5]^R, T_q^R)=0.150$ & Yes       &	2	& 0.150           & $T[1,5]$   \\ \hline
$p_2$         & $\Theta(T[2, 2], T_q)=0.236$ & $\Theta(T[2, 5]^R, T_q^R)=0.227$ & Yes       &	3	& 0.236           & $T[2,2]$   \\ \hline
$p_3$         & $\Theta(T[3, 3], T_q)=0.183$ & $\Theta(T[3, 5]^R, T_q^R)=0.215$ & No        &	3	& 0.236           & $T[2,2]$   \\ \hline
$p_4$         & $\Theta(T[3, 4], T_q)=0.236$ & $\Theta(T[4, 5]^R, T_q^R)=0.215$ & No        &	3	& 0.236           & $T[2,2]$   \\ \hline
$p_5$         & $\Theta(T[3, 5], T_q)=0.215$ & $\Theta(T[5, 5]^R, T_q^R)=0.152$ & No        &	3	& 0.236           & $T[2,2]$   \\ \hline
Output    & \multicolumn{6}{c|}{$T_{best} = T[2,2]$ with $\Theta_{best}=0.236$}                                        \\ \hline
  \end{tabular}
  \label{table:pss}
  \end{minipage}
\end{tabular}
\end{figure*}

The ExactS algorithm is costly since it explores $O(n^2)$ subtrajectories. The SizeS algorithm runs faster than ExactS since it explores about $O(\xi \cdot n)$ subtrajectories ($\xi << n$). Thus, an intuitive idea to push the efficiency further up is to explore fewer subtrajectories. In the following, we design a series of three approximate algorithms, which all share the idea of splitting a data trajectory into several subtrajectories and returning the one that is the most similar to the query trajectory. These algorithms differ from each other in using different heuristics for deciding where to split the data trajectory. With this splitting strategy, the number of subtrajectories that would be explored is bounded by $n$ and in practice, much smaller than $n$. We describe these algorithms as follows.

\smallskip\noindent\textbf{(1) Prefix-Suffix Search (PSS).} The PSS algorithm is a greedy one,
which maintains a variable $T_{best}$ storing the subtrajectory that is the most similar to the query trajectory found so far. Specifically, it scans the points of the data trajectory $T$ in the order of $p_1, p_2, ..., p_n$.
When it scans $p_i$, it computes the similarities between the two subtrajectories that would be formed if it splits $T$ at $p_i$, i.e., $T[h, i]$ and $T[i, n]$, and the query trajectory $T_q$, where $p_h$ is the point following the one, at which the last split was done if any and $p_h$ is the first point $p_1$ otherwise.
In particular, we replace the part of computing the similarity between the suffix $T[i, n]$ and the query trajectory with that between their reversed versions, denoted by $T[i, n]^R$ and $T_q^R$, respectively.
This is because (1) $\Theta(T[i, n]^R, T_q^R)$ could be computed incrementally based on $\Theta(T[i+1, n]^R, T_q^R)$ and
(2) $\Theta(T[i, n]^R, T_q^R)$ and $\Theta(T[i, n], T_q)$ are equal for some similarity measurements such as DTW and Frechet and positively correlated for others such as t2vec as we found via experiments.
If any of these two similarities are larger than the best-known similarity, it performs a split operation at $p_i$ and updates $T_{best}$ accordingly; otherwise, it continues to scan the next point $p_{i+1}$.
%
At the end, it returns $T_{best}$. The procedure of PSS is presented in Algorithm~\ref{alg:PSS}.

To illustrate, consider an example shown in Figure~\ref{fig:input}, where $T$ is a data trajectory with 5 points $p_{1:5}$ and $T_q$ is a query trajectory with 3 points $q_{1:3}$. Suppose that we measure the similarity between two trajectories using the ratio of 1 over the DTW distance between them. Consider the process of the PSS algorithm, which is depicted in Table~\ref{table:pss}. When it scans $p_1$, it considers two subtrajectories, namely $T[1,1]$ and $T[1,5]$, which have the similarities, i.e., $0.124$ and $0.150$, both larger than the best-known so far, i.e., 0. Therefore, it performs a split operation at $p_1$ and updates $h$, $\Theta_{best}$ and $T_{best}$ accordingly, as shown in the $3^{rd}$ row of the table. It continues to scan $p_2$ and considers two subtrajectories $T[2,2]$ and $T[2,5]$. Since their similairities are larger than the best-known one, it performs a split at $p_2$ and updates $h$, $\Theta_{best}$ and $T_{best}$. It then scans $p_3$, $p_4$, and $p_5$ and does not perform a split at any of them. Therefore, $h$, $\Theta_{best}$ and $T_{best}$ are kept unchanged. Finally, it returns $T_{best}$, i.e., $T[2,2]$, which has the similarity of $0.236$. Note that $T[2,4]$ has the largest similarity to $T_q$, which is $1/3 = 0.333$.

We analyze the time complexity of PSS as follows. When it scans a specific point $p_i$, the time costs include that of computing $\Theta(T[h, i], T_q)$ and also that of computing $\Theta(T[i, n]^R, T_q^R)$.
Consider the former part.
If $i = h$, it is $\Phi_{ini}$.
If $i \ge h + 1$, it is $\Phi_{inc}$
since $\Theta(T[h, i], T_q)$ could be computed based on $\Theta(T[h, i-1], T_q)$ incrementally.
Consider the latter part. It is simply $O(\Phi_{inc})$.
In conclusion, the time complexity of PSS is $O(n_1 \cdot (\Phi_{ini} + \Phi_{inc}) + (n-n_1) \cdot \Phi_{inc}) = O(n_1\cdot \Phi_{ini} + n\cdot \Phi_{inc})$, where $n_1$ is the number of points where splits are done.
For example, when DTW or Frechet is used, the time complexity of PSS is $O(n_1 \cdot m + n\cdot m) = O(n \cdot m)$ and when t2vec is used, it is $O(n_1 \cdot 1 + n \cdot 1) = O(n)$.


\begin{algorithm}[t!]
	\caption{Prefix-Suffix Search (PSS)}\label{alg:PSS}
		\footnotesize{
        \KwIn{
		A data trajectory $T$ and query trajectory $T_q$;}
		\KwOut{
		A subtrajectory of $T$ that is similar to $T_q$;}
		$T_{best} \leftarrow \emptyset$; $\Theta_{best} \leftarrow 0$;\\
		compute $\Theta(T[n,n]^R, T_q^R)$;\\
		compute $\Theta(T[n-1,n]^R, T_q^R)$, $\Theta(T[n-2,n]^R, T_q^R)$, ..., $\Theta(T[1,n]^R, T_q^R)$ incrementally;\\
		$h\leftarrow 1$;\\
		\ForAll{$1 \leq i \leq |T|$}{
		compute $\Theta(T[h, i], T_q)$ incrementally if possible;\\ 
		\If{$\max\{\Theta(T[h, i], T_q), \Theta(T[i, n]^R, T_q^R)\} > \Theta_{best}$}{
		$\Theta_{best} \leftarrow \max\{\Theta(T[h, i], T_q), \Theta(T[i, n]^R, T_q^R)\}$;\\
		\eIf{$\Theta(T[h, i], T_q) > \Theta(T[i, n]^R, T_q^R)$}
        {
		$T_{best} \leftarrow T[h, i]$;
        }
		{
		$T_{best} \leftarrow T[i, n]$;
		}
		$h\leftarrow i+1$;\\
		}
		}
		\textbf{return} $T_{best}$;
	}
\end{algorithm}

\smallskip\noindent\textbf{(2) Prefix-Only Search (POS).}
In PSS, when it scans a point $p_i$, it considers two subtrajectories, namely $T[h, i]$ and $T[i, n]$.
An alternative is to consider the prefix $T[h, i]$ only - one argument is that the suffix $T[i, n]$ might be destroyed when further splits are conducted.
A consequent benefit is that the time cost of computing $\Theta(T[i, n], T_q)$ would be saved. We call this algorithm the POS algorithm.
As could be verified, POS has the same time complexity as PSS though the former runs faster in practice.

\smallskip\noindent\textbf{(3) Prefix-Only Search with Delay (POS-D).}
POS performs a split operation whenever a prefix, which is better than the best subtrajectory known so far, is found. This looks a bit rush and may prevent a better subtrajectory to be formed by extending it with a few more points. Thus, we design a variant of POS, called \emph{Prefix-Only Search with Delay} (POS-D). Whenever a prefix is found to be more similar to the query trajectory than the best subtrajectory known so far, POS-D continues to scan $D$ more points and splits at one of these $D+1$ points, which has the corresponding prefix the most similar to the query trajectory. It could be verified that with this delay mechanism, the time complexity of the algorithm does not change though in practice, it would be slightly higher.

While these splitting-based algorithms including PSS, POS and POS-D, return reasonably good solutions in practice, they may return solutions that are arbitrarily worse than the best one in theory. We illustrate this in the technical report version~\cite{TR} due to the page limit.

%% file: rl_v2.tex
\section{Reinforcement Learning Based Algorithm}
\label{section:RLS}

A splitting-based algorithm has its effectiveness rely on the quality of the process of splitting a data trajectory. In order to find a solution of high quality, it requires to perform split operations at appropriate points such that some subtrajectories that are similar to a query trajectory are formed and then explored. The three splitting-based algorithms, namely PSS, POS and POS-D, mainly use some hand-crafted heuristics for making decisions on whether to perform a split operation at a specific point.
This process of splitting a trajectory into subtrajectories is a typical \emph{sequential decision making} process. Specifically, it scans the points sequentially and for each point, it makes a decision on whether or not to perform a split operation at the point.
In this paper, we propose to model this process as a \emph{Markov decision process} (MDP)~\cite{puterman2014markov} (Section~\ref{subsec:mdp-modelling}), adopt a \emph{deep-$Q$-network} (DQN)~\cite{mnih2013playing} for learning an optimal policy for the MDP (Section~\ref{subsec:dqn}), and then develop an algorithm called \emph{reinforcement learning based search} (RLS), which corresponds to a splitting-based algorithm that uses the learned policy for the process of splitting a data trajectory (Section~\ref{subsec:rls}) and an augmented version of RLS, called RLS-Skip, with better efficiency (Section~\ref{subsec:rls_skip})

\subsection{Trajectory Splitting as a MDP}
\label{subsec:mdp-modelling}
A MDP consists of four components, namely \emph{states}, \emph{actions}, \emph{transitions}, and \emph{rewards}, where (1) a state captures the \emph{environment} that is taken into account for decision making by an \emph{agent}; (2) an action is a possible decision that could be made by the agent; (3) a transition means that the state changes from one to another once an action is taken; and (4) a reward, which is associated with a transition, corresponds to some feedback indicating the quality of the action that causes the transition.
%
We model the process of splitting a data trajectory as a MDP as follows.

\smallskip\noindent\textbf{(1) States.} We denote a state by $s$. Suppose it is currently scanning point $p_t$. $p_h$ denotes the point following the one, at which the last split operation happens if any and $p_1$ otherwise. We define the state of the current environment as a triplet $(\Theta_{best}, \Theta_{pre}, \Theta_{suf})$, where $\Theta_{best}$ is the largest similarity between a subtrajectory found so far and the query trajectory $T_q$, $\Theta_{pre}$ is $\Theta(T[h, t], T_q)$ and $\Theta_{suf}$ is $\Theta(T[t, n]^R, T_q^R)$. As could be noticed, a state captures the information about the query trajectory, the data trajectory, the point at which the last split happens, and the point that is being scanned, etc. Note that the state space is a three-dimensional continuous one.

\smallskip\noindent\textbf{(2) Actions.} We denote an action by $a$. We define two actions, namely $a = 1$ and $a = 0$. The former means to perform a split operation at the point that is being scanned and the latter means to move on to scan the next point.

\smallskip\noindent\textbf{(3) Transitions.}
In the process of splitting a trajectory, given a current state $s$ and an action $a$ to take, the probability that we would observe a specific state $s'$ is unknown. We note that the method that we use for solving the MDP in this paper is a \emph{model-free} one and could solve the MDP problem even with its transition information unknown.

\smallskip\noindent\textbf{(4) Rewards.}
We denote a reward by $r$.
We define the reward associated with the transition from state $s$ to state $s'$ after action $a$ is taken as $(s'.\Theta_{best} - s.\Theta_{best})$, where the $s'.\Theta_{best}$ is the first component of state $s'$ and $s.\Theta_{best}$ is the first component of state $s$. With this reward definition, the goal of the MDP problem, which is to maximize the accumulative rewards, is consistent with that of the process of splitting a data trajectory, which is to form a subtrajectory with the greatest possible similarity to the query trajectory. To see this, consider that the process goes through a sequence of states $s_1, s_2, ..., s_N$ and ends at $s_N$. Let $r_1, r_2, ..., r_{N-1}$ denote the rewards received at these states except for the termination state $s_N$. Then, when the future rewards are not discounted, we have $$\Sigma_{t}r_t = \Sigma_{t}(s_t.\Theta_{best} - s_{t-1}.\Theta_{best}) = s_N.\Theta_{best} - s_1.\Theta_{best}$$ where $s_N.\Theta_{best}$ corresponds to the similarity between the best subtrajectory found and the query trajectory $T_q$ and $s_1.\Theta_{best}$ corresponds to the best known similarity at the beginning, i.e., 0. Therefore, maximizing the accumulative rewards is equivalent to maximizing the similarity between the subtrajectory to be found and $T_q$ in this case.

\subsection{Deep-$Q$-Network (DQN) Learning}
\label{subsec:dqn}
The core problem of a MDP is to find an optimal \emph{policy} for the agent, which corresponds to a function $\pi$ that specifies the action that the agent should choose when at a specific state so as to maximize the accumulative rewards.
One type of methods that are commonly used is those \emph{value-based} methods~\cite{sutton2018reinforcement,mnih2013playing}. The major idea is as follows. First, it defines an optimal action-value function $Q^*(s, a)$ (or $Q$ function), which represents the maximum amount of expected accumulative rewards it would receive by following any policy after seeing the state $s$ and taking the action $a$.
Second, it estimates $Q^*(s, a)$ using some methods such as $Q$-learning~\cite{watkins1992q} and deep-$Q$-network (DQN)~\cite{mnih2013playing}.
Third, it returns the policy, which always chooses for a given state $s$ the action $a$ that maximizes $Q^*(s, a)$.

\begin{algorithm}[t]
	\caption{Deep-$Q$-Network (DQN) Learning with Experience Replay}\label{alg:dqn}
    \LinesNumbered
	\footnotesize{
		\KwIn{A database $\mathcal{D}$ of data trajectories and a set of $\mathcal{D}_q$ of query trajectories;}
        \KwOut{Learned action-value function $Q(s,a;\theta)$;}
		initialize the reply memory $M$;\\
		initialize the main network $Q(s, a; \theta)$ with random weights $\theta$;\\
		initialize the target network $\hat{Q}(s, a; \theta^-)$ with weights $\theta^- = \theta$;\\
			\For{$episode = 1, 2, 3,...$}{
			sample a data and query trajectory $T$, $T_q$;\\
			$h \leftarrow 1$;\\
			$\Theta_{best} \leftarrow 0$; $\Theta_{pre} \leftarrow \Theta(T[h,h], T_q)$; $\Theta_{suf} \leftarrow \Theta(T[h,n]^R, T_q^R)$;\\
			observe the first state $s_1 = (\Theta_{best}, \Theta_{pre}, \Theta_{suf})$;\\
			\For{each step $1 \leq t \leq |T|$}{
			select a random action $a_t$ with probability $\epsilon$ and select action $a_t = \arg\max_{a}Q(s_t,a;\theta)$ with probablity ($1-\epsilon$);\\
			\If{$a_t = 1$}{
			$h\leftarrow t+1$;}
			$\Theta_{best} \leftarrow \max\{s_t.\Theta_{best}, s_t.\Theta_{pre}, s_t.\Theta_{suf}\}$;\\
			\If{$t=|T|$}{
			\textbf{break};}
			$\Theta_{pre} \leftarrow \Theta(T[h,t+1], T_q)$; $\Theta_{suf} \leftarrow \Theta(T[t+1,n]^R, T_q^R)$;\\
			observe the next state $s_{t+1} = (\Theta_{best}, \Theta_{pre}, \Theta_{suf})$;\\
			observe the reward $r_t = s_{t+1}.\Theta_{best} - s_t.\Theta_{best}$;\\
			store the experience $(s_t,a_t,r_t,s_{t+1})$ in the replay memory $\mathcal{M}$;\\
			sample a random minibatch of experiences from $\mathcal{M}$ uniformly;\\
		    perform a gradient descent step on the loss as computed by Equatioin~(\ref{equ:dqn-loss}) wrt $\theta$;\\
			}
			copy the main network $Q(s,a;\theta)$ to $\hat{Q}(s, a; \theta^-)$;
}}
\end{algorithm}

\begin{figure}[t]
	\centering
	\includegraphics[width=8.5cm]{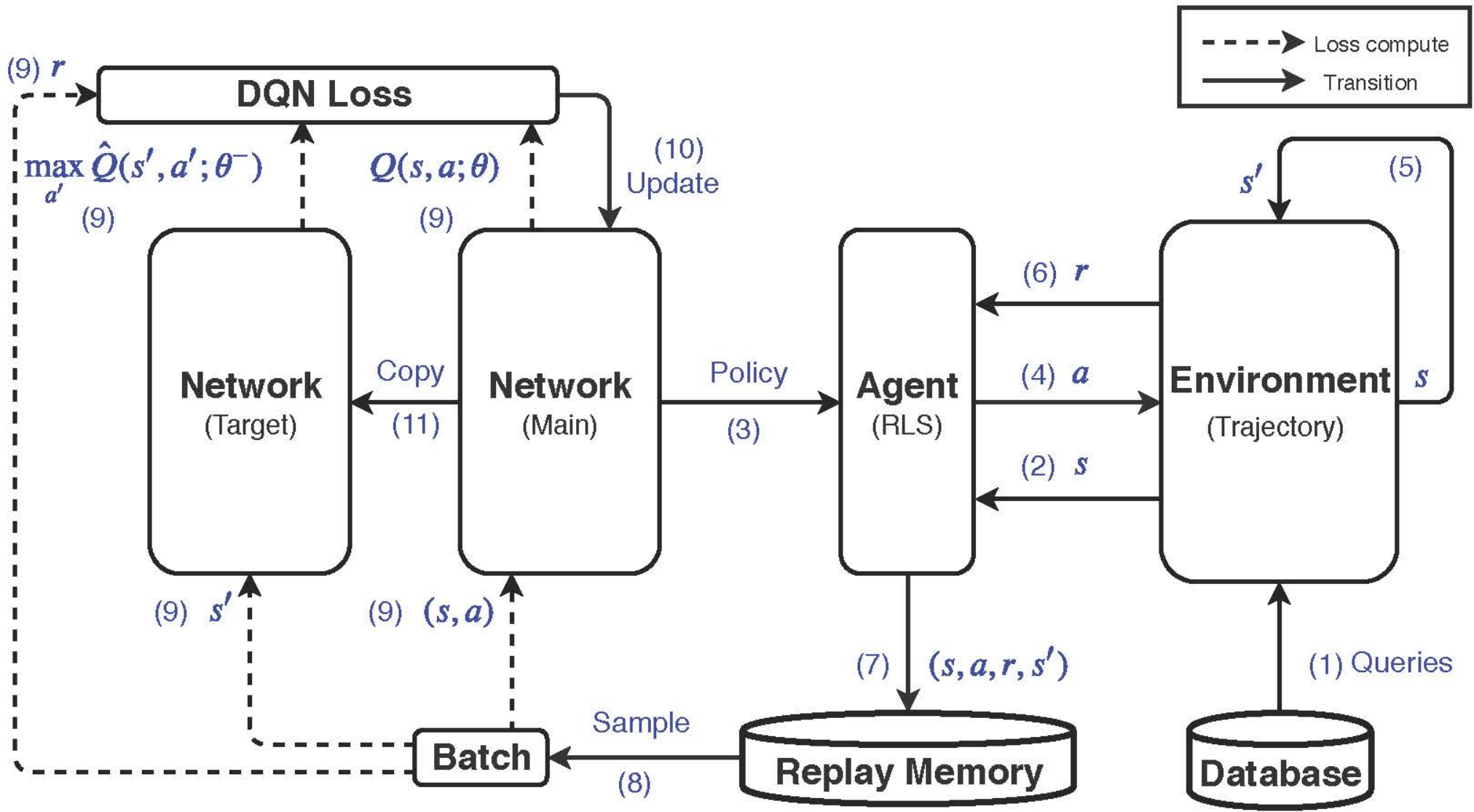}\\
	\caption{Deep $Q$ learning with experience replay}
	\label{fig:dqn}
\end{figure}

In our MDP, the state space is a three dimensional continuous one, and thus we adopt the DQN method. Specifically, we use the \emph{deep $Q$ learning with replay memory}~\cite{mnih2015human} for learning the $Q$ functions. This method maintains two neural networks. One is called the \emph{main network} $Q(s,a;\theta)$, which is used to estimate the $Q$ function. The other is called the \emph{target network} $\hat{Q}(s,a;\theta^-)$, which is used to compute some form of loss for training the main network. Besides, it maintains a fixed-size pool called \emph{replay memory}, which contains the latest transitions that are sampled uniformly and used for training the main network. The intuition is to avoid the correlation among consecutive transitions. The detailed procedure of DQN for our MDP is presented in Algorithm~\ref{alg:dqn}, which we go through as follows. We maintain a database $\mathcal{D}$ of data trajectories and a set of $\mathcal{D}_q$ of query trajectories. It first initializes the reply memory $\mathcal{M}$ with some capacity, the main network $Q(s, a; \theta)$ with random weights, and the target network $\hat{Q}(s,a;\theta^-)$ by copying $Q(s,a;\theta)$ (Lines 1 - 3). Then, it involves a sequence of many episodes. For each episode, it samples a data trajectory $T$ from $\mathcal{D}$ and a query trajectory $T_q$ from $\mathcal{D}_q$, both uniformly (Lines 4 - 5). It initializes a variable $h$ such that $p_h$ corresponds to the point following the one, at which the last split operation is performed if any and $p_1$ otherwise (Line 6). It also initializes the state $s_1$ (Lines 7 - 8). Then, it proceeds with $|T|$ time steps. At the $t^{th}$ time step, it scans point $p_t$ and selects an action using the $\epsilon$-greedy strategy based on the main network, i.e., it performs a random action $a_t$ with the probability $\epsilon$ ($0 < \epsilon < 1$) and $a_t = \arg\max_{a}Q(s_t,a;\theta)$ with the probability $(1-\epsilon)$ (Lines 9 - 10).
If $a_t = 1$, it splits the trajectory at point $p_t$ and updates $h$ to be $t+1$ (Lines 11 - 13).
It then updates $\Theta_{best}$ if possible (Line 14).
If the current point being scanned is the last point $p_n$, it terminates (Lines 15 - 17).
Otherwise, it observes a new state $s_{t+1}$ and the reward $r_t$ (Lines 18 - 20).
It then stores the experience $(s_t, a_t, r_t, s_{t+1})$ in the reply memory, samples a minibatch of experiences, and uses it to perform a gradient descent step for updating $\theta$ wrt a loss function (Lines 21 - 23). The loss function for one experience $(s, a, r, s')$ is as follows.
\begin{equation}
L(\theta) = (y - Q(s, a; \theta))^2
\label{equ:dqn-loss}
\end{equation}
where $y$ is equal to $r$ if $s'$ is a termination step and $r + \gamma\cdot \max_{a'}\hat{Q}(s', a'; \theta^-)$ otherwise.
Finally, it updates the target network $\hat{Q}(s, a;\theta^-)$ with the main network $Q(s,a;\theta)$ at the end of each episode (Line 25).
A graphical illustration of the method is shown in Figure~\ref{fig:dqn}.

\subsection{Reinforcement Learning based Search Algorithm (RLS)}
\label{subsec:rls}
Once we have estimated the $Q$ functions $Q(s, a; \theta)$ via the deep $Q$ learning with experience replay, we use the policy, which always takes for a given state $s$ the action that maximizes $Q(s,a; \theta)$, for the process of splitting a data trajectory. Among all subtrajectories that are formed as a result of the process, we return the one with the greatest similarity to the query trajectory $T_q$. We call this algorithm \emph{reinforcement learning based search} (RLS). Essentially, it is the same as PSS except that it uses a policy learned via DQN instead of human-crafted heuristics for making decisions on how to split a data trajectory.

RLS has the same time complexity as PSS
since
both RLS and PSS make decisions based on the similarities of the subtrajectories that are being considered when scanning a point and the best-known similarity:
(1) RLS constructs a state involving them and goes through the main network of DQN with the state information, which is $O(1)$ given that the network is small-size (e.g., a few layers); and
(2) PSS simply conduct some comparisons among the similarities, which is also $O(1)$.
In terms of effectiveness,
RLS provides consistently better solutions than PSS as well as POS and POS-D,
as will be shown in the empirical studies,
and the reason is possibly that RLS is based on a learned policy,
which makes decision more intelligent than simple heuristics that are human-crafted.

\subsection{Reinforcement Learning based Search with Skipping (RLS-Skip)}
\label{subsec:rls_skip}

\begin{table*}[htb]
\centering
\scriptsize
\caption{Illustration of RLS-Skip with the DTW distance.}
\label{table:rls_skip}
\begin{tabular}{|c|c|c|c|c|c|}
\hline
Initial      & \multicolumn{5}{c|}{$h = 1$, $T_{best} = \emptyset$ and  $\Theta_{best}=0$}                                                                       \\ \hline
Point   & State & Action & $h$ & $\Theta_{best}$ & $T_{best}$ \\ \hline
$p_1$           & $s_1=(\Theta_{best}=0, \Theta_{pre=T[1,1]}=0.124, \Theta_{suf=T[1,5]}=0.150)$     & $a_1 = \arg\max_{a}Q(s_1,a;\theta)=1: split$ & $2$ & $0.150$  & $T[1,5]$   \\ \hline
$p_2$           & $s_2=(\Theta_{best}=0.150, \Theta_{pre=T[2,2]}=0.236, \Theta_{suf=T[2,5]}=0.227)$ & $a_2 = \arg\max_{a}Q(s_2,a;\theta)=2: skip$ & $2$ & $0.236$  & $T[2,2]$   \\ \hline
$p_3$ (skipped) & -         & -        & -     & -  & -   \\ \hline
$p_4$           & $s_3=(\Theta_{best}=0.236, \Theta_{pre=T[2,4]}=0.333, \Theta_{suf=T[4,5]}=0.215)$ & $a_3 = \arg\max_{a}Q(s_3,a;\theta)=1: split$ & $5$ & $0.333$  & $T[2,4]$   \\ \hline
$p_5$           & $s_4=(\Theta_{best}=0.333, \Theta_{pre=T[5,5]}=0.152, \Theta_{suf=T[5,5]}=0.152)$ & $a_4 = \arg\max_{a}Q(s_4,a;\theta)=0: no-split$ & $5$ & $0.333$      & $T[2,4]$   \\ \hline
Output      & \multicolumn{5}{c|}{$T_{best} = T[2,4]$ with $\Theta_{best}=0.333$}                                                                       \\ \hline
\end{tabular}
\end{table*}

In the RLS algorithm, each point is considered as a candidate for performing a split operation. While this helps to attain a reasonably large space of subtrajectories for exploration and hence achieving good effectiveness, it is somehow conservative and incurs some cost of decision marking for \emph{each} point. An alternative is to go a bit more optimistic and \emph{skip} some points from being considered as places for split operations. The benefit would be immediate, i.e., the cost of making decisions at these points is saved. Motivated by this, we propose to augment the MDP that is used by RLS by introducing $k$ more actions (apart from two existing ones: scanning the next point and performing a split operation), namely skipping $1$ point, skipping $2$ points, ..., skipping $k$ points. Here, $k$ is a hyperparameter, and by skipping $j$ points ($j = 1, 2, ..., k$), it means to skip points $p_{i+1}, p_{i+2}, ..., p_{i+j}$ and scan point $p_{i+j+1}$ next, where $p_i$ is the point that is being scanned. All other components of the MDP are kept the same as that for RLS. Note that when $k = 0$, this MDP reduces to the original one for RLS. We call the algorithm based on this augmented MDP \emph{RLS-Skip}.

To illustrate, consider again the example shown in Figure~\ref{fig:input}. Suppose that it has learned a policy using the DQN method, which is captured by the main network $Q(s, a; \theta)$. The process of RLS-Skip is depicted in Table~\ref{table:rls_skip}. Suppose the parameter $k$ is equal to 1, which implies that there are three possible actions 0 (no split), 1 (split), and 2 (no split and skip of 1 point).
In addition, we write $\Theta_{pre} = \Theta(T[i,j], T_q)$ as $\Theta_{pre=T[i,j]}$ and
$\Theta_{suf} = \Theta(T[i,j], T_q)$ as $\Theta_{suf=T[i,j]}$ for simplicity. At the very beginning, it initializes $h$, $\Theta_{best}$ and $T_{best}$. It then scans point $p_1$, observes the first state $s_1$ as $(\Theta_{best} = 0, \Theta_{pre = T[1,1]} = 0.124, \Theta_{suf=T[1,5]} = 0.150)$ and finds the action $a_1 = \arg\max_{a}Q(s_1,a;\theta) = 1$, meaning to perform a split operation at $p_1$. It then updates $h$, $\Theta_{best}$, and $T_{best}$ as shown in the $3^{rd}$ row of the table. It continues to scan point $p_2$, observes the second state $s_2$ as $(\Theta_{best} = 0.150, \Theta_{pre = T[2,2]} = 0.236, \Theta_{suf=T[2,5]} = 0.227)$ and finds the action $a_2 = \arg\max_{a}Q(s_2,a;\theta) = 2$, meaning to skip the next 1 point, i.e., $p_3$. It keeps $h$ unchanged (since no splits are done) but updates $\Theta_{best}$ and $T_{best}$ to be $0.236$ and $T[2,2]$, respectively, since $T[2,2]$ is the subtrajectory with the largest similarity among all subtrajectories that have been considered. As a result of the skipping, it scans point $p_4$ next and proceeds similarly. It performs a split operation when scanning point $p_4$ and terminates after scanning point $p_5$. At the end, it returns $T[2,4]$, which has the similarity of $0.333$. 

While the cost of making decisions at those points that are skipped (i.e., that of going through the main network of the DQN) could be saved in RLS-Skip, the cost of constructing the states at those points that are not skipped would be more or less that of constructing the states at \emph{all} points since the state at a point involves some similarities, which are computed incrementally based on the similarities computed at those points before the point. Thus, by applying the skipping strategy alone would not help much in reducing the time cost since the cost of maintaining the states dominates that of making decisions. To fully unleash the power of the skipping strategy, we propose to ignore those points that have been skipped when maintaining the states. That is, to maintain the state $(\Theta_{best}, \Theta_{pre}, \Theta_{suf})$ at a point $p_i$, we compute $\Theta_{best}$ and $\Theta_{suf}$ in the same way as we do in RLS and $\Theta_{pre}$ as the similarity between the query trajectory and the subtrajectory consisting of those points that are before $p_i$ and \emph{have not been skipped}. Here, the prefix subtrajectory corresponds to a \emph{simplification} of that used in RLS~\cite{long2013direction}. While RLS-Skip has the same worse-case time complexity as RLS, e.g., it reduces to RLS when no skipping operations happen, the cost of maintaining the states for RLS-Skip would be much smaller. As shown in our empirical studies, RLS-Skip runs significantly faster than RLS as well as PSS, POS and POS-D.
In addition,
RLS-Skip and RLS do not provide theoretical guarantees on the approximation quality due to their learning nature. The proofs can be found in the appendix of the technical report~\cite{TR}. Nevertheless, they work quite well in practice (e.g., RLS has the approximation ratio smaller than 1.1 for all similarity measurements and on all datasets (Figure~\ref{effect_result})). In addition, the problem instances that we constructed for proving the negative results in fact rarely happen in practice, which are confirmed by the effectiveness results on real datasets.

%% file: effectiveness.tex
\section{EXPERIMENTS}
\label{experiment}

We present the experimental set-up in Section~\ref{exper_setup} and then the experimental results in Section~\ref{effectiveness}.

\subsection{Experimental Setup}
\label{exper_setup}

\smallskip
\noindent\textbf{Dataset.} Our experiments are conducted on three real-world trajectory datasets. The first dataset, denoted by Porto, is collected from the city of Porto~\footnote{https://www.kaggle.com/c/pkdd-15-predict-taxi-service-trajectory-i/data}, Portugal, which consists around 1.7 million taxi trajectories over 18 months with a sampling interval of 15 seconds and a mean length around 60. 
The second dataset, denoted by Harbin, involves around 1.2 million taxi trajectories collected from 13,000 taxis over 8 months in Harbin, China with non-uniform sampling rates and a mean length around 120.
The third dataset, denoted by Sports, involves around 0.2 million soccer player and ball trajectories collected from STATS Sports~\footnote{https://www.stats.com/artificial-intelligence (STATS, copyright 2019)} with a uniform sampling rate of 10 times per second and a mean length around 170.

\smallskip
\noindent\textbf{Parameter Setting.} For training t2vec model, we follow the original paper~\cite{li2018deep} by excluding those trajectories that are short and use their parameter settings.
For SizeS, we use the setting $\xi = 5$ (with the results of its effect shown later on).
For POS-D, we vary the parameter $D$ from 4 to 7, and since the results are similar, we use the setting $D = 5$.
For the neural networks involved in the RL-based algorithms, i.e., RLS and RLS-Skip, we use a feedforward neural network with 2 layers. In the first layer, we use the ReLu function with 20 neurons, and in the second layer, we use the sigmoid function with $2+k$ neurons as the output corresponding to different actions, where for RLS we use $k=0$ and for RLS-Skip, we use $k = 3$ by default.
In the training process, the size of replay memory $\mathcal{M}$ is set at 2000. We train our model on 25k random trajectory pairs, using Adam stochastic gradient descent with an initial learning rate of 0.001. The minimal $\epsilon$ is set at 0.05 with decay 0.99 for the $\epsilon$-greedy strategy, and the reward discount rate $\gamma$ is set at 0.95.

\smallskip
\noindent\textbf{Compared Methods.} We compare RL-based Search (RLS), RL-based Search with skipping (RLS-Skip) and the proposed non-learning based algorithms (Section~\ref{sec:non-learning}), namely ExactS, SizeS, PSS, POS, and POS-D.
For RLS and RLS-Skip, when t2vec is adopted, we ignore the $\Theta_{suf}$ component of a state based on empirical findings.

In addition, we consider three competitor methods, namely UCR~\cite{mueen2016extracting,rakthanmanon2012searching,rakthanmanon2013addressing}, Spring~\cite{sakurai2007stream},
and Random-S.
UCR was originally developed for searching subsequences of a time series, which are the most similar to a query time series and the similarity is based on the DTW distance. UCR enumerates all subsequences that are of the same length of the query time series and employs a rich set of techniques for pruning many of the subsequences.
We adapt UCR for our similar subtrajectory search problem (details of adaptions are provided in the appendix of the technical report version~\cite{TR}).
We note that UCR only works for DTW, but not for Frechet or t2vec.
Spring is an existing algorithm for searching a subsequence of a time series, which is the most similar to a query time series. It is designed for DTW.
Random-S randomly samples a certain number of subtrajectories of data trajectory and among them, returns the one with the highest similarity to the query trajectory. Since these methods are either not general for all similarity measurements (e.g., UCR) or involve some parameter that is difficult to set (e.g., Random-S with a parameter of sample size), we compare these two competitor methods with our RLS-Skip algorithm only in terms of effectiveness and efficiency.
%

%
%
\begin{figure}[t]
	\centering
	\begin{tabular}{c c c}
		\centering
		\begin{minipage}{2.6cm}
			\includegraphics[width=2.8cm]{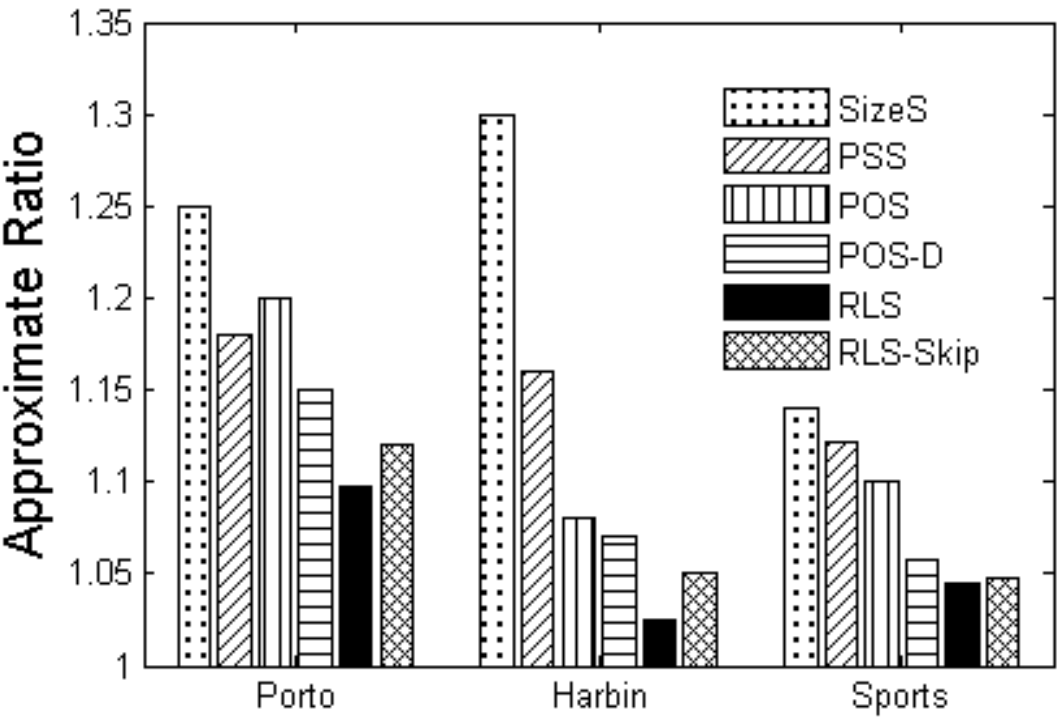}
		\end{minipage}
		&
		\begin{minipage}{2.6cm}
			\includegraphics[width=2.8cm]{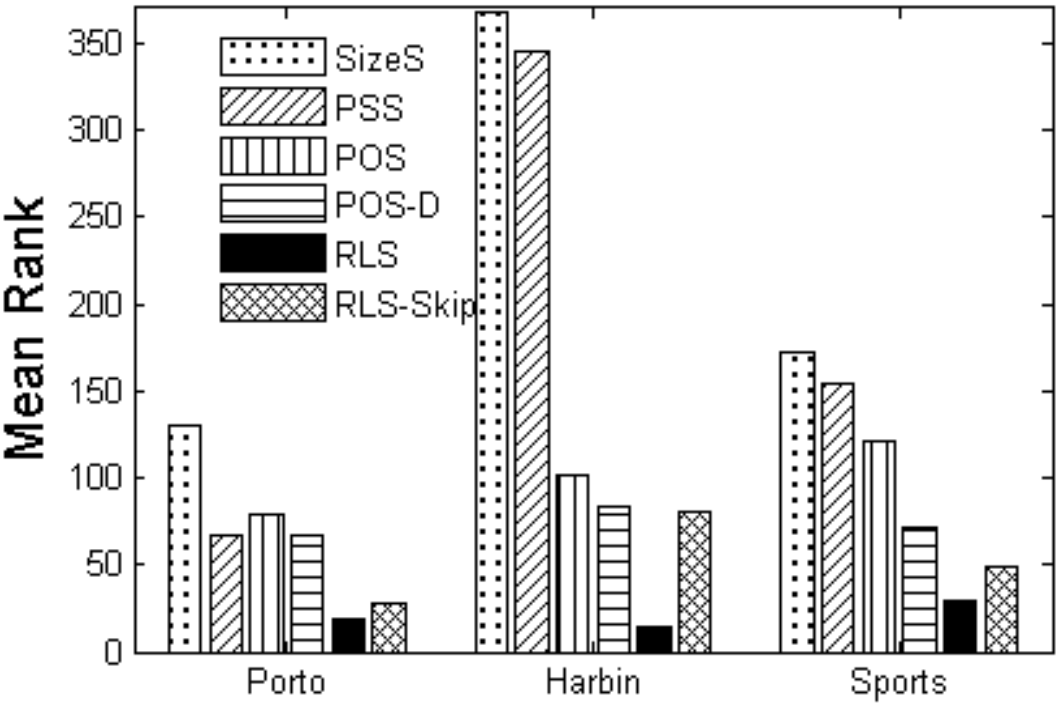}
		\end{minipage}
		&
		\begin{minipage}{2.6cm}
			\includegraphics[width=2.8cm]{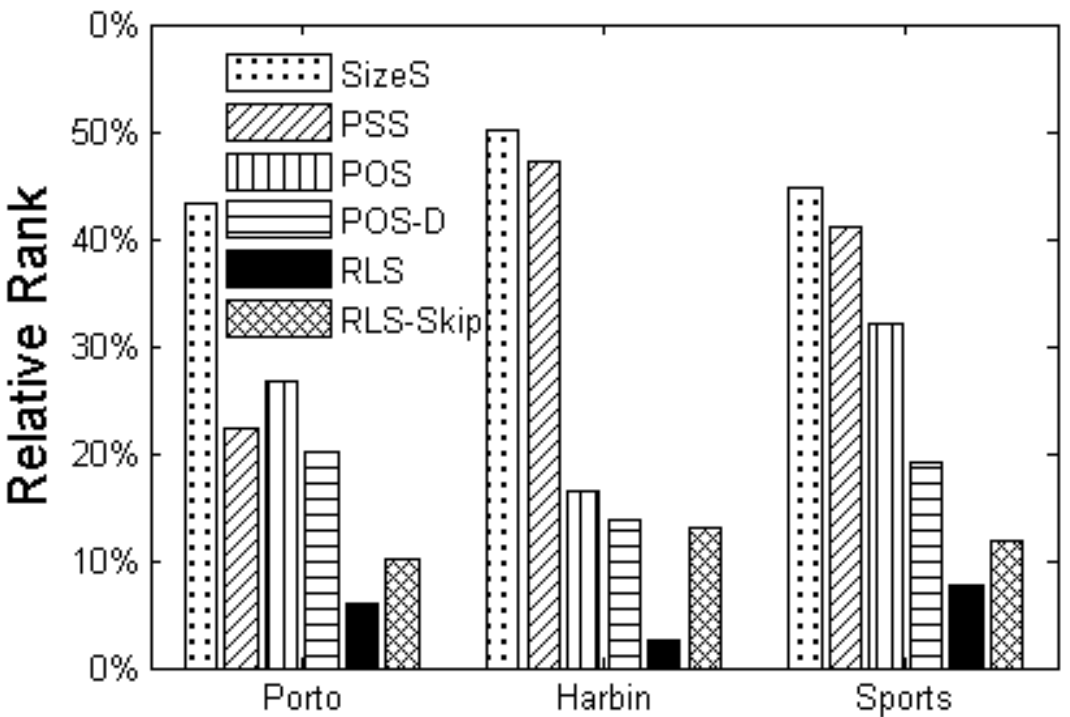}
		\end{minipage}
		\\
		\scriptsize{(a) AR (t2vec)}
		&
		\scriptsize{(b) MR (t2vec)}
		&
		\scriptsize{(c) RR (t2vec)}
		\\
		\begin{minipage}{2.6cm}
			\includegraphics[width=2.8cm]{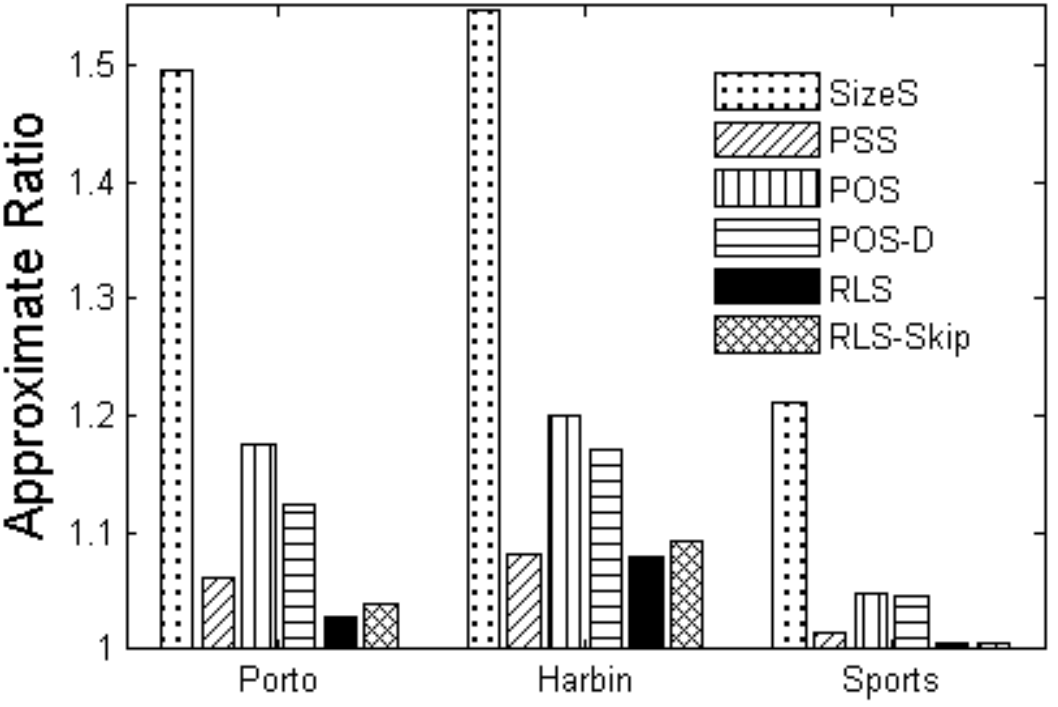}
		\end{minipage}
		&
		\begin{minipage}{2.6cm}
			\includegraphics[width=2.8cm]{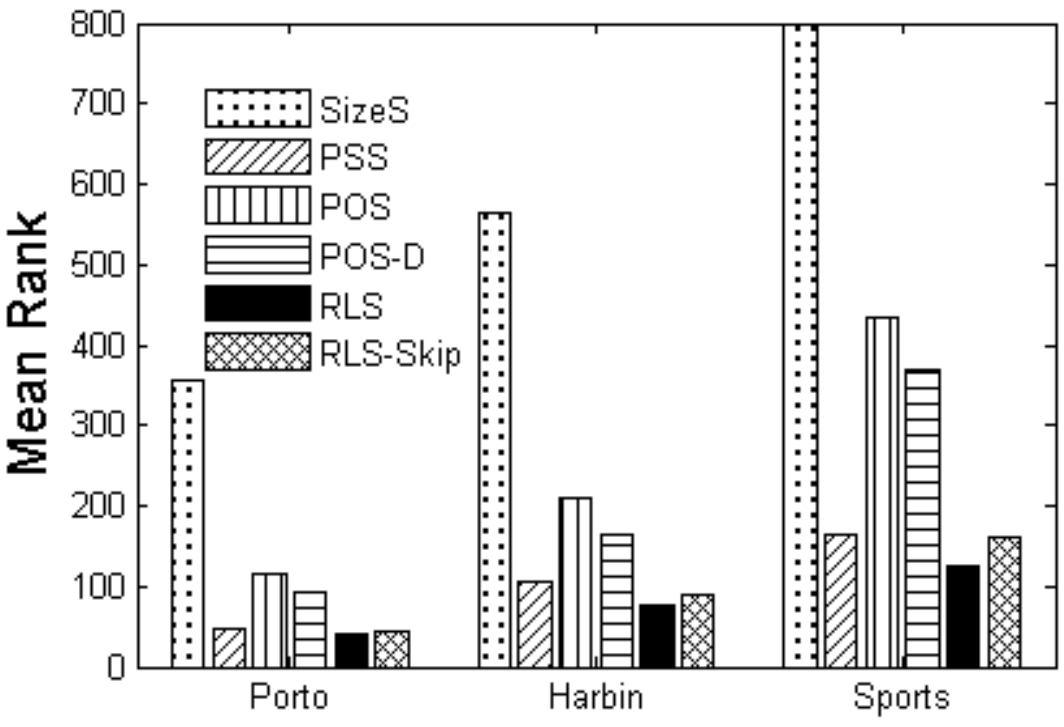}
		\end{minipage}
		&
		\begin{minipage}{2.6cm}
			\includegraphics[width=2.8cm]{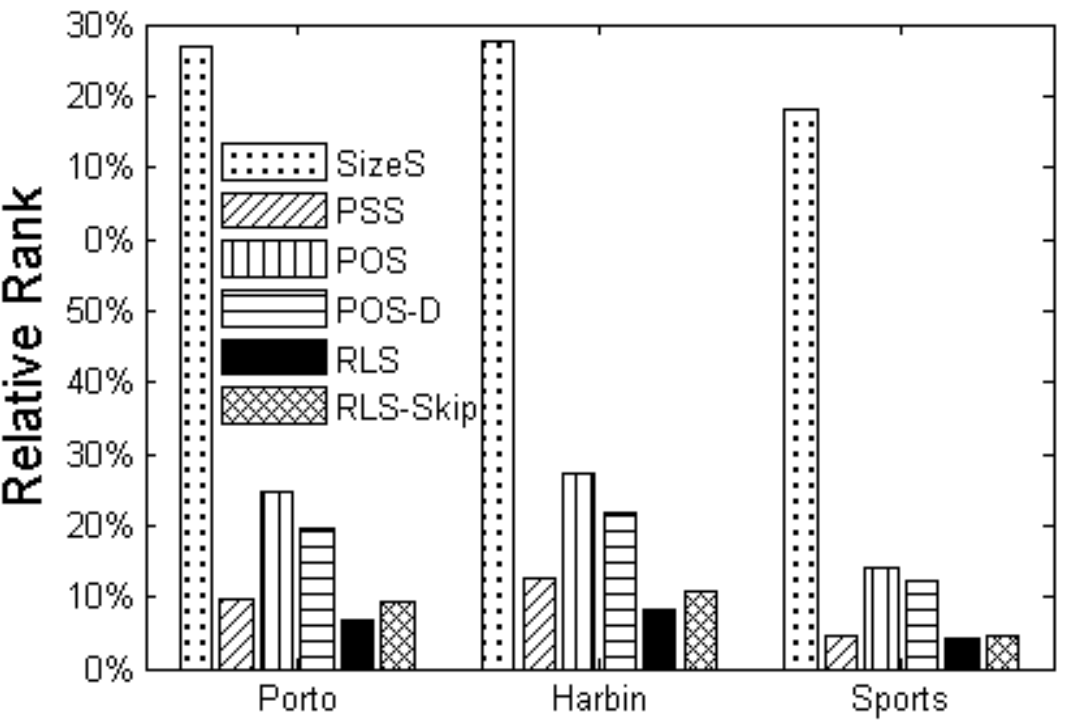}
		\end{minipage}
		\\
		\scriptsize{(d) AR (DTW)}
		&
		\scriptsize{(e) MR (DTW)}
		&
		\scriptsize{(f) RR (DTW)}
		\\
		\begin{minipage}{2.6cm}
			\includegraphics[width=2.8cm]{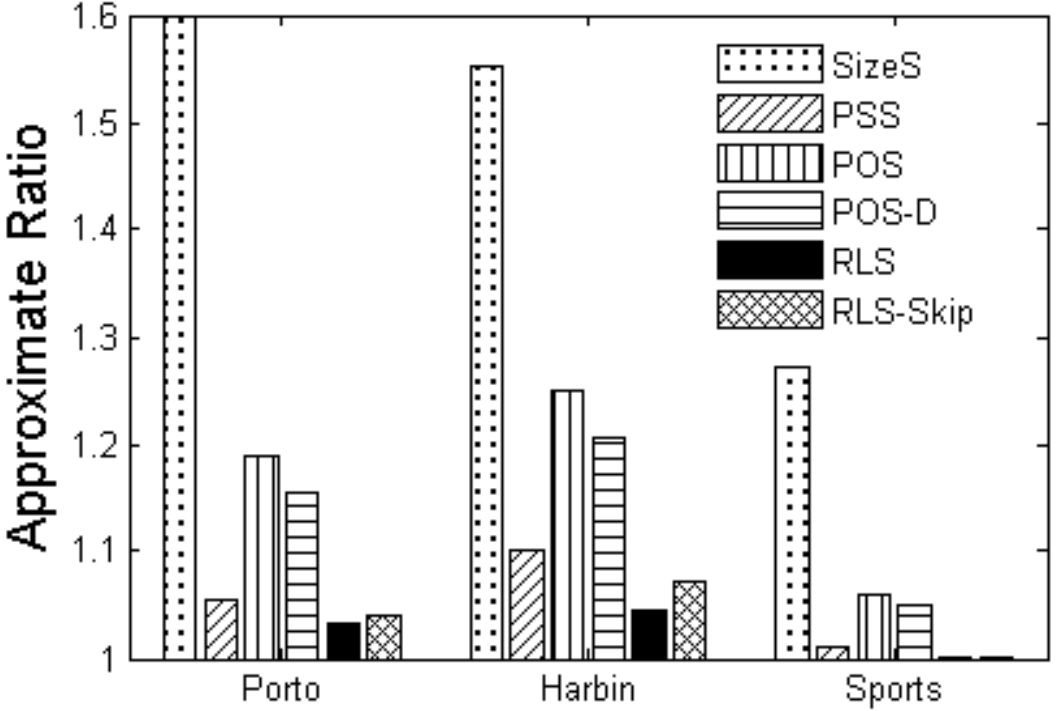}
		\end{minipage}
		&
		\begin{minipage}{2.6cm}
			\includegraphics[width=2.8cm]{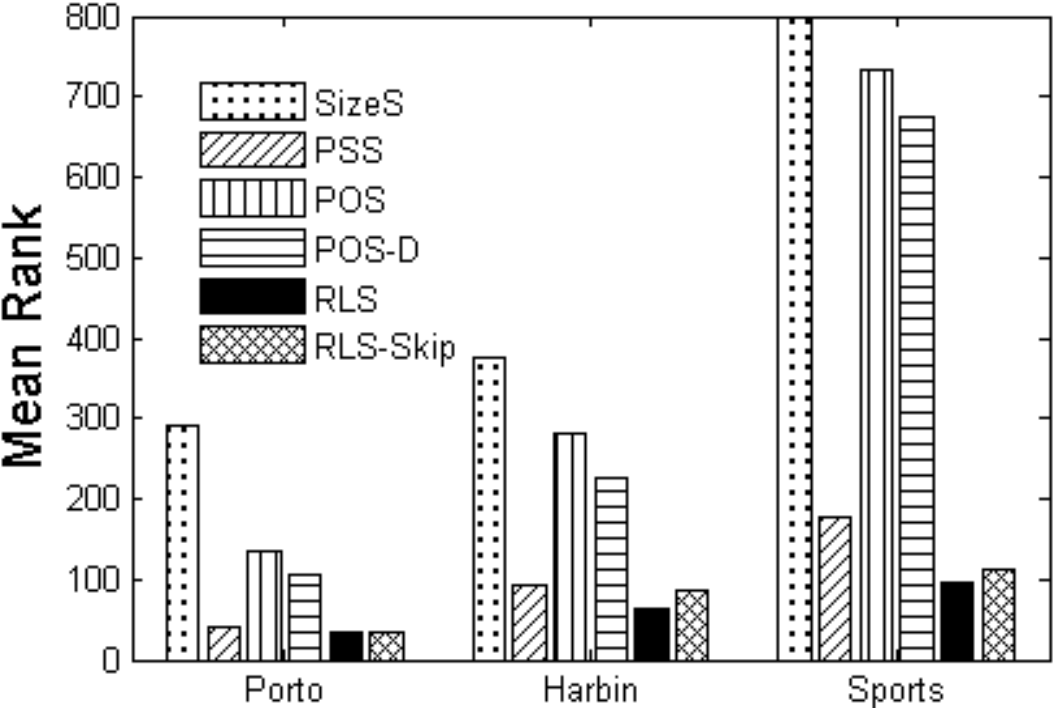}
		\end{minipage}
		&
		\begin{minipage}{2.6cm}
			\includegraphics[width=2.8cm]{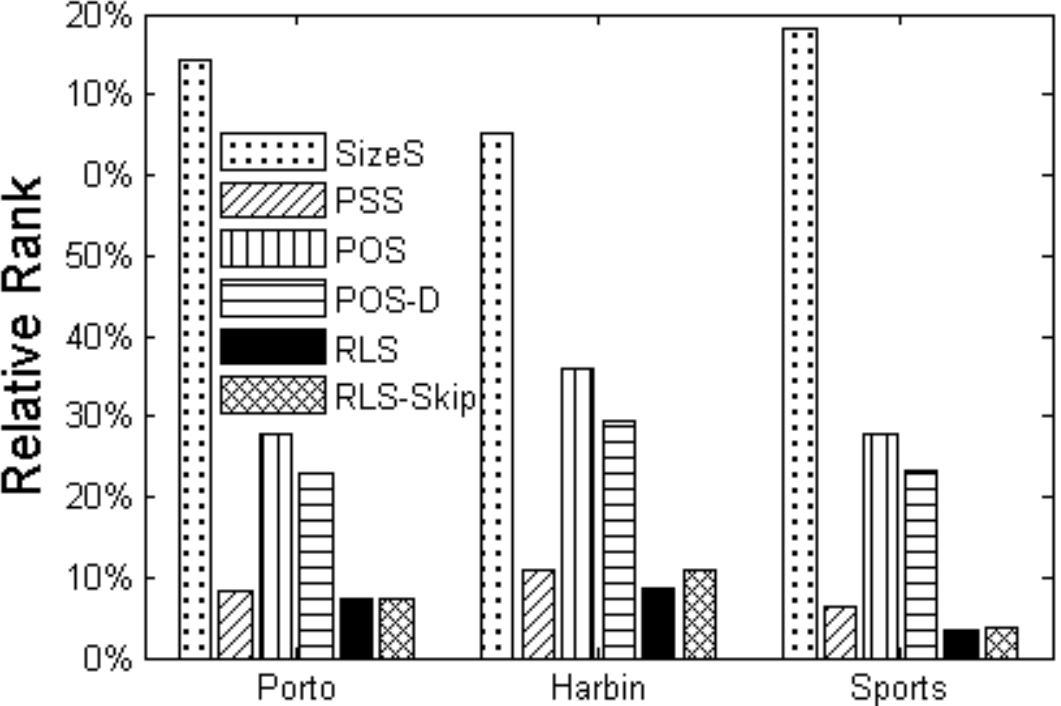}
		\end{minipage}
		\\
		\scriptsize{(g) AR (Frechet)}
		&
		\scriptsize{(h) MR (Frechet)}
		&
		\scriptsize{(i) RR (Frechet)}
	\end{tabular}
	\caption{Effectiveness for t2vec (a)-(c), DTW (d)-(f) and Frechet (g)-(i).}
	\label{effect_result}
\end{figure}
\begin{figure}[t]
	\centering
	\begin{tabular}{c c c}
		\begin{minipage}{2.6cm}
			\includegraphics[width=2.85cm]{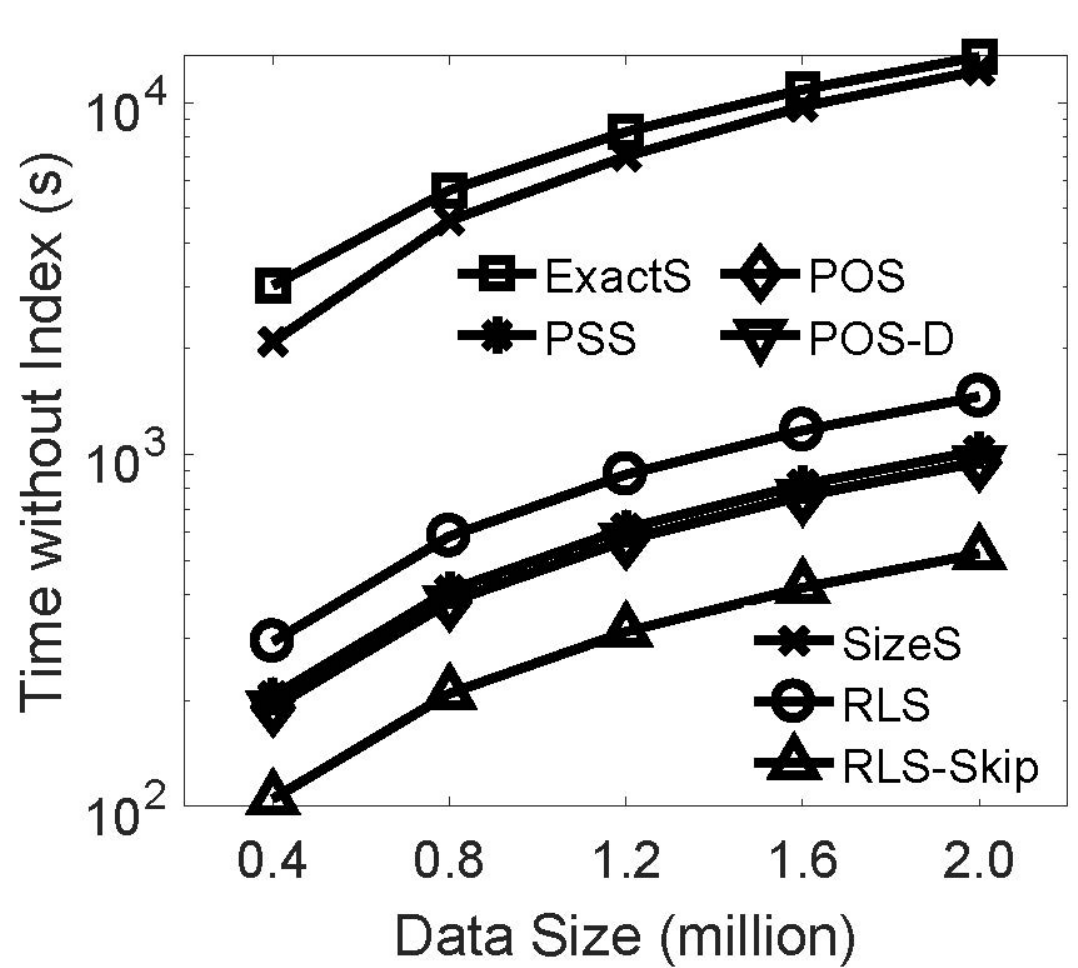}
		\end{minipage}
		&
		\begin{minipage}{2.6cm}
			\includegraphics[width=2.85cm]{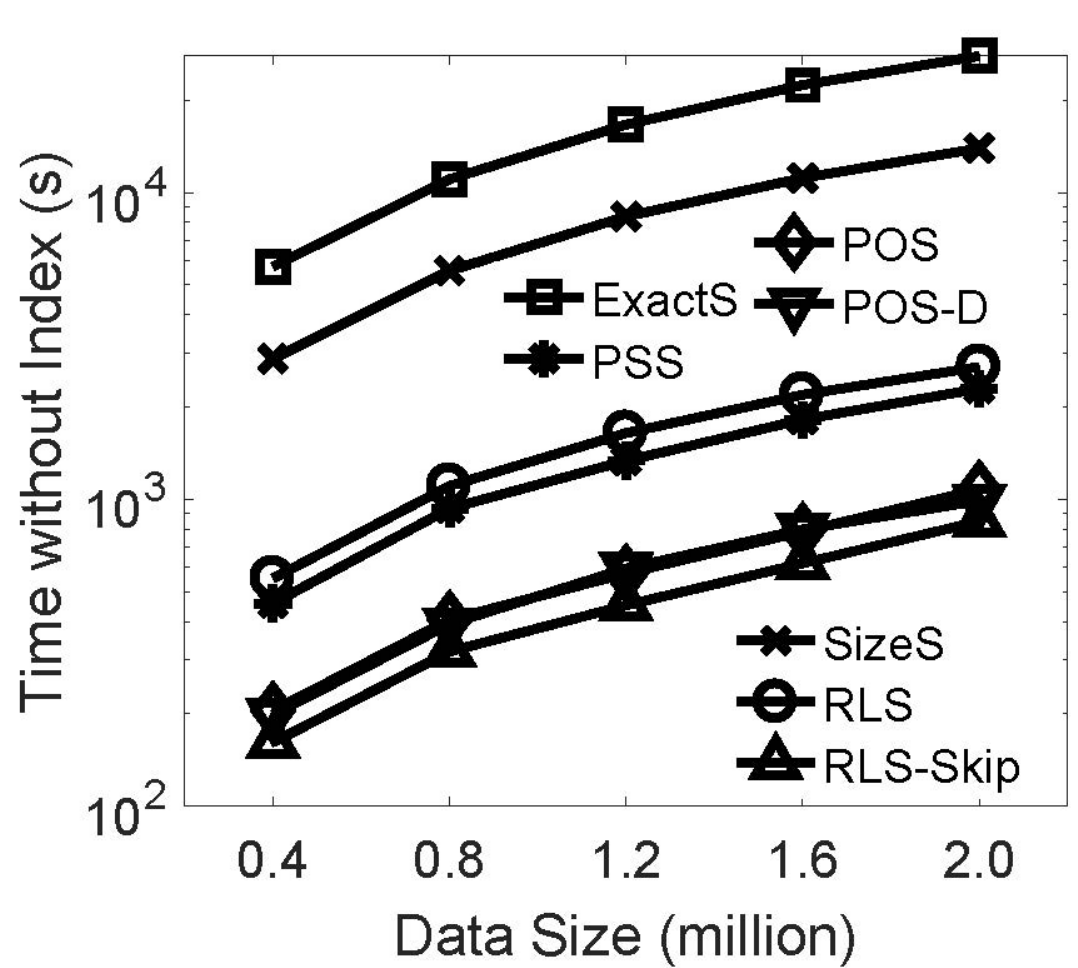}
		\end{minipage}
		&
		\begin{minipage}{2.6cm}
			\includegraphics[width=2.85cm]{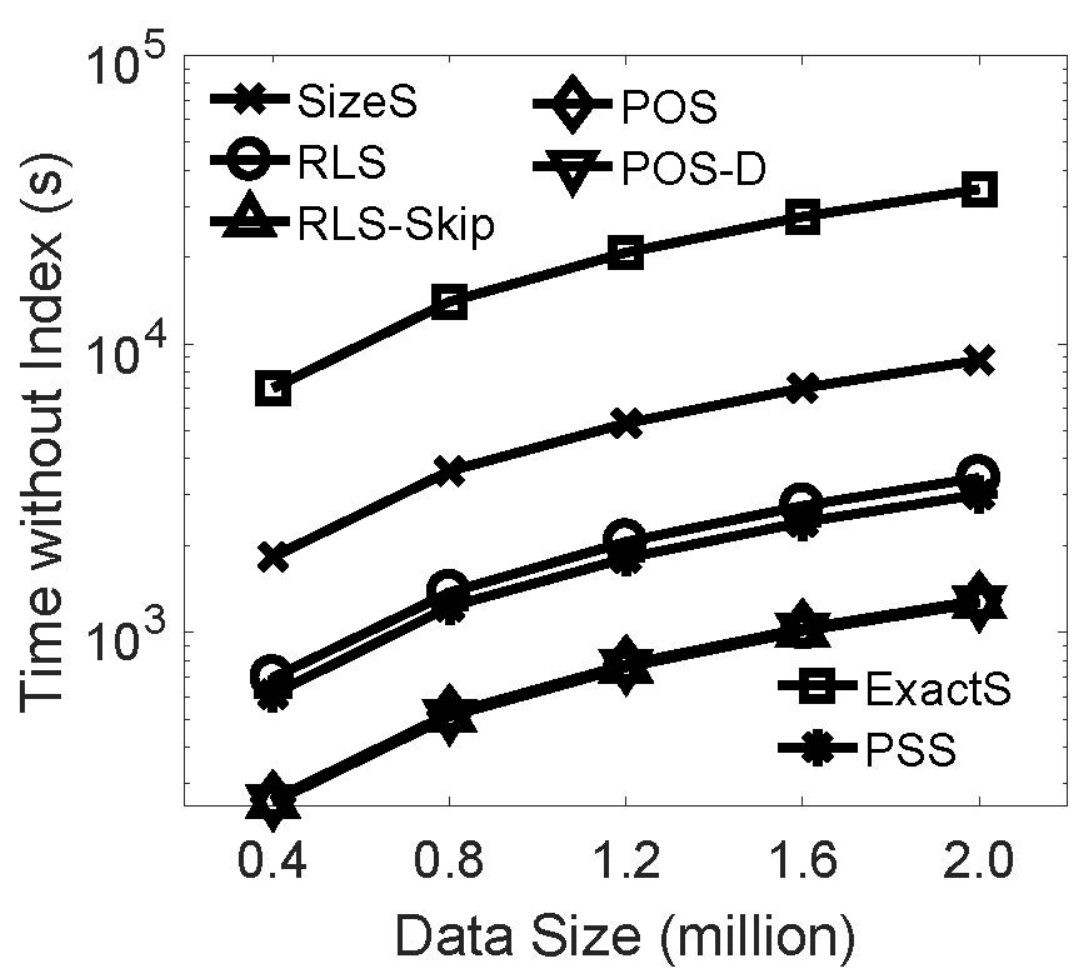}
		\end{minipage}
		\\
		\scriptsize (a) Porto (t2vec)
		&
		\scriptsize (b) Porto (DTW)
		&
		\scriptsize (c) Porto (Frechet)
		\\
		
		\begin{minipage}{2.6cm}
			\includegraphics[width=2.85cm]{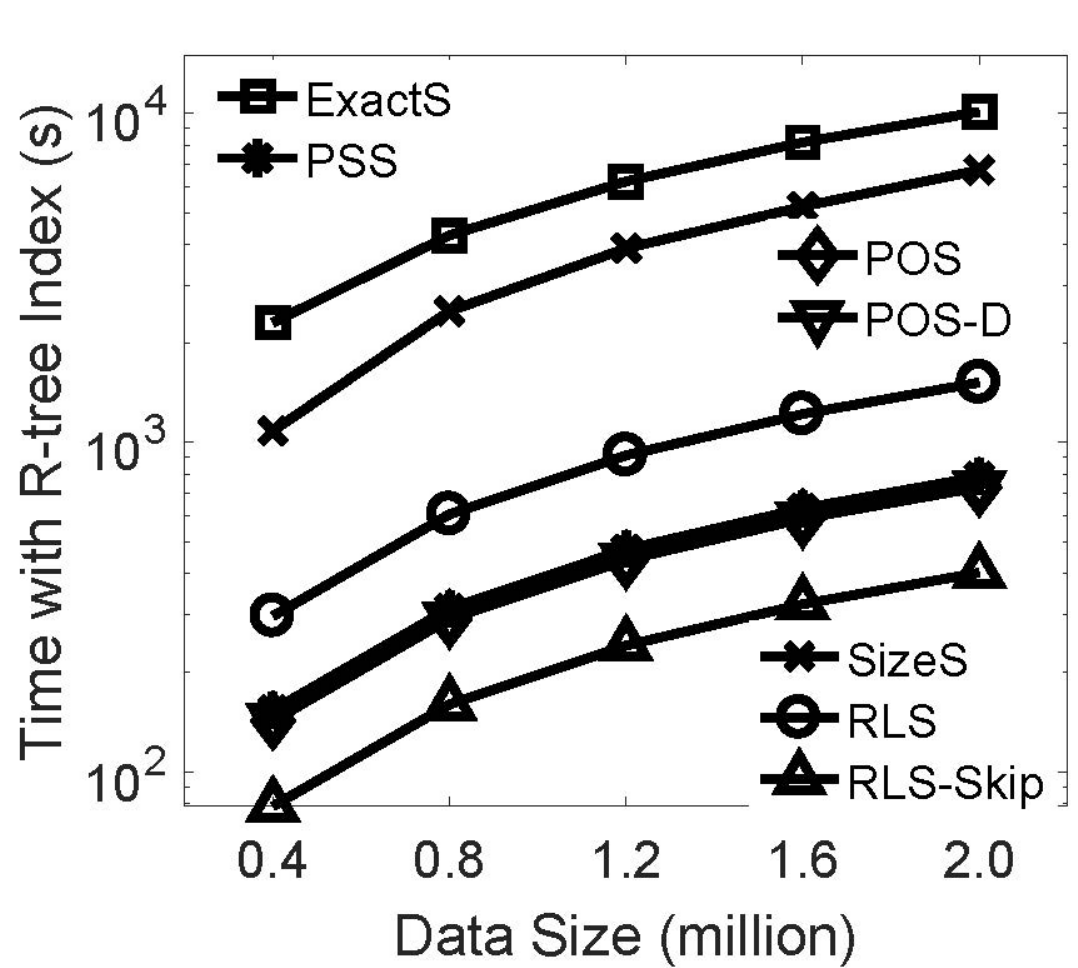}
		\end{minipage}
		&
		\begin{minipage}{2.6cm}
			\includegraphics[width=2.85cm]{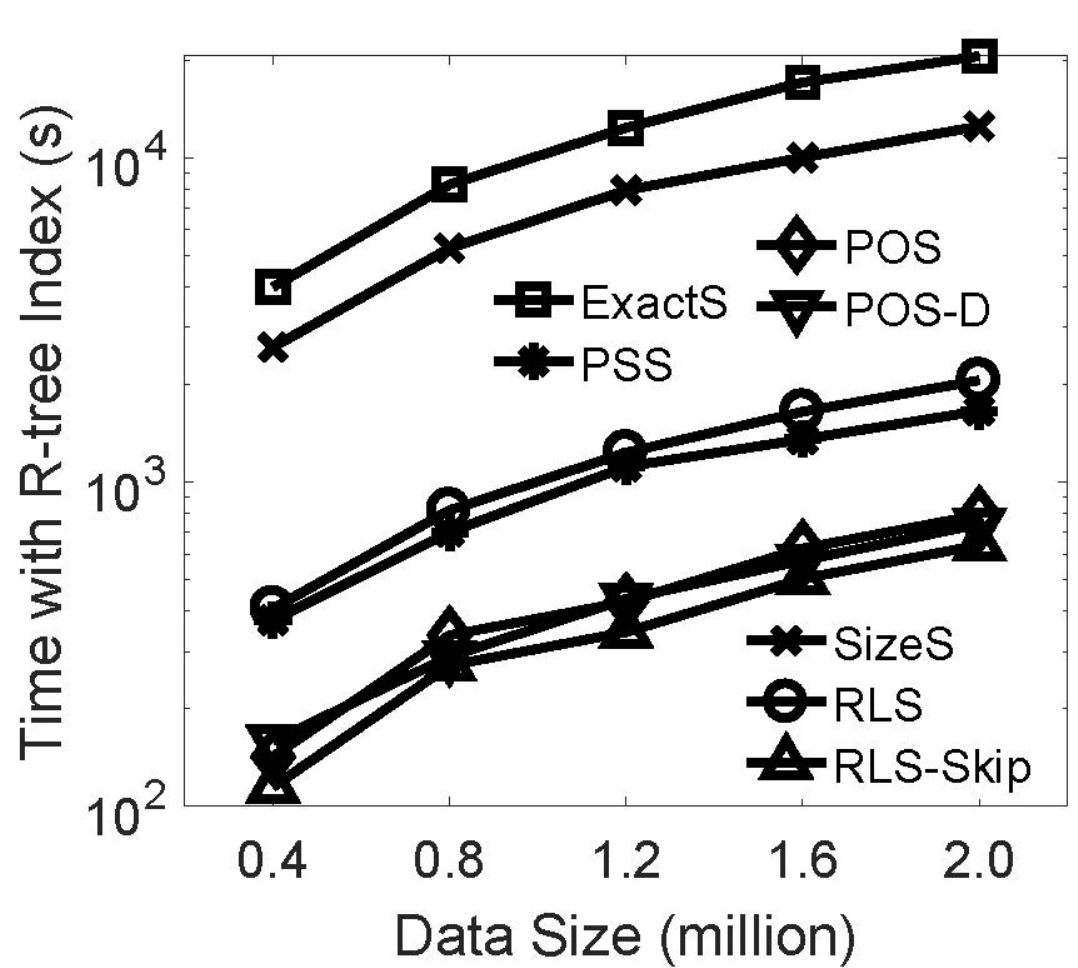}
		\end{minipage}
		&
		\begin{minipage}{2.6cm}
			\includegraphics[width=2.85cm]{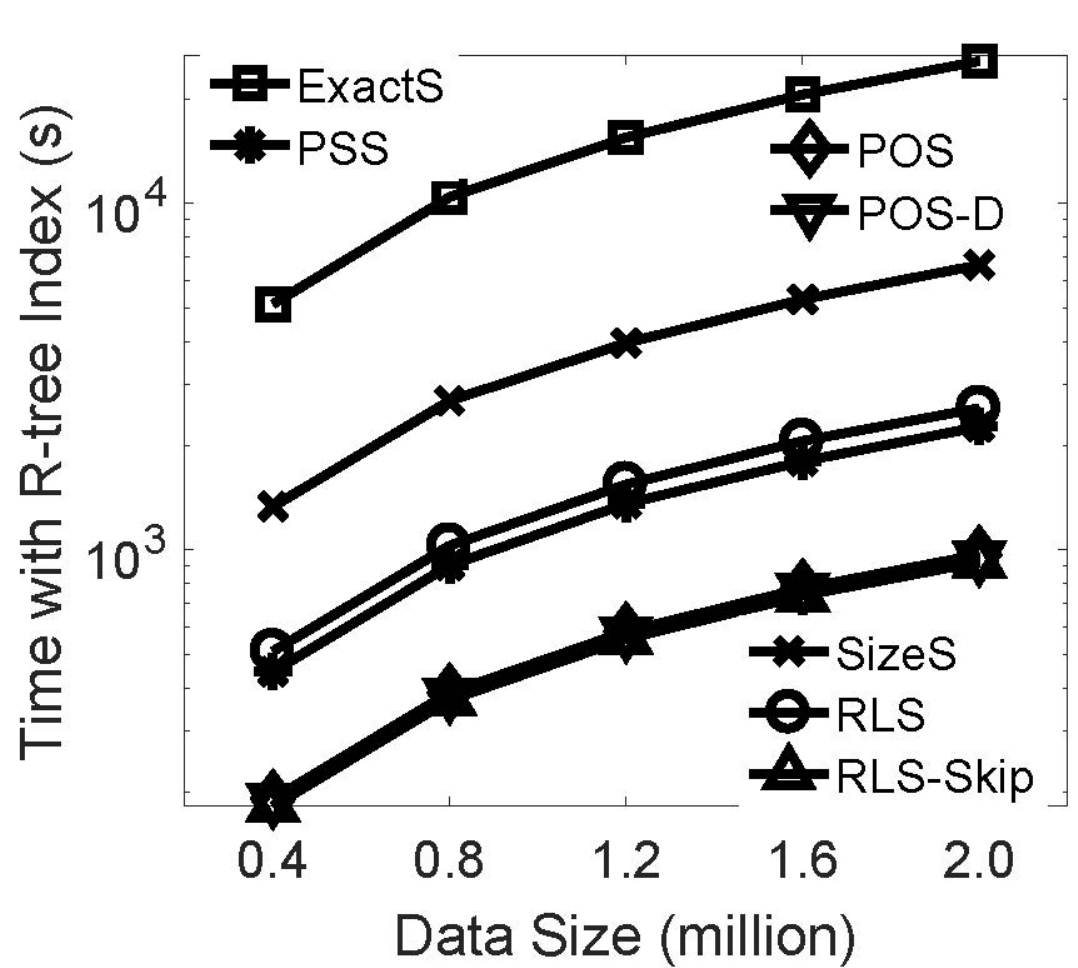}
		\end{minipage}
		\\
		\scriptsize (d) Porto (t2vec)
		&
		\scriptsize (e) Porto (DTW)
		&
		\scriptsize (f) Porto (Frechet)
		
	\end{tabular}
	\caption{Efficiency without index (a)-(c) and with R-tree index (d)-(f) on Porto.}
	\label{fig:efficiencyRLS}
\end{figure}

\smallskip
\noindent\textbf{Evaluation Metrics.} We use three metrics to evaluate the effectiveness of an approximate algorithm. (1) Approximate Ratio (AR): It is defined as 
the ratio between the dissimilarity of the solution wrt a query trajectory, which is returned by an approximate algorithm, and that of the solution returned by an exact algorithm.
A smaller AR indicates a better algorithm.
(2) Mean Rank (MR): 
We sort all the subtrajectories of a data trajectory in ascending order of their dissimilarities wrt a query trajectory. 
MR is defined as the rank of the solution returned by an approximate algorithm.
(3) Relative Rank (RR): 
RR is a normalized version of MR by the total number subtrajectories of a data trajectory.
A smaller MR or RR indicates a better algorithm.


\smallskip
\noindent\textbf{Evaluation Platform.} All the methods are implemented in Python 3.6. The implementation of RLS is based on Keras 2.2.0. The experiments are conducted on a server with 32-cores of Intel(R) Xeon(R) Gold 6150 CPU @ 2.70GHz 768.00GB RAM and one Nvidia Tesla V100-SXM2 GPU.

\subsection{Experimental Results}
\label{effectiveness}

\noindent\textbf{(1) Effectiveness results.}
We randomly sample 10,000 trajectory pairs from a dataset, and for each pair we use one trajectory as the query trajectory to search the most similar subtrajectory from the other one.
%
Figure~\ref{effect_result} shows the results.
%
The results clearly show that
RLS and RLS-Skip consistently outperform all other non-learning based approximate algorithms in terms of all three metrics on both datasets and under all three trajectory similarity measurements.
For example, RLS outperforms POS-D, the best non-learning algorithm when using t2vec, by 70\% (resp. 83\%) in terms of RR on Porto (resp. Harbin); RLS outperforms PSS, the best non-learning based algorithm when using DTW, by 25\% (resp. 20\%) in terms of MR on Porto (resp. Harbin); RLS outperforms PSS, the best non-learning based algorithm when using Frechet, by 25\% (resp. 20\%) in terms of MR on Porto (resp. Harbin).
%
Among PSS, POS, and POS-D, PSS performs the best for DTW and Frechet; However, for t2vec, PSS provides similar accuracy as POS and POS-D on Porto, but performs much worse on Harbin.
The reason is that for DTW and Frechet, PSS computes exact similarity values for suffix subtrajectories, while for t2vec, it computes only approximate ones. Therefore, PSS has a relatively worse accuracy when used for t2vec.
%
%
%
%
%
We also observe that SizeS is not competitive compared with other approximate algorithms.
%
In addition, RLS-Skip has its effectiveness a bit worse than RLS, but still better than those non-learning based algorithms due to the fact that it is based on a learned policy for decision making.
%



\smallskip
\noindent\textbf{(2) Efficiency results.}
We prepare different databases of data trajectories by including different amounts of trajectories from a dataset and vary the total number of points in a databases.
For each database, we randomly sample 10 query trajectories from the dataset, run for each query trajectory a query for finding the top-50 similar subtrajectories, and then collect the average running time over 10 queries.
The results of running time on the Porto dataset are shown in Figure~\ref{fig:efficiencyRLS}, and those on the other datasets could be found in the technical report~\cite{TR}.
%
%
%
RLS-Skip runs the fastest
since on those points that have been skipped,
the cost of maintaining the states and making decisions is saved.
In contrast, none of the other algorithms skip points.
ExactS has the longest running time, e.g., ExactS is usually around 7-15 times slower than PSS, POS, POS-D, RLS and 20-30 times slower than RLS-Skip.
%
RLS is slightly slower than PSS, POS, POS-D.
This is because RLS makes the splitting decision via a learning model while the other three use a simple similarity comparison.

\smallskip
\noindent\textbf{(3) Scalability.} We investigate the scalability of all the algorithms based on the results reported in Figure~\ref{fig:efficiencyRLS}.
All those splitting-based algorithms including PSS, POS, POS-D, RLS and RLS-Skip scale well. 

\smallskip
\noindent\textbf{(4) Working with indexes.}
Following two recent studies~\cite{yao2019computing, wang2018torch} on trajectory similarity search, we employ the Bounding Box R-tree Index for boosting the efficiency. 
%
It indexes the MBRs of data trajectories and prunes all those data trajectories whose MBRs do not interact with the MBR of a given query trajectory.
%
%
We note that in theory, exact solutions might be filtered out by the index (e.g., the most similar subtrajectory may be part of a trajectory, whose MBR does not interact with that of a query one), but in practice, this rarely happen. For example, as found in our experiments on the Porto dataset, when DTW and Frechet are used, the results returned when using the index and those when using no indexes are exactly the same, i.e., no results are missed out. When t2vec is used, at most 20\% results are missed.  Furthermore, for cases of finding approximate solutions, as most of the proposed algorithms do, missing some potential solutions for better efficiency is acceptable.
%
%
Compared with the results without indexes in Figure~\ref{fig:efficiencyRLS}(a)-(c), the results using the R-tree index as shown in Figure~\ref{fig:efficiencyRLS}(d)-(f) are lower by around 20--30\%.


\begin{figure}[!t]
	\centering
	\begin{tabular}{c c c}
		\begin{minipage}{2.6cm}
			\includegraphics[width=2.85cm]{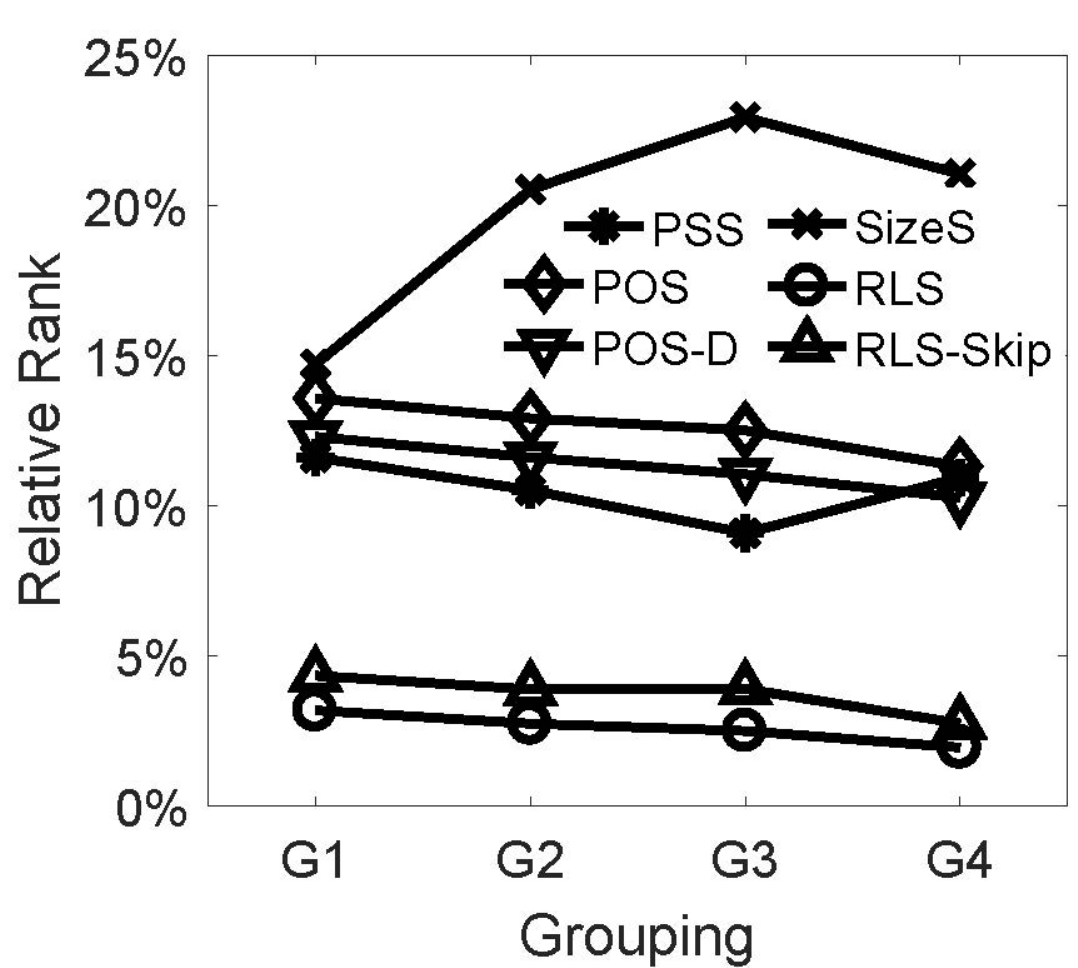}
		\end{minipage}
		&
		\begin{minipage}{2.6cm}
			\includegraphics[width=2.85cm]{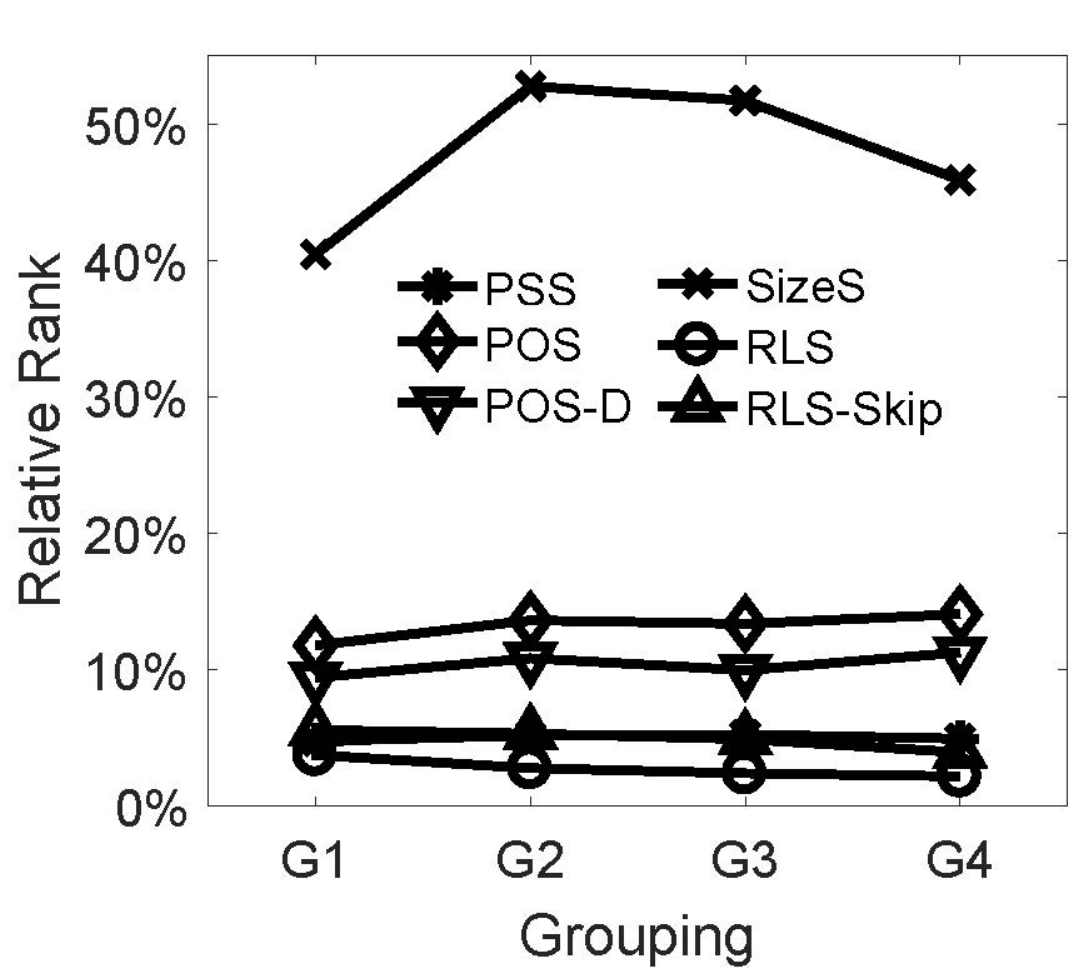}
		\end{minipage}
		&
		\begin{minipage}{2.6cm}
			\includegraphics[width=2.85cm]{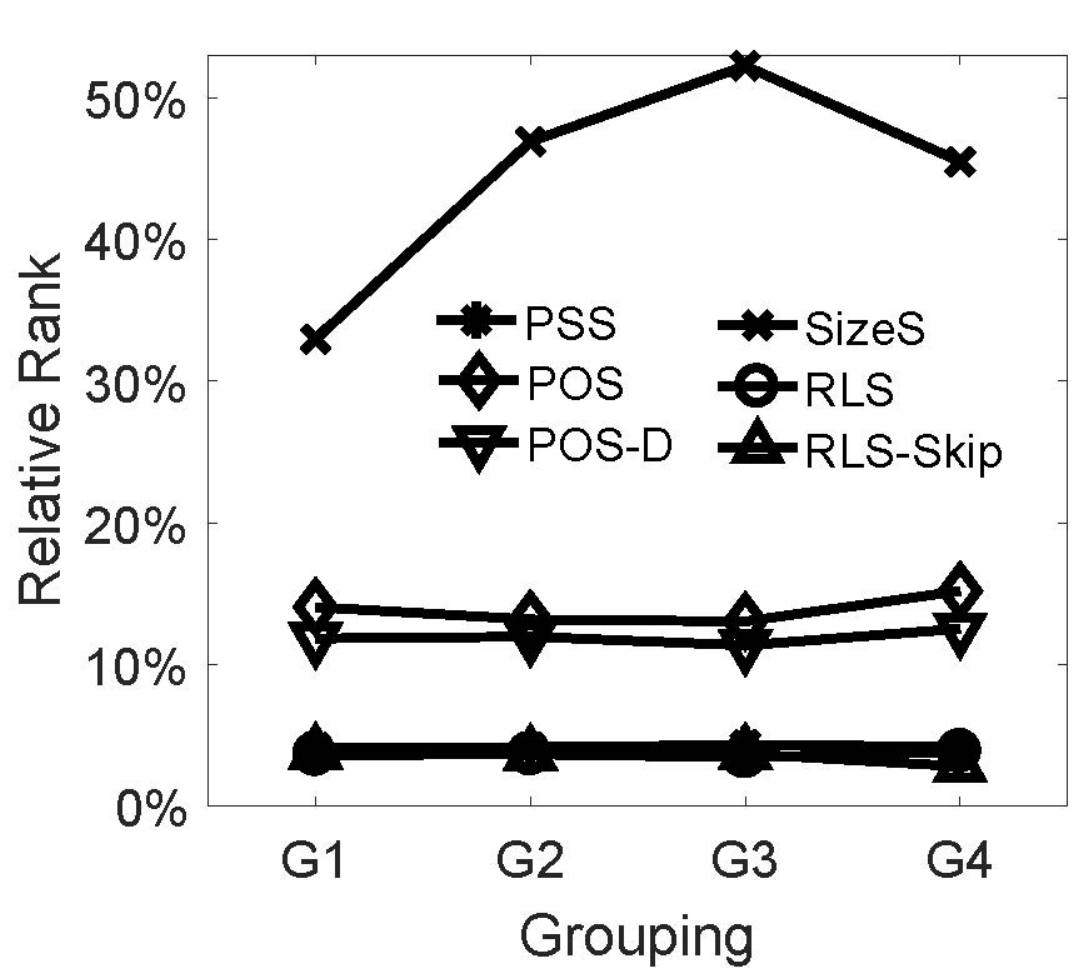}
		\end{minipage}
		\\
		\scriptsize (a) Porto (t2vec)
		&
		\scriptsize (b) Porto (DTW)
		&
		\scriptsize (c) Porto (Frechet)
	\end{tabular}
	\caption{Effectiveness with varying query lengths.}
	\label{RR:querylength}
\end{figure}

\begin{figure}[!t]
	\centering
	\begin{tabular}{c c c}
		\begin{minipage}{2.6cm}
			\includegraphics[width=2.85cm]{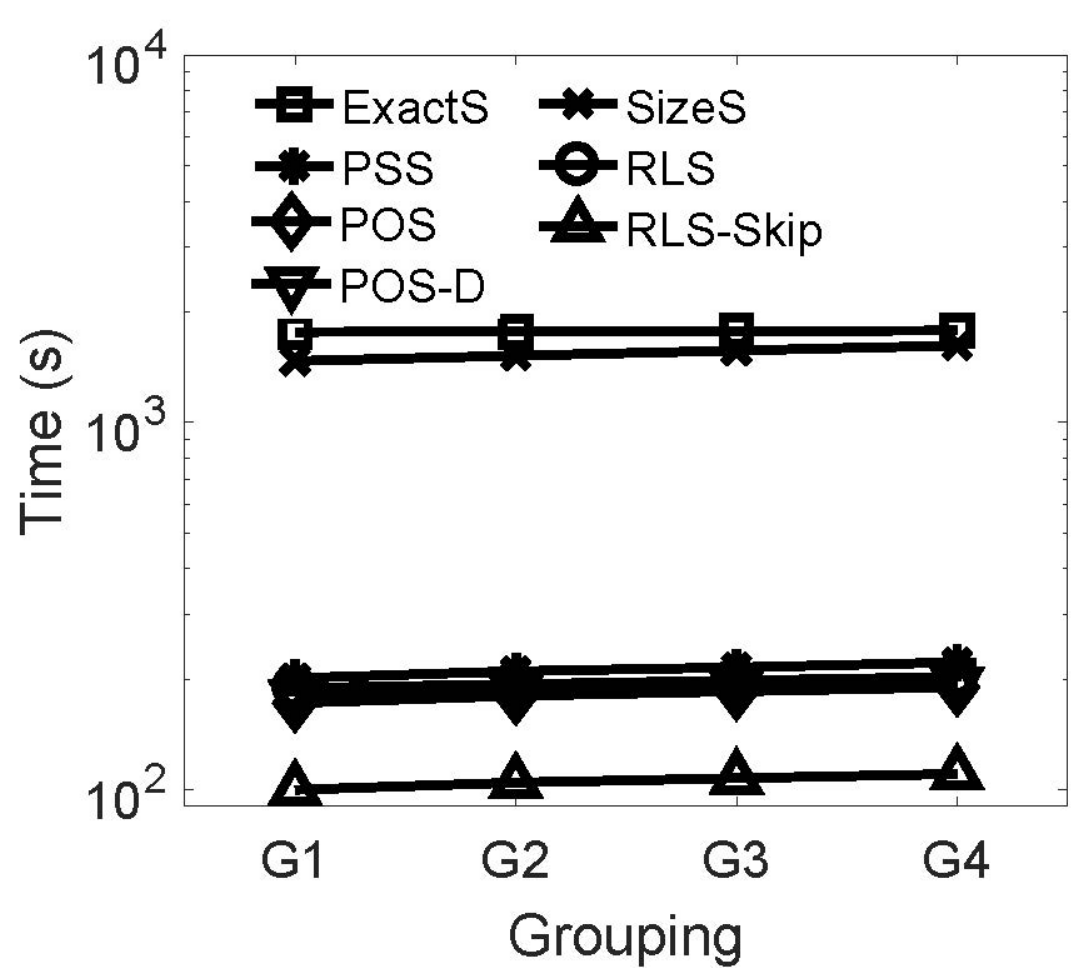}
		\end{minipage}
		&
		\begin{minipage}{2.6cm}
			\includegraphics[width=2.85cm]{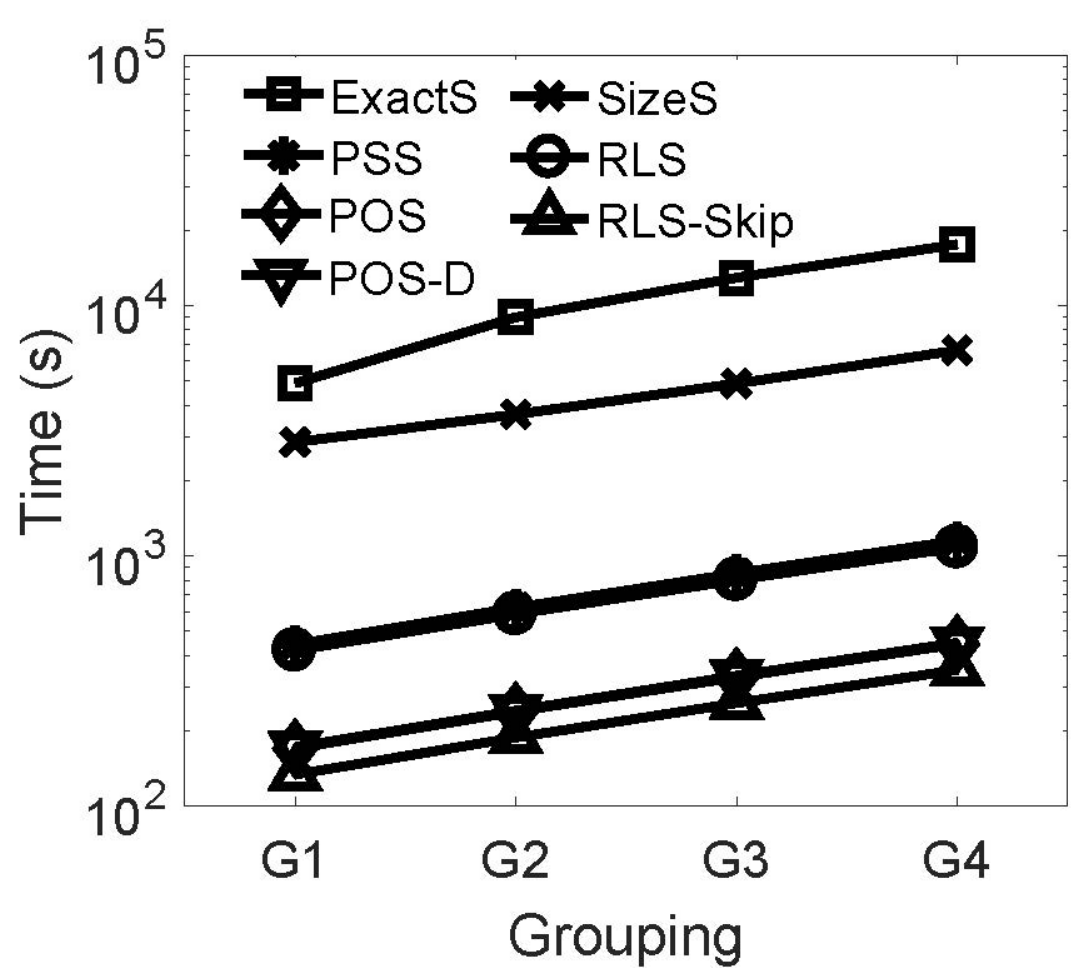}
		\end{minipage}
		&
		\begin{minipage}{2.6cm}
			\includegraphics[width=2.85cm]{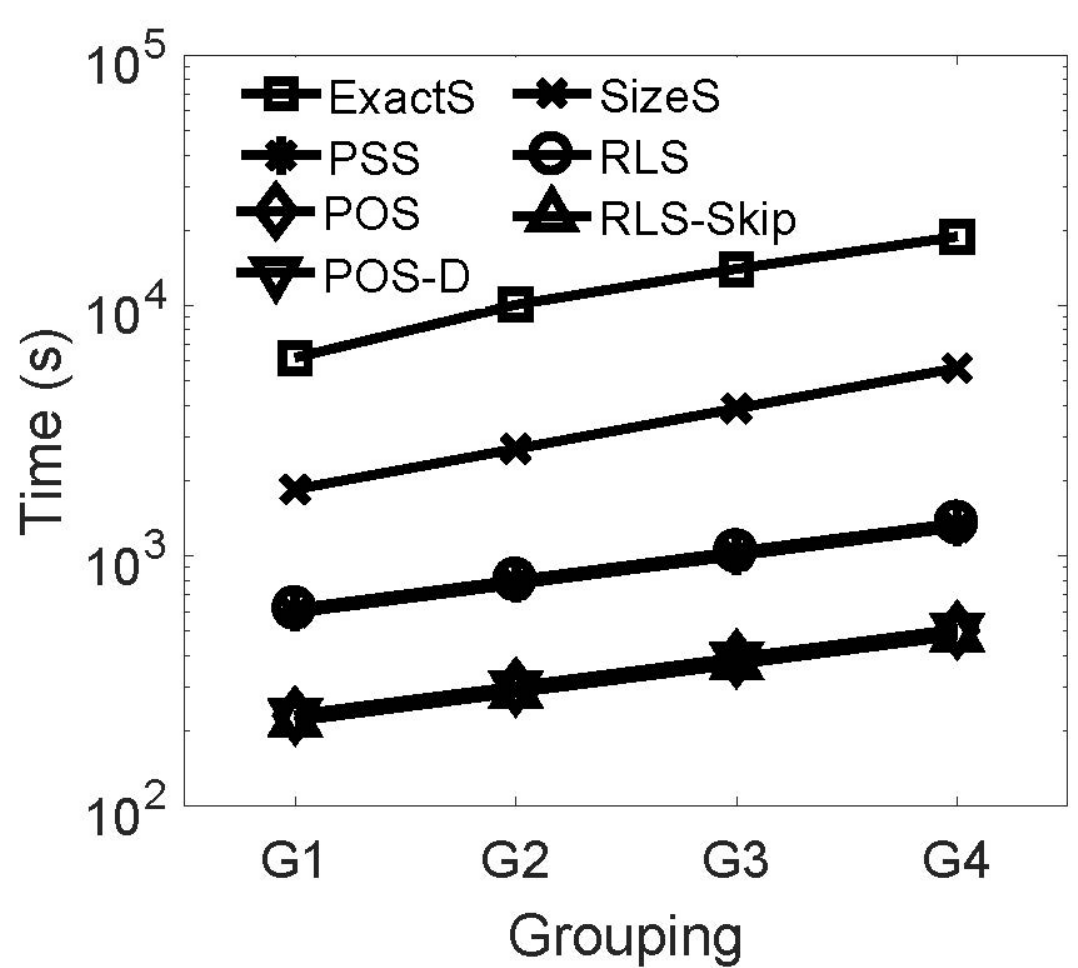}
		\end{minipage}
		\\
		\scriptsize (a) Porto (t2vec)
		&
		\scriptsize (b) Porto (DTW)
		&
		\scriptsize (c) Porto (Frechet)
	\end{tabular}
	\caption{Efficiency with varying query lengths.}
	\label{time:querylength}
\end{figure}

\smallskip \noindent\textbf{(5) The effect of query trajectory length.}
We prepare four groups of query trajectories from a dataset, namely $G1$, $G2$, $G3$, and $G4$, each with 10,000 trajectories, such that the lengths of the trajectories in a group are as follows: $G1=[30,45)$, $G2=[45,60)$, $G3=[60,75)$ and $G4=[75,90)$. Then, for each query trajectory, we prepare a data trajectory from the dataset. Note that for a query trajectory and a corresponding data trajectory, the latter may be longer than or shorter than the former.
For each group, we report the average results.
The results of RR on Porto are shown in Figure~\ref{RR:querylength}.
%
%
We observe that the RRs of all algorithms except for SizeS remain stable when the query length grows.
For SizeS, the RRs fluctuate with the change of the query length. This is because the length of the most similar subtrajectory may not have similar length as the query trajectory, and thus it may miss high-quality results when the search space is constrained by the parameter $\xi$.
%
The results of running time on Porto are shown in Figure~\ref{time:querylength}.
We notice that for t2vec, the running times of all the algorithms are almost not affected by the query length. This is because for t2vec, the time complexity of computing a similarity is constant once the vector of the query trajectory is learned. For DTW, Frechet, the time complexity of computing a similarity increase with the query length, as shown in Table~\ref{tab:algorithm-complexities}.
\begin{table}[!t]
	\centering
	\scriptsize
	\caption{The effect of skipping steps $k$ for RLS-Skip.}
	\begin{tabular}{|c|c|c|c|c|c|c|c|}
		\hline
		Metrics & $k=0$ &$k=1$ & $k=2$ & $k=3$ & $k=4$ & $k=5$\\ \hline
		AR  &1.028 &1.039 &1.042 &1.044  &1.055 &1.069 \\
		MR  &41.138 &56.633 &58.077 &64.741 &70.281 &94.356 \\
		RR  &3.5\% &5.4\% &5.6\% &5.8\% &6.3\% &8.9\% \\
		Time (ms)    &55.2 &39.8 &38.5 &35.8 &31.8 &22.9 \\ \hline
		Skip Pts &0\% &3.1\% &13.1\% &17.7\% &29.5\% &47.6\% \\ \hline
	\end{tabular}
	\label{para_steps}
\end{table}
\begin{table*}[t!]
	\centering
	\scriptsize
	\caption{Comparison with Trajectory Similarity Computation and Subtrajectory Similarity Computation.}
	\begin{tabular}{|c|c|c|c|c|c|c|c|c|c|c|c|c|c|}
		\hline
		\multicolumn{2}{|c|}{Similarity} & \multicolumn{4}{c|}{t2vec} & \multicolumn{4}{c|}{DTW} & \multicolumn{4}{c|}{Frechet} \\ \hline
		Dataset                    & Problem    & AR  & MR  & RR  & Time (ms) & AR  & MR & RR & Time (ms) & AR   & MR  & RR  & Time (ms)  \\ \hline
		\multirow{2}{*}{Porto}     & SimTra        &1.313     &156.153     &23.3\%     &28.5          &2.100     &752.831    &70.7\%    &18.1          &1.883      &559.462     &56.5\%     &19.2           \\ \cline{2-14}
		& SimSub        &1.098     &18.323     &3.0\%     &39.6          &1.028     &41.138    &3.5\%    &55.2          &1.034      &34.162     &3.6\%     &69.6           \\ \hline
		\multirow{2}{*}{Harbin}    & SimTra        &1.293     &678.311     &46.9\%     &31.7          &2.326     &1218.908    &72.2\%    & 27.1          &1.891      &854.042     &53.9\%     &28.6        \\ \cline{2-14}
		& SimSub        &1.025     &14.945     &1.3\%     &62.6          &1.081     &75.324    &4.1\%    & 114.4         &1.045      &64.729     &4.4\%     &130.6       \\ \hline
		\multirow{2}{*}{Sports}    & SimTra        &1.221     &345.488    &43.4\%     &46.1          &1.659    &4291.666   &59.8\%   &107.5         &1.403      &3272.743     &48.2\%    &133.3        \\ \cline{2-14}
		& SimSub        &1.045     &28.761     &3.8\%     &210.3          &1.005    &126.334   &2.1\%   &254.7         &1.002      &95.280     &1.7\%    &302.3       \\ \hline
	\end{tabular}
	\label{tscssc}
\end{table*}

\smallskip
\noindent\textbf{(6) The effect of skipping steps $k$.}
According to the results, a general trend is that with larger settings of $k$, RLS-Skip has its effectiveness drop but its efficiency grow because RLS-Skip tends to skip more points.
We present in Table~\ref{para_steps} the results on Porto for DTW only due to the page limit. We also report the portion of skipped points in the Porto dataset with 10,000 trajectories. Note that when $k$ is set to 0, RLS-Skip degrades to RLS. For other experiments, we choose $k=3$ as a reasonable trade-off between effectiveness and efficiency.

\begin{figure}[t]
	\hspace{-0.4cm}
	\centering
	\begin{tabular}{c c}
		\begin{minipage}{4cm}
			\includegraphics[width=4.15cm]{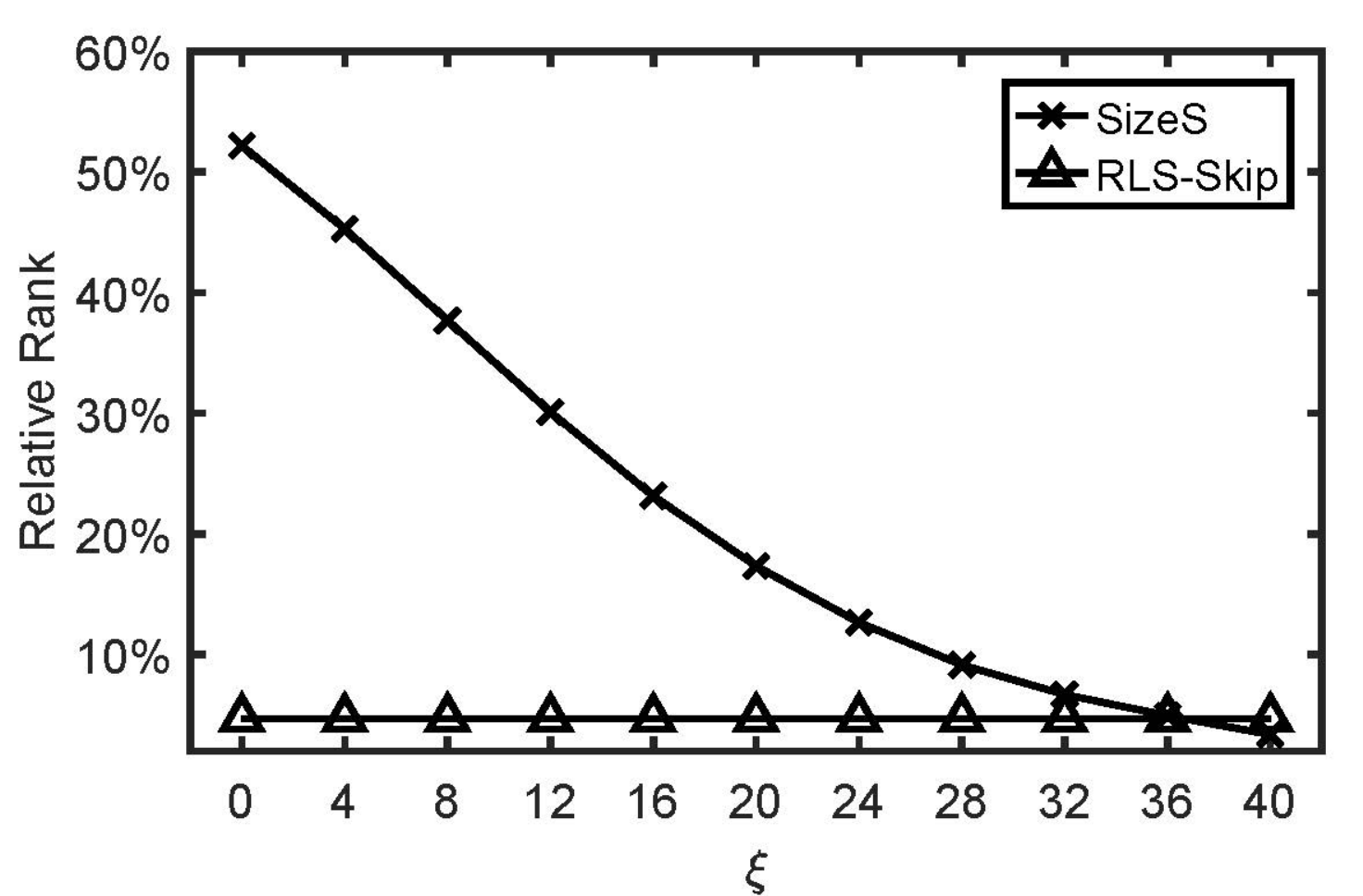}
		\end{minipage}
		&
		\begin{minipage}{4cm}
			\includegraphics[width=4.15cm]{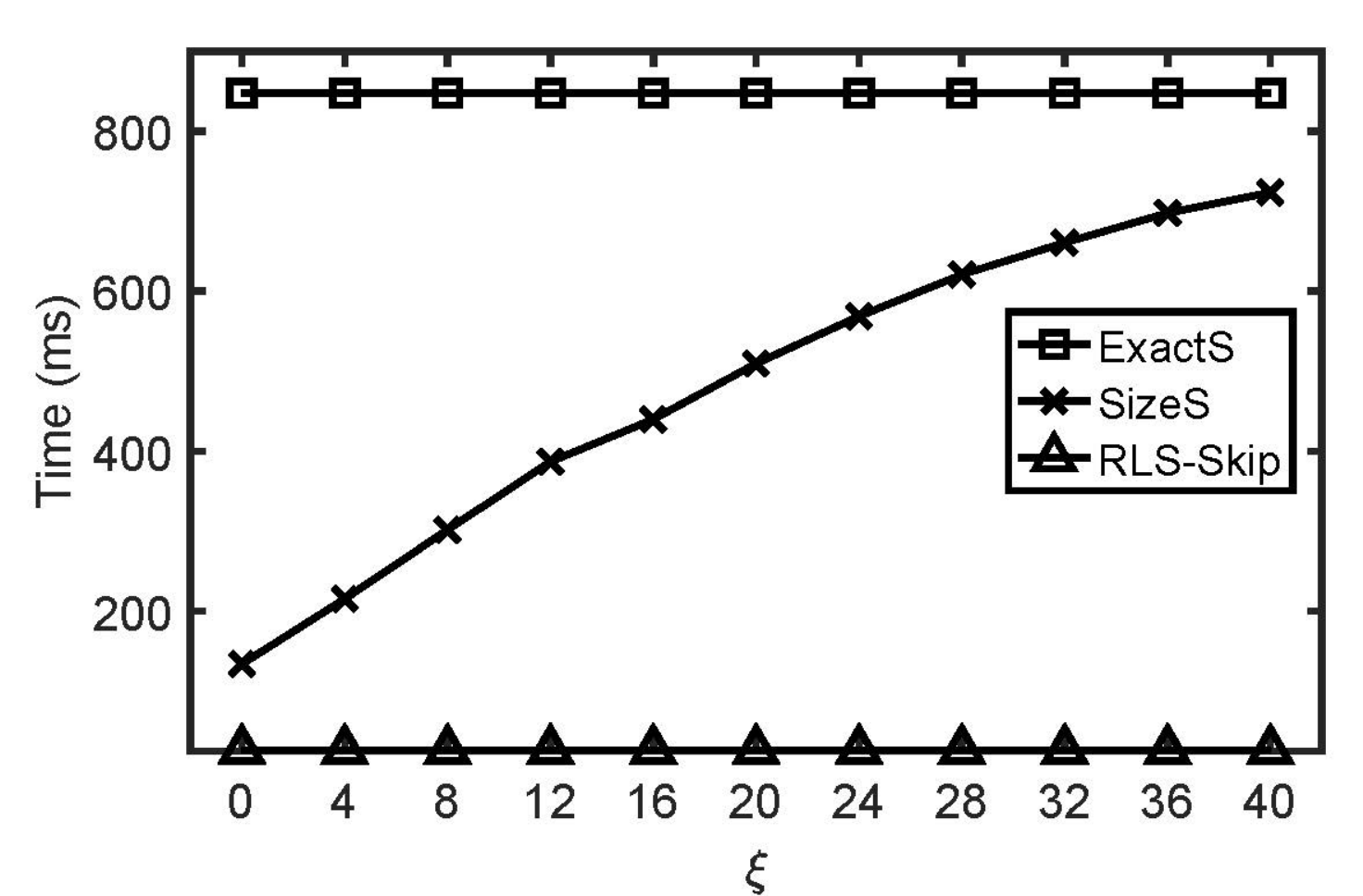}
		\end{minipage}
		\\
		\small (a) Relative Rank (DTW)
		&
		\small (b) Time Cost (DTW)
	\end{tabular}
	\caption{The effect of soft margin $\xi$ for SizeS.}
	\label{para_fls}
\end{figure}

\smallskip
\noindent\textbf{(7) The effect of parameter $\xi$.} 
Figure~\ref{para_fls} shows SizeS's RR and running time averaged on 10,000 trajectory pairs from the Porto dataset. 
As expected, as $\xi$ grows, the RR of SizeS becomes better, but running time increases and approaches to that of ExactS.

\smallskip
\noindent\textbf{(8) Comparison with similar trajectory search (SimTra).}
The solution of the similar trajectory search (SimTra) could be regarded as an approximate solution of the SimSub problem because a data trajectory by itself is a subtrajectory.
%
We compare this approximate solution by SimTra and that by the RLS algorithm.
We report the average results over 10,000 trajectory pairs.
The results are shown in Table~\ref{tscssc}.
The MR and RR of SimTra are around 10 times larger than those of SimSub for t2vec and 20 times for DTW and Frechet,
which shows that SimTra is not a good approximation for SimSub, though SimTra runs faster than SimSub.

\begin{figure}[t]
	\centering
	\begin{tabular}{c c c c}
		\begin{minipage}{4cm}
			\includegraphics[width=4.15cm]{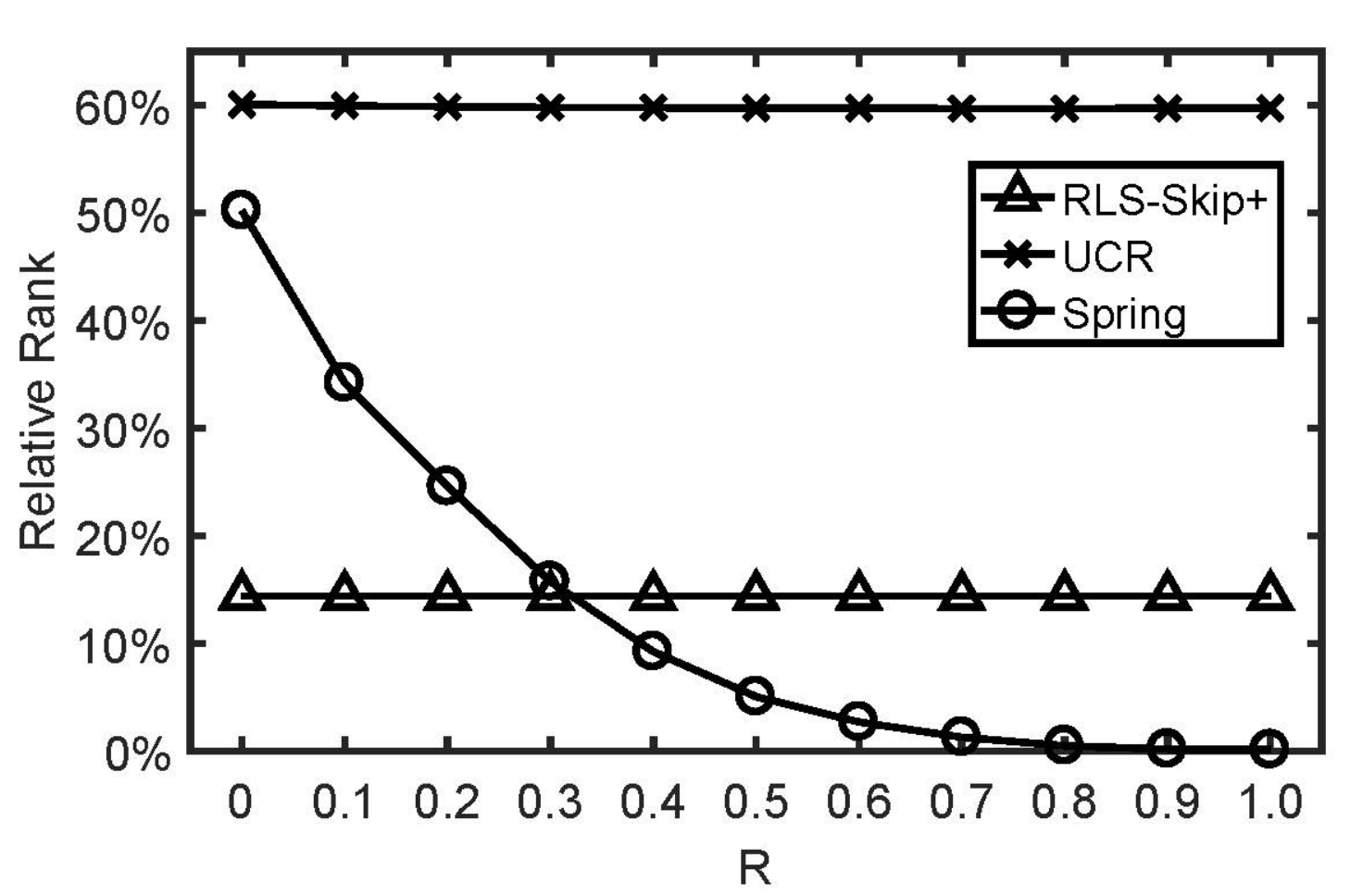}
		\end{minipage}
		&
		\begin{minipage}{4cm}
			\includegraphics[width=4.15cm]{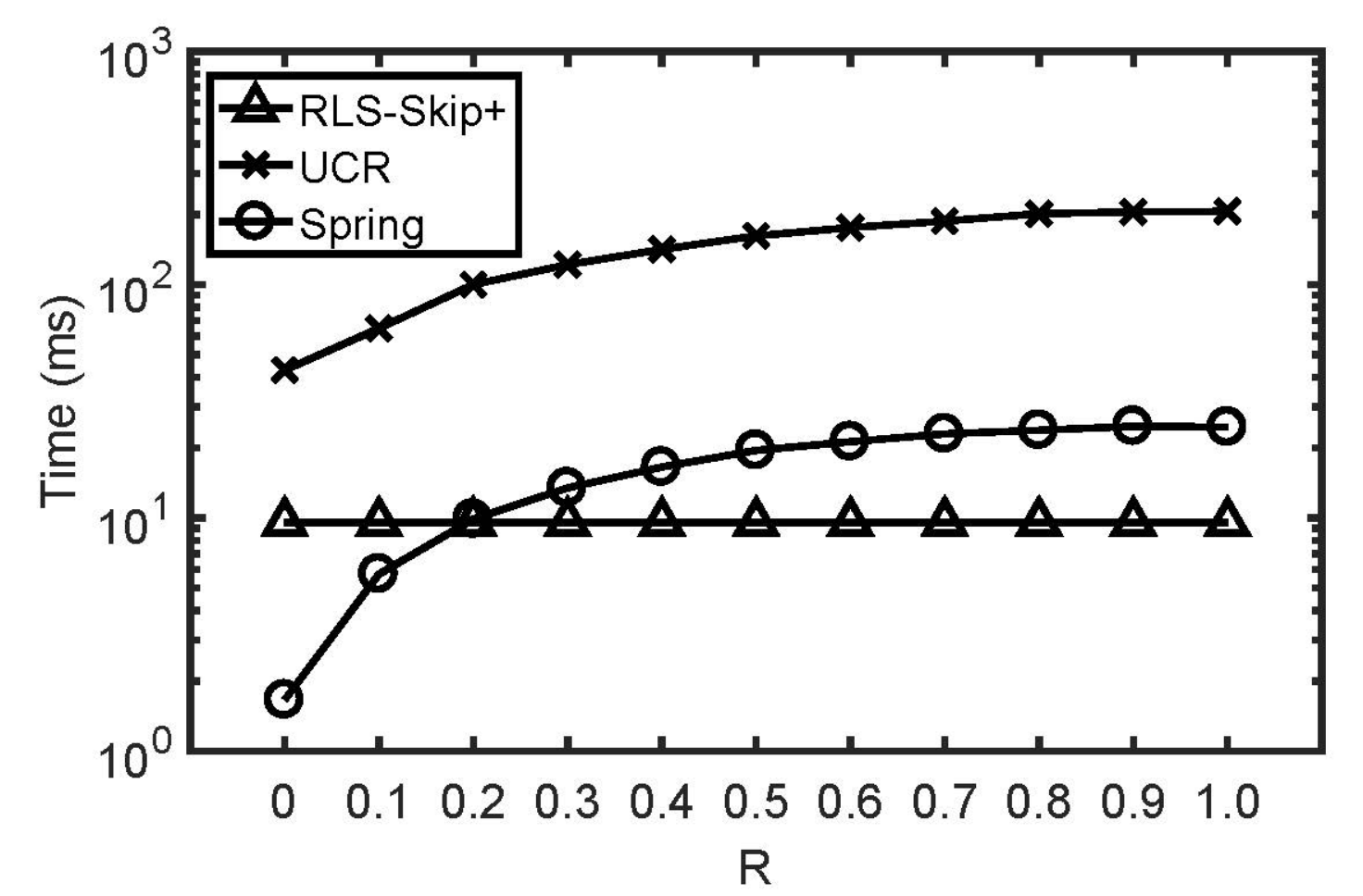}
		\end{minipage}
		\\
		\small (a) Relative Rank (DTW)
		&
		\small (b) Time Cost (DTW)
	\end{tabular}
	\caption{Comparison with UCR and Spring.}
	\label{fig:UCR}
\end{figure}

\smallskip
\noindent\textbf{(9) Comparison with algorithms for specific measurements (UCR and Spring).}
In UCR and Spring, a point $q_i$ from the query trajectory can be aligned with only those points $p_j$ from the data trajectory $T$ with $j\in [i-R\cdot |T|, i+R\cdot |T|]$. 
We vary the parameter $R$ in this experiment.
When $R = 1$, it reduces to the unconstrained DTW that is used in this paper.
Essentially, $R$ controls how accurately the DTW distance is computed: the higher $R$ is, the more accurate (but also more costly) the computation is.
We note that even when $R=1$,
UCR does not return exact solutions since it considers subtrajectories of the same size of the query one only.
For this part of experiment,
we drop the component $\Theta_{suf}$ when defining the MDP of RLS-Skip for better efficiency and call the resulting algorithm \emph{RLS-Skip+}.
The results are shown in Figure~\ref{fig:UCR}, where we vary the parameter $R$ from 0 to 1. 
We notice that (1) RLS-Skip+ dominates UCR in terms of both efficiency and effectiveness;
(2) the RR of UCR changes slightly from 60.1\% (when $R=0$) to 59.7\% (when $R=1$),
which shows that the performance of UCR is insensitive to the parameter $R$;
(3) under settings of $0.2\le R\le 0.3$, RLS-Skip+ dominates Spring in terms of both effectiveness and efficiency; and (4) under other settings, RLS-Skip+ and Spring provide different trade-offs between effectiveness and efficiency.

\begin{figure}[t]
	\hspace{-4mm}
	\centering
	\begin{tabular}{c c c c}
		\begin{minipage}{4cm}
			\includegraphics[width=4.15cm]{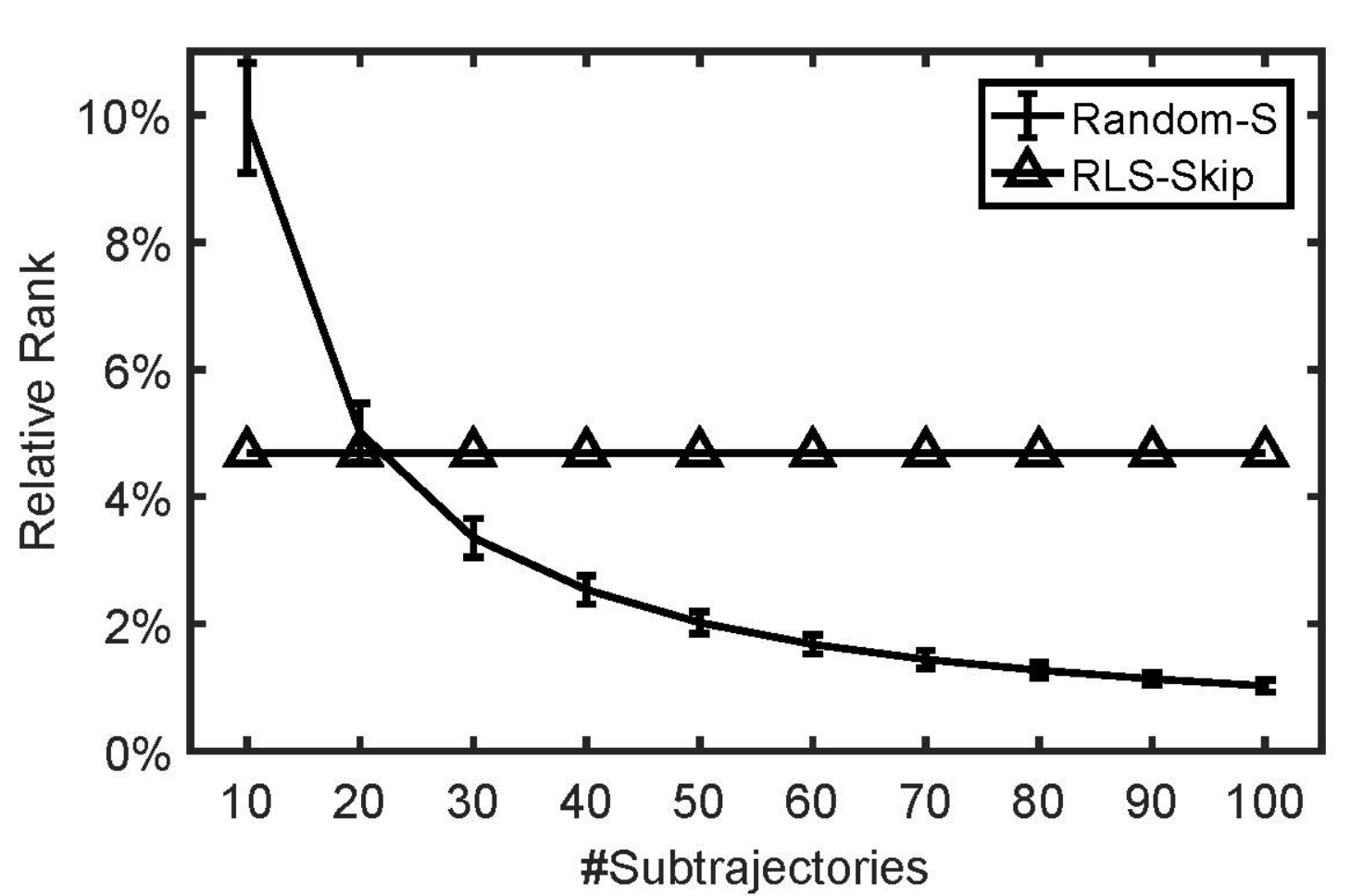}
		\end{minipage}
		&
		\begin{minipage}{4cm}
			\includegraphics[width=4.15cm]{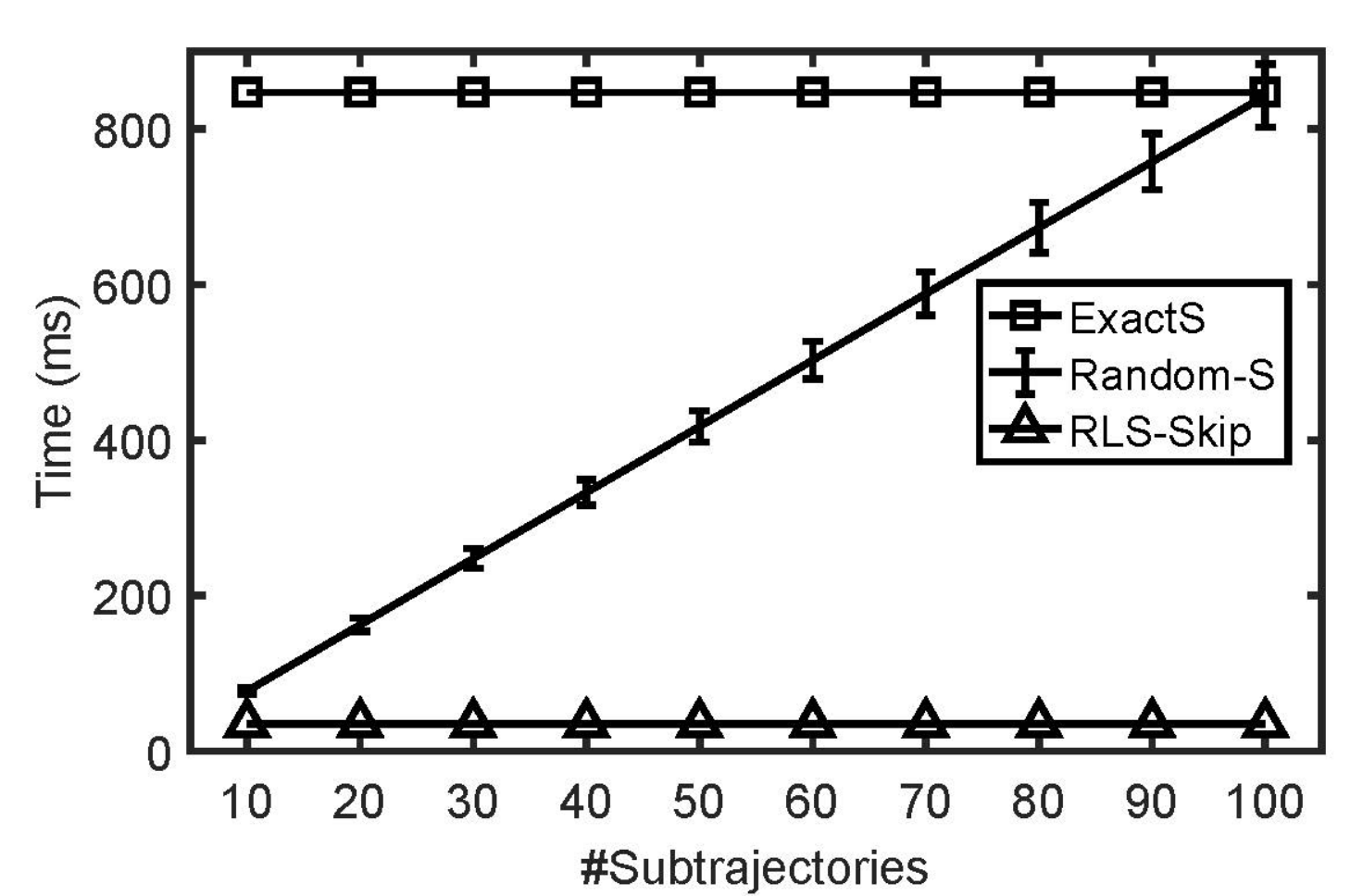}
		\end{minipage}
		\\
		\small (a) Relative Rank (DTW)
		&
		\small (b) Time Cost (DTW)
	\end{tabular}
	\caption{Comparison with Random-S.}
	\label{fig:random-s}
\end{figure}

\smallskip
\noindent\textbf{(10) Comparison with Random-S.}
The results are shown in Figure~\ref{fig:random-s}, where we vary the sample size from 10 to 100 and for each sample size, we run the algorithm 100 times and collect the average and standard deviations of the metrics of RR and running time. We notice that for a relatively small sample size, e.g., 100, the running time of Random-S is almost that of ExactS and significantly larger than that of RLS-Skip (25 times higher). This is because for Random-S, the subtrajectories that are considered could be quite different, and thus it is not possible to compute their similarities \emph{incrementally} as it does for ExactS. Whereas when the sample size is small, e.g., below 20, Random-S has its effectiveness significantly degraded, which is clearly worse than that of RLS-Skip.

\smallskip
\noindent\textbf{(11) Training time.}
The training times of the RLS and RLS-Skip models on different datasets are shown in Table~\ref{learntime}. It normally takes a couple of hours to train a reinforcement learning model for RLS and RLS-Skip. It takes less time to train RLS-Skip than RLS since we use the same number of trajectory pairs and epochs for training both algorithms and RLS-Skip runs faster.

\begin{table}[t]
\label{learntime}
	\centering
	\scriptsize
	\captionof{table}{Training time (hours).}
	\hspace{-2mm}
	\begin{tabular}{|c|c|c|c|c|c|c|}
		\hline
		\parbox{1cm}{Similarity} & \multicolumn{2}{c|}{t2vec} & \multicolumn{2}{c|}{DTW} & \multicolumn{2}{c|}{Frechet} \\ \hline
		Algorithms             & RLS       & RLS-Skip       & RLS      & RLS-Skip      & RLS      & RLS-Skip      \\ \hline
		Porto                  &7.2           &4.4                &10.1          &4.8               &10.6          &5.5               \\ \hline
		Harbin                 &9.7           &5.4                &13.9          &5.7               &14.2          &8.3               \\ \hline
		Sports                 &12.5           &7.6                &83.3          &46.6               &104.1          &52.4               \\ \hline
	\end{tabular}
    \label{learntime}
\end{table}

%% file: conclusion.tex
\section{CONCLUSION}
\label{conclusion}
In this paper, we study the similar subtrajectory search (SimSub) problem and develop a suite of algorithms including an exact algorithm, an approximate algorithm providing a controllable trade-off between efficiency and effectiveness, and a few splitting-based algorithms, among which some are based on pre-defined heuristics and some are based on deep reinforcement learning called RLS and RLS-Skip.
We conducted extensive experiments on real datasets, which verified that among the approximate algorithms, learning based algorithms achieve the best effectiveness and efficiency.
%
In the future, we plan to explore some more similarity measurements for the SimSub problem, e.g., the constrained DTW distance and other similarity measurements reviewed in Section~\ref{related}.

\smallskip
\noindent\textbf{Acknowledgments.} This research is supported by the Nanyang Technological University Start-UP Grant from the College of Engineering under Grant M4082302 and by the Ministry of Education, Singapore, under its Academic Research Fund Tier 1 (RG20/19 (S)).
Gao Cong acknowledges the support by Singtel Cognitive and Artificial Intelligence Lab for Enterprises (SCALE@NTU), which is a collaboration between Singapore Telecommunications Limited (Singtel) and Nanyang Technological University
(NTU) that is funded by the Singapore Government through the Industry Alignment Fund - Industry Collaboration Projects Grant, and a Tier-1 project RG114/19.
The authors would like to thank Eamonn Keogh for pointing out some references to the time series literature and also the anonymous reviewers for their constructive comments.

%% file: appendix.tex
	\section{Quality Analysis of the SizeS Algorithm}
	\label{sec:sizes-proof}
	There usually exists a space, within which the objects move. Therefore, we assume the points of trajectories are all located in a $d_{max} \times d_{max}$ rectangle, where $d_{max}$ is a large number that captures the extent of the space. We use a coordinate system whose origin is at the middle of this rectangle.

	\smallskip\noindent\textbf{Case 1: DTW.}
	Consider a problem input with a query trajectory with $m$ points $T_q = <p_1', p_2', ..., p_m'>$ and a data trajectory with $n = m^2$ points $T = <p_{1,1}, p_{1,2}, ..., p_{1,m}, p_{2,1}, p_{2,2}, ..., p_{2,m}, ..., p_{m,1}, p_{m,2}, p_{m,m}>$. We assume that $m$ is an even number and $l = m / 2$. In addition, we let $d = d_{max} / m$.
	The locations of the points in these trajectories are provided as follows.
	\begin{itemize}
		\item $p_i' = ( - ( l - i + 1/2) \cdot d, 0)$ for $i = 1, 2, ..., l$;
		\item $p_i' = ( (i - l - 1/2) \cdot d, 0)$ for $i = l+1, l+2, ..., m$;
		\item $p_{i,j}$'s ($1\le j \le m$) are evenly located at the circle with its center at $p_i'$ and its radius equal to $\epsilon$, where $\epsilon$ is a very small real number for $i = 1, 2, ..., m$;
	\end{itemize}
	
	Consider the optimal solution. Its DTW distance, denoted by $D_o$, is at most the distance between $T$ and $T_q$, which is equal to $m^2\cdot \epsilon$ (points $p_{i,j}$ for $1\le j\le m$ are aligned with point $p_i'$ for $1\le i\le m$). That is, we have $D_o \le m^2 \cdot \epsilon$.
	
	Consider the approximate solution returned by SizeS.  Suppose that $\xi = 0$. That is, we only consider subtrajectories with the length exactly equal to $m$. We further know that each such sub-trajectory consists of points, either all located at a circle with the center at one of the point of $T_q$ or some located at a circle with the center at a point $p_i'$ and others located at a circle with the center at a point $p_{i+1}'$ ($1\le i < m$). It could be verified that among these subtrajectories, the one with some located at the circle with the center at $p_l'$ and others at $p_{l+1}'$ has the smallest DTW distance from $T_q$ and would be returned. We denote its DTW distance to $T_q$ by $D_a$. It could be verified that $D_a > 2\times (\Sigma_{i=1}^{l-1} ((l-i)\cdot d - \epsilon) + \epsilon)$, where the lower bound of the Euclidean distance from a point of $T_q$ to its aligned point of the returned subtrajectory is (1) $((l-i)\cdot d - \epsilon)$ for $1\le i\le l-1$ and (2) $\epsilon$ for $i = l$; and (3) that of the point symmetric to $p_i'$ (w.r.t. the origin) for $l+1\le i\le m$ due to the symmetry. Therefore, we have
	\begin{align*}
	\frac{D_a}{D_o}
	& \ge \frac{2\times (\Sigma_{i=1}^{l-1} ((l-i)\cdot d - \epsilon) + \epsilon)}{m^2 \cdot \epsilon}\\
	& = \frac{m^2/4\cdot d - m/2\cdot d - m\cdot \epsilon + 4\cdot\epsilon}{m^2\cdot \epsilon} \\
	&= \frac{1/4\cdot d - 1/(2m)\cdot d - \epsilon/m + 4/m^2\cdot \epsilon}{\epsilon}
	\end{align*}
	which approaches infinity when m approaches infinity and $\epsilon$ approaches 0.
	In summary, the solution returned by SizeS could be arbitrarily worse than the optimal one when DTW distance is used.
	
	\smallskip\noindent\textbf{Case 2: Frechet and t2vec.}
	Consider the case when Frechet distance is used. The optimal solution and the approximate solution returned would be the sames as the case when DTW distance is used, but the distances would be different. $D_o$ would be equal to $\epsilon$ and $D_a$ would be at least $((l-1)\cdot d - \epsilon)$. Therefore, we have
	$$
	\frac{D_a}{D_o} \ge \frac{(l-1)\cdot d - \epsilon}{\epsilon}
	$$
	which approaches infinity when $\epsilon$ approaches 0.
	
	Consider the case when t2vec is used. Since it is a learning-based distance - in theory, it may reduce to any possible distance metric such as DTW and Frechet. Thus, the analysis for DTW or Frechet could be carried over for t2vec.

	\section{Quality Analysis of Spliting-based Algorithms}
	\label{sec:splitting-proof}
	
	\smallskip\noindent\textbf{Case 1: PSS with DTW.}
	Consider a problem input with a data trajectory with $n+3$ points $T=<p_1', p_2', p_1, p_2, ..., p_n, p_3'>$ ($n$ is a positive integer) and a query trajectory $T_q = <p'>$. Let $d = d_{max}/2$. The locations of the points in these trajectories are provided as follows.
	\begin{itemize}
		\item $p_1' = (-d/2, 0)$, $p_2' = (-d, 0)$;
		\item $p_i = (0, 0)$ for $i = 1, 2, ..., n$;
		\item $p_3' = (d, 0)$;
		\item $p' = (0, \epsilon)$, where $\epsilon$ is a very small non-negative real number;
	\end{itemize}
	
	Consider the optimal solution. It could be any subtrajectory $<p_i>$ ($1\le i \le n$) and the corresponding DTW distance is equal to $\epsilon$, which we denote by $D_o$.
	
	Consider the approximate solution returned by PSS. It is the subtrajectory $<p_1'>$, which is explained as follows. When it scans the first point $p_1'$, it would split the trajectory at $p_1'$ and update the best-known subtrajectory to be $<p_1'>$ with the DTW distance equal to $\sqrt{d^2/4 + \epsilon^2}$. It then continues to scan the following points $p_2', p_1, ..., p_n, p_3'$ and would not perform any split operations at these points due to the fact that $p_2'$ and $p_3'$ are farther away from $p'$ than $p_1'$. As a result, $<p_1'>$ would be returned as a solution. We denote the DTW distance of this solution by $D_a$, i.e., $D_a = \sqrt{d^2/4 + \epsilon^2}$.
	
	Consider the approximation ratio (AR). We have
	$$
	\frac{D_a}{D_o} = \frac{\sqrt{d^2/4 + \epsilon^2}}{\epsilon} > \frac{d/2}{\epsilon}
	$$
	which approaches infinity when $\epsilon$ approaches zero.
	
	Consider the mean rank (MR). We know that the rank of the approximate solution $<p_1'>$ is at least $\frac{n(n+1)}{2}+1$ since any subtrajectory of $<p_1, p_2, ..., p_n>$ has a smaller DTW distance than $<p_1'>$ (assuming $\epsilon = 0$). Therefore, the mean rank would approach infinity when $n$ approaches infinity.
	
	Consider the relative rank (RR). Based on the analysis of mean rank, we know that the relative rank of the approximate solution $<p_1'>$ is at least $\frac{\frac{n(n+1)}{2}+1}{\frac{(n+3)(n+4)}{2}}$, which approaches 1 when $n$ approaches infinity.
	
	In conclusion, the solution returned by PSS could be arbitrarily worse than the optimal one in terms of AR, MR, and RR, when the DTW distance is used.
	
	\smallskip\noindent\textbf{Case 2: Other Algorithms and Similarity Measurements}
	Consider the other algorithms, i.e., POS and POS-D. It could be verified that they would run exactly in the same way as PSS on the problem input provided in the Case 1. Therefore, the conclusion would be carried over. Consider the other measurements, namely, Frechet and t2vec. For Frechet, it could be verified that the optimal solution and the approximate solution returned by the algorithms are the same as those in the Case 1 and their Frechet distances and DTW distances to $T_q$ are equal since both solutions and $T_q$ involve one single point. As a result, the conclusion for DTW distance could be carried over for Frechet distance. For t2vec, since it is a learning-based distance - in theory, it may reduce to any possible distance metric such as DTW and Frechet. Thus, the analysis for DTW in the Case 1 could be carried over for t2vec.

\section{Adaption of UCR}
\label{section:ucr-and-spring}
UCR~\cite{rakthanmanon2012searching} was originally developed for searching subsequences of a time series, which are the most similar to a query time series and the similarity is based on the DTW distance. Specifically, UCR enumerates all subsequences that are of the same length of the query time series and employs a rich set of techniques for pruning many of the subsequences.
We use UCR for our ``similar subtrajectory search'' problem by adapting UCR's pruning techniques for trajectories. UCR involves seven techniques used for pruning, organized in two groups.
Let $T = <p_1, p_2, ..., p_n>$ be a data trajectory and $T_q = <q_1, q_2, ..., q_m>$ be a query trajectory. Suppose we are considering a subtrajectory of $T$, denoted by $T'=<p_1, p_2, ..., p_m>$ without loss of generality. We describe the adaptions of the techniques involved in UCR, which are used to prune $T'$ from being considered if possible as follows.
\\
Group 1: Known Optimizations
\begin{itemize}
  \item Early Abandoning of $LB_{Keogh}$.
  This is to compute a lower bound called $LB_{Keogh}$ of the DTW distance between $T'$ and $T_q$, denoted by $LB_{Keogh}(T', T_q)$, and prune $T'$ if the lower bound is larger than the best-known DTW distance. Specifically, let (1) $R\in[0,1]$ be a real number, (2)
  $q_{i-\lfloor R\cdot m\rfloor: i+\lfloor R\cdot m\rfloor}$ be the set involving $\lfloor R\cdot m\rfloor$ points before $q_i$, $\lfloor R\cdot m\rfloor$ points after $q_i$, and point $p_i$, and (3) $MBR(\cdot)$ be the minimum bounding box of a set of points.
  We compute $LB_{Keogh}(T', T_q)$ as follows.
  \begin{small}
  	\begin{equation*}
  	LB_{Keogh}(T', T_q)= \sum_{i=1}^{m} \left\{
  	\begin{aligned}
  	& d(p_i, MBR(q_{i-\lfloor R\cdot m\rfloor: i+\lfloor R\cdot m\rfloor})) \\
  	& \text{ if $p_i$ outside $MBR(q_{i-\lfloor R\cdot m\rfloor: i+\lfloor R\cdot m\rfloor})$}\\
  	& 0 \text{ otherwise}
  	\end{aligned}
  	\right.
  	\end{equation*}
  \end{small}
  where $d(p_i, MBR(\cdot))$ denotes shortest distance between $p_i$ and $MBR(\cdot)$.

  \item Early Abandoning of DTW. During the process of computing the DTW distance between $T'$ and $T_q$, when the accumulated DTW distance exceeds the best-known DTW distance, we abandon the computation.

  \item Earlier Early Abandoning of DTW using $LB_{Keogh}$.
  For $i = 1, 2, ..., m$, if the sum of the DTW distance between $T_q[1:i]$ and any prefix of $T'[1:i]$ (or between $T'[1:i]$ and any prefix of $T_q[1:i]$) and the $LB_{Keogh}$ bound between $T_q[i:m]$ and $T'[i:m]$ is larger than the best-known DTW distance, we prune $T'$. Note that DTW distances used in this pruning are maintained in the process of computing the DTW distance between $T'$ and $T_q$.
\end{itemize}
Group 2: Novel Optimizations in UCR Suite.
\begin{itemize}
  \item Just-in-time Z-normalizations. We do not adapt the Z-normalization technique since it is designed for one-dimensional data and cannot be used for trajectory data that is two-dimensional.
  \item Reordering Early Abandoning.
  We consider the points of $T_q$ in a descending order of their distances to the y-axis for computing the bound of $LB_{Keogh}$.

  \item Reversing $LB_{Keogh}$.
  We reverse the roles of $T'$ and $T_q$ and compute another $LB_{Koegh}$ bound. We then use the larger one among the two $LB_{Koegh}$ bounds to act as a tighter $LB_{Keogh}$ bound when necessary.

  \item Cascading Lower Bounds.
  We first compute the $LB_{Kim}FL$ bound, which is another simple lower bound of the DTW distance between $T'$ and $T_q$, denoted by $LB_{Kim}FL(T', T_q)$. Specifically, $LB_{Kim}FL(T', T_q)$ is defined as  $(d(T_q[1], T'[1]) + d(T_q[m], T'[m]))$. The time complexity of this step is simply $O(1)$. If this bound does not help to prune $T'$, we cascade the techniques of early abandoning of $LB_{Keogh}$, early abandoning of DTW, and earlier early abandoning of DTW using $LB_{Keogh}$. Finally, if $T'$ has still not been pruned, we compute the DTW distance between $T'$ and $T_q$.
\end{itemize}
We note that UCR is designed specifically for the DTW distance and cannot be used for the problem when the Frechet or t2vec distance is used.

\section{Additional Experimental Results}
\begin{figure*}[!t]
	\hspace*{-.5cm}
	\centering
	\begin{tabular}{c c c c c c}
		\begin{minipage}{2.6cm}
			\includegraphics[width=2.85cm]{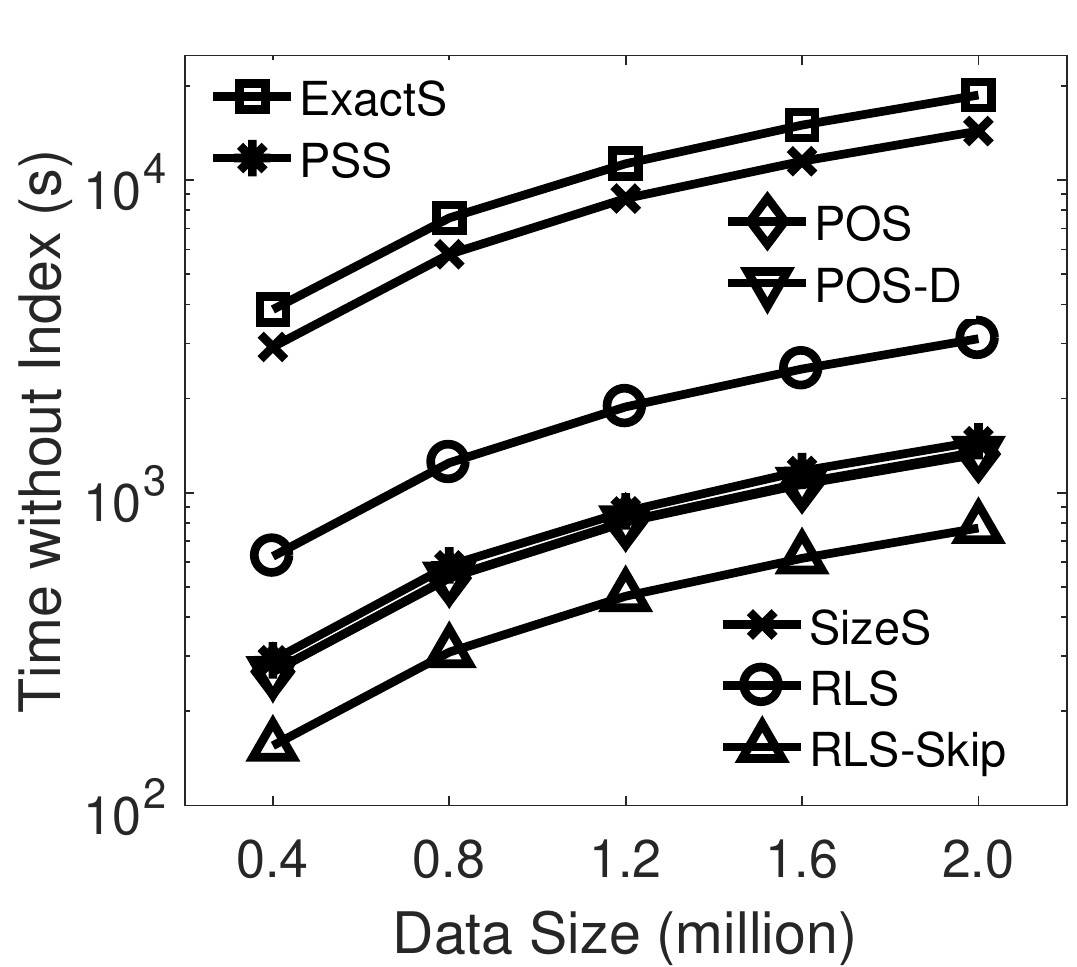}
		\end{minipage}
		&
		\begin{minipage}{2.6cm}
			\includegraphics[width=2.85cm]{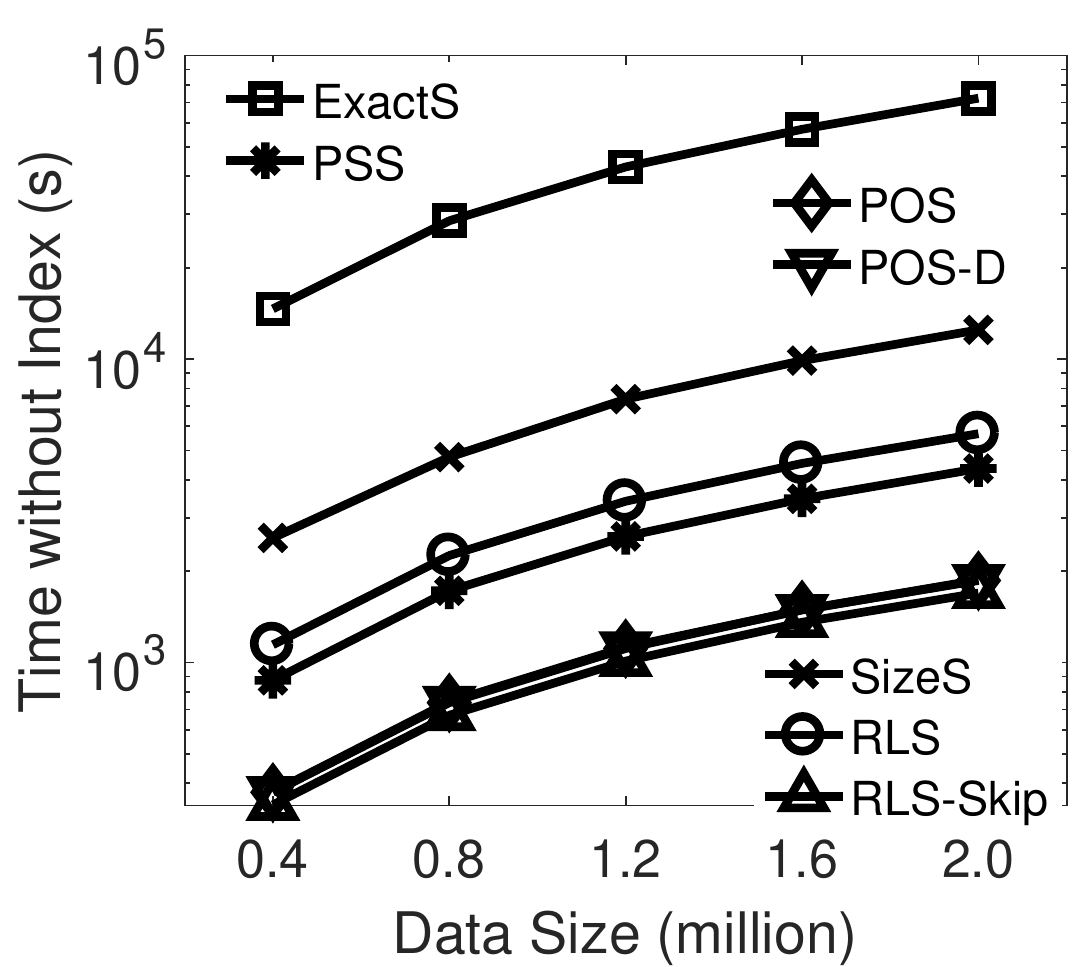}
		\end{minipage}
		&
		\begin{minipage}{2.6cm}
			\includegraphics[width=2.85cm]{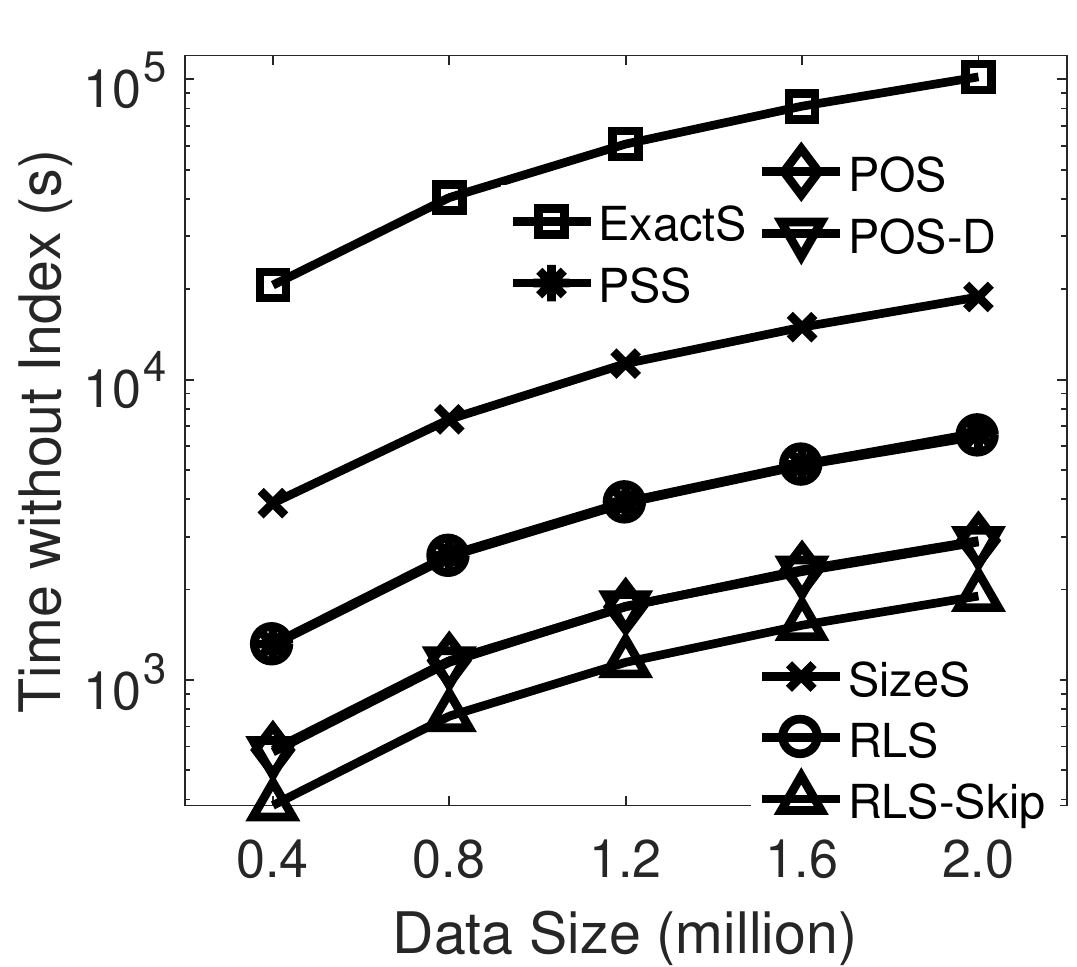}
		\end{minipage}
        &
        \begin{minipage}{2.6cm}
			\includegraphics[width=2.85cm]{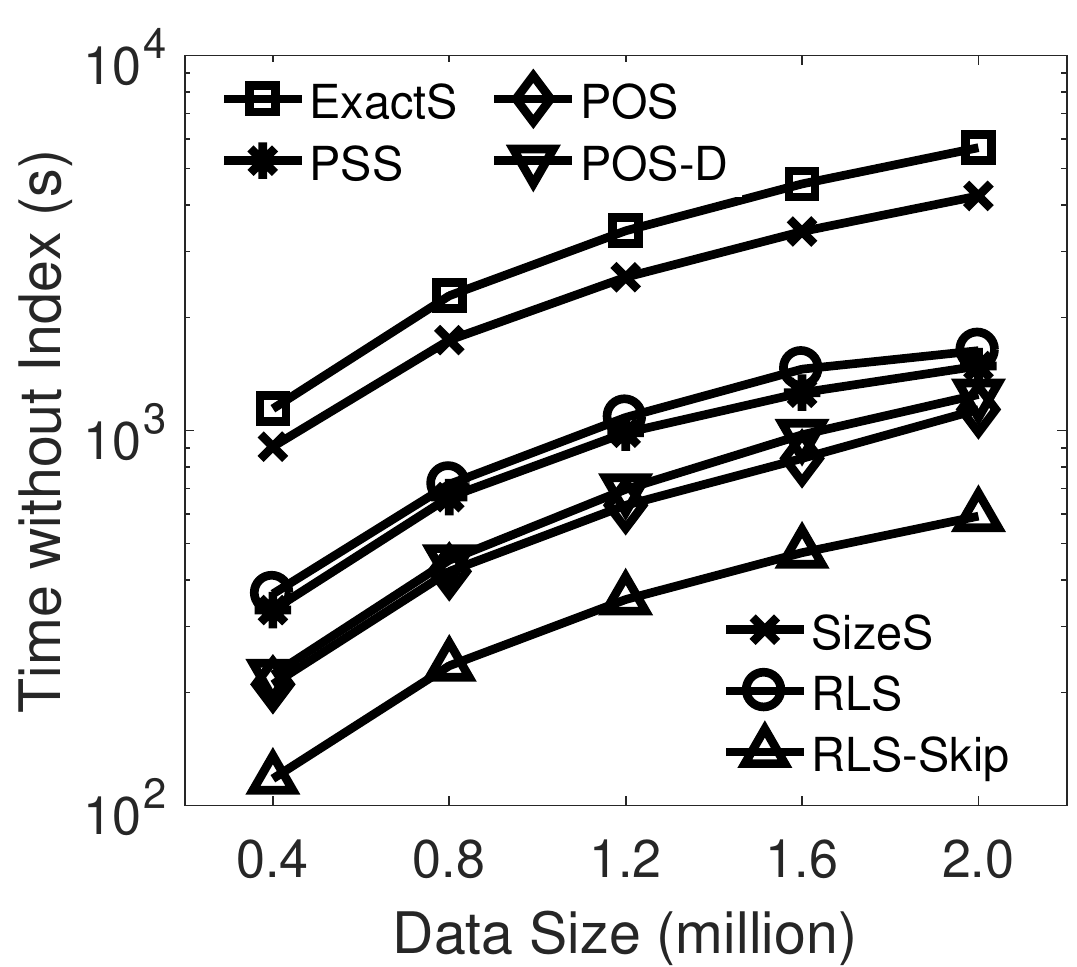}
		\end{minipage}
		&
		\begin{minipage}{2.6cm}
			\includegraphics[width=2.85cm]{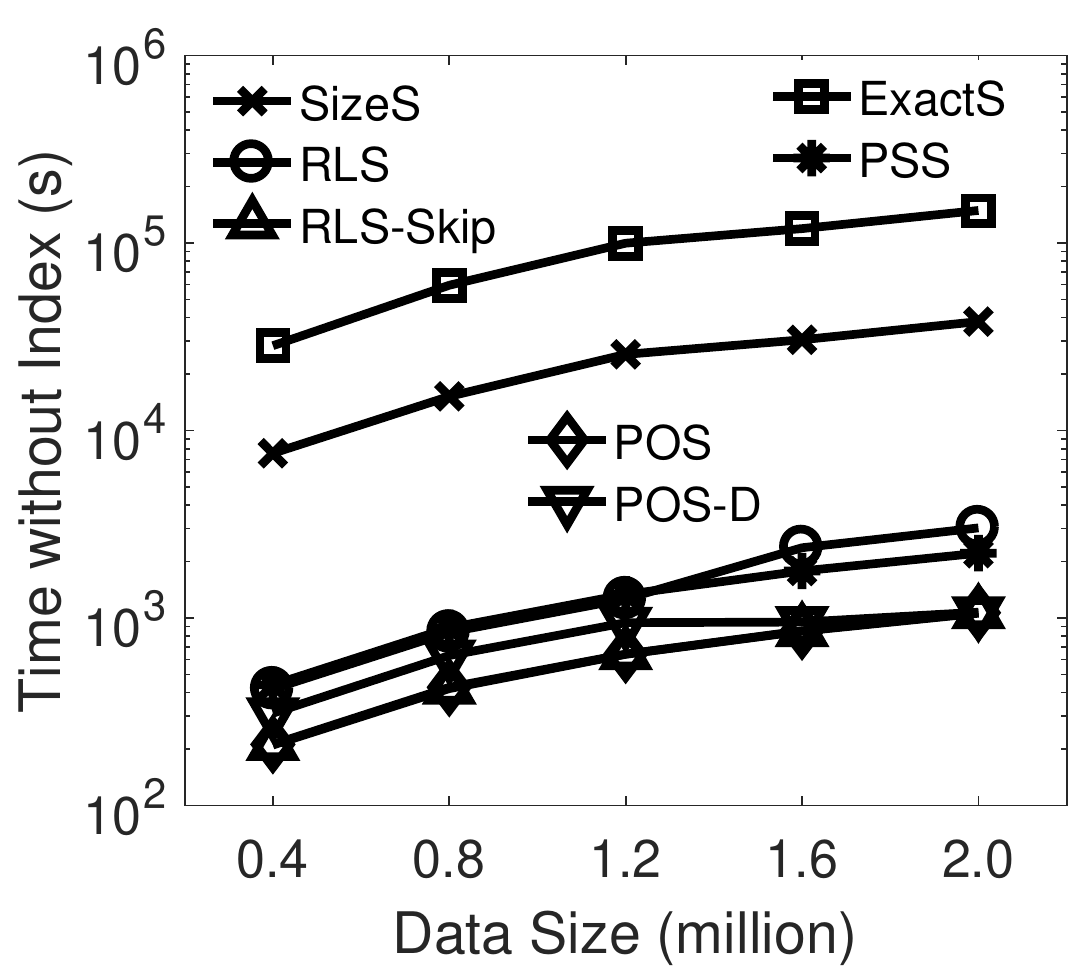}
		\end{minipage}
		&
		\begin{minipage}{2.6cm}
			\includegraphics[width=2.85cm]{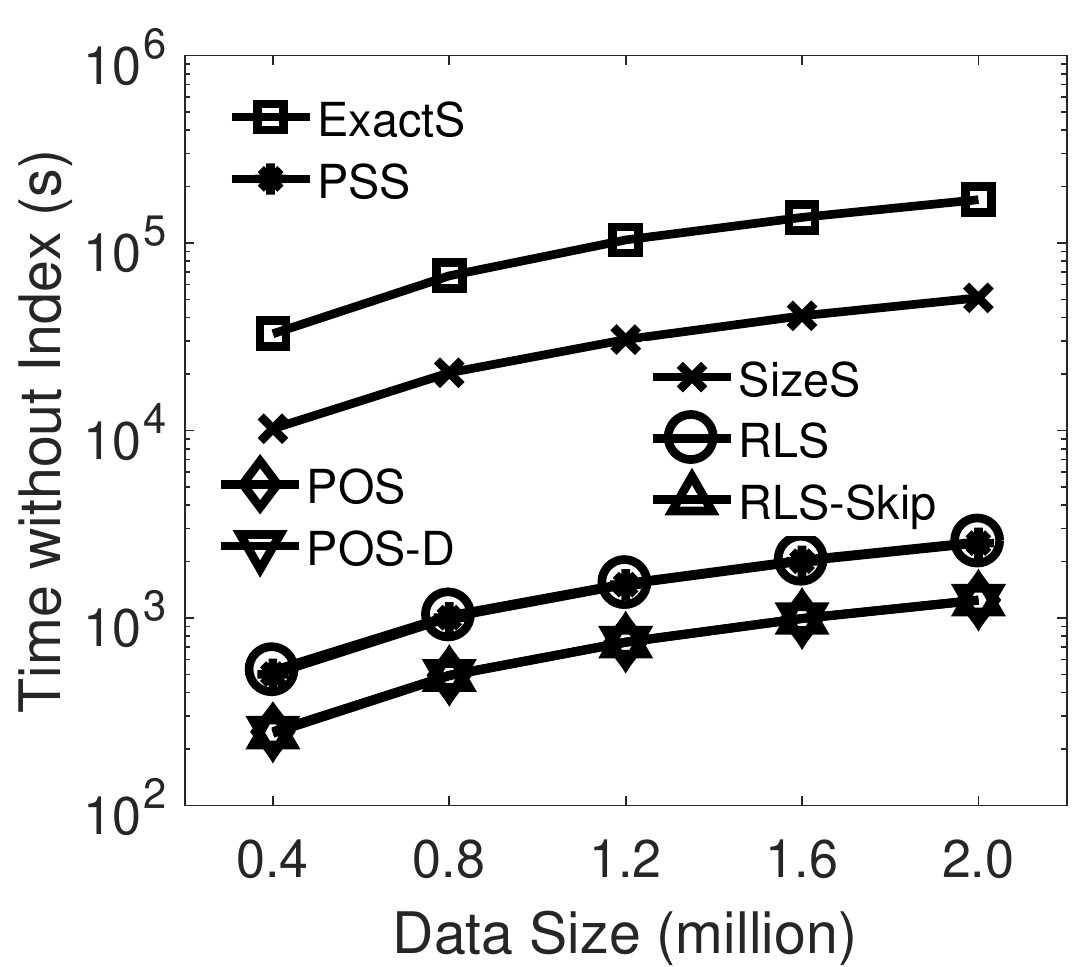}
		\end{minipage}
		\\
		\scriptsize (a) Harbin (t2vec)
        &
		\scriptsize (b) Harbin (DTW)
		&
		\scriptsize (c) Harbin (Frechet)
        &
		\scriptsize (d) Sports (t2vec)
        &
		\scriptsize (e) Sports (DTW)
		&
		\scriptsize (f) Sports (Frechet)
		\\
		\begin{minipage}{2.6cm}
			\includegraphics[width=2.85cm]{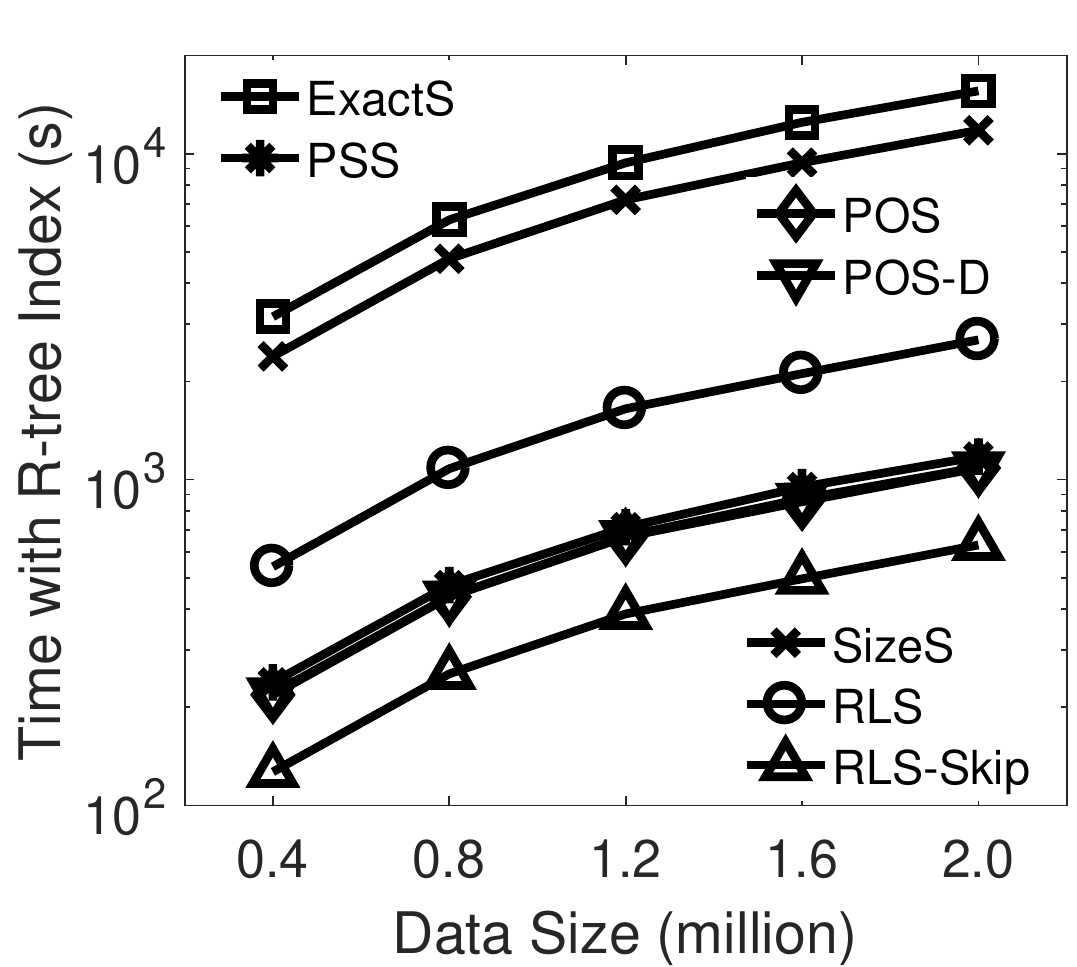}
		\end{minipage}
		&
		\begin{minipage}{2.6cm}
			\includegraphics[width=2.85cm]{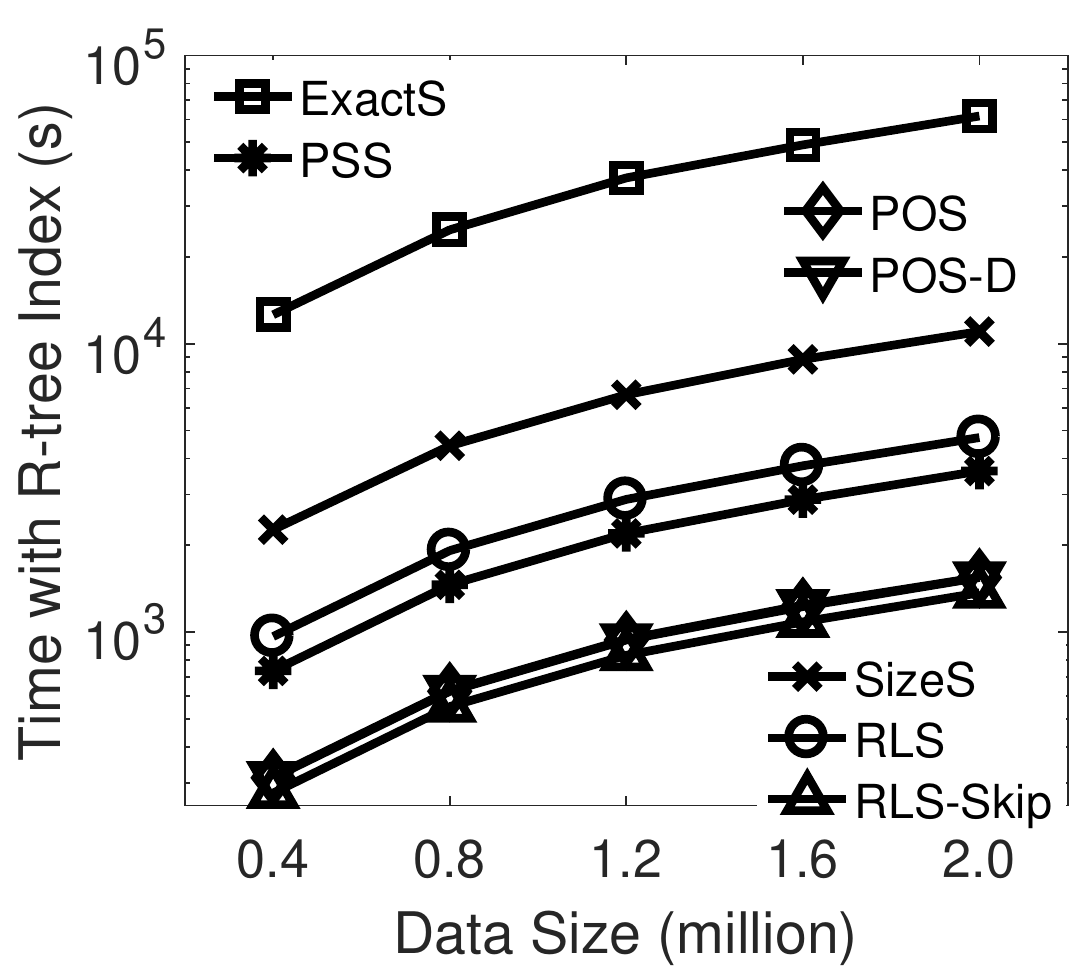}
		\end{minipage}
		&
		\begin{minipage}{2.6cm}
			\includegraphics[width=2.85cm]{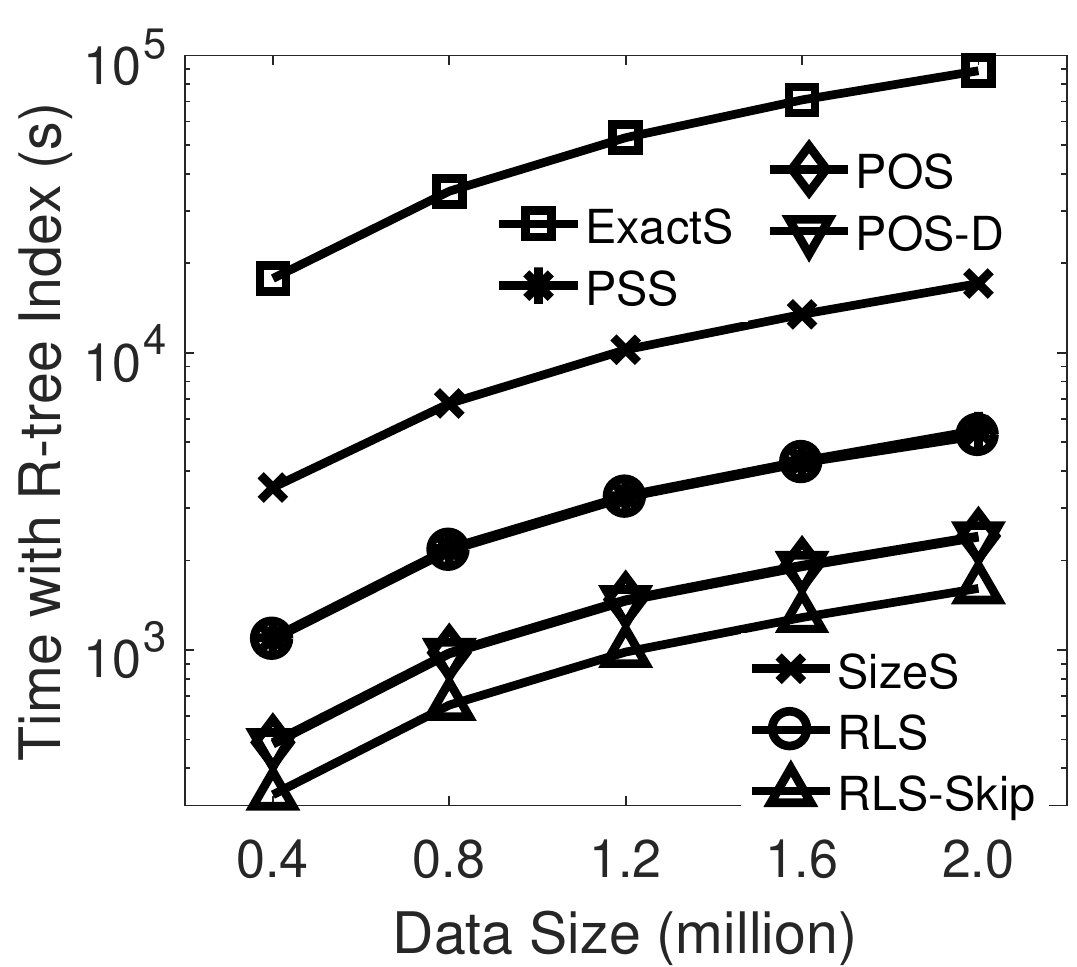}
		\end{minipage}
        &
        \begin{minipage}{2.6cm}
			\includegraphics[width=2.85cm]{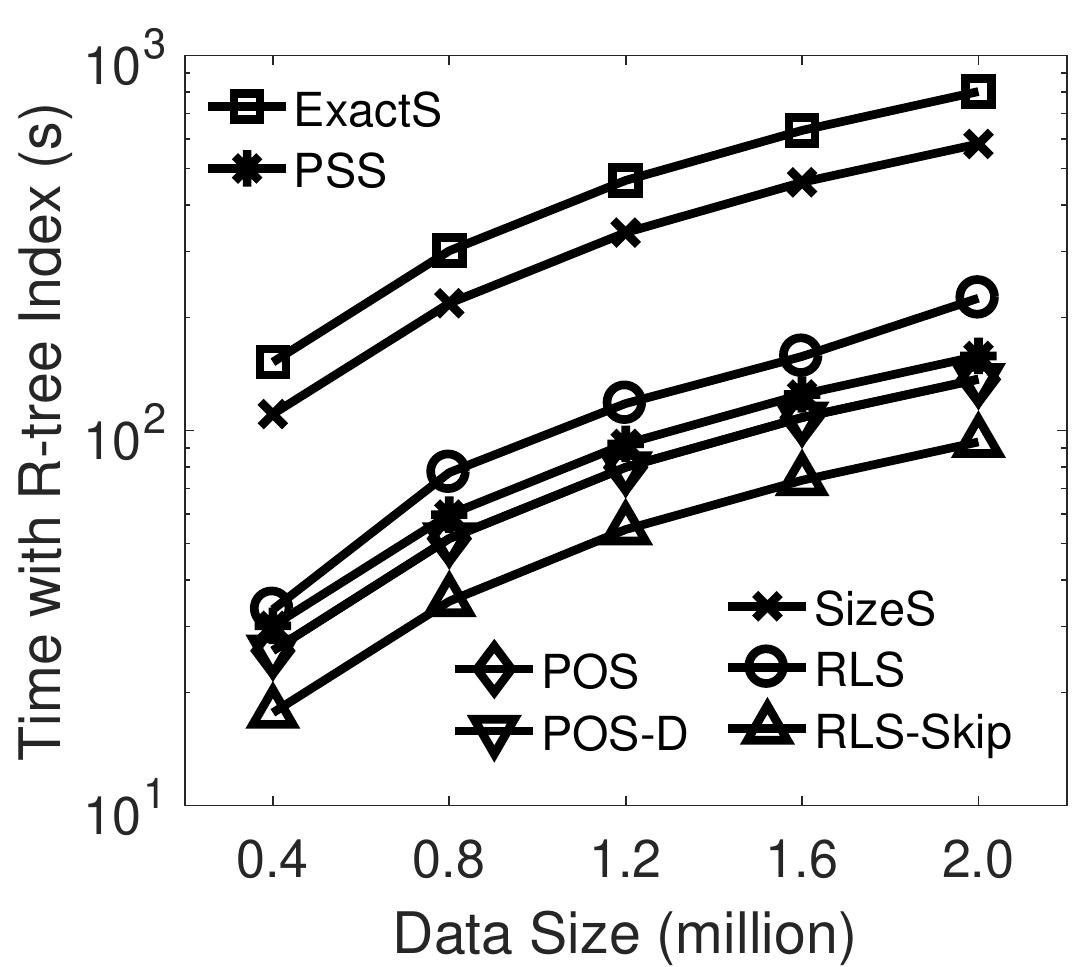}
		\end{minipage}
		&
		\begin{minipage}{2.6cm}
			\includegraphics[width=2.85cm]{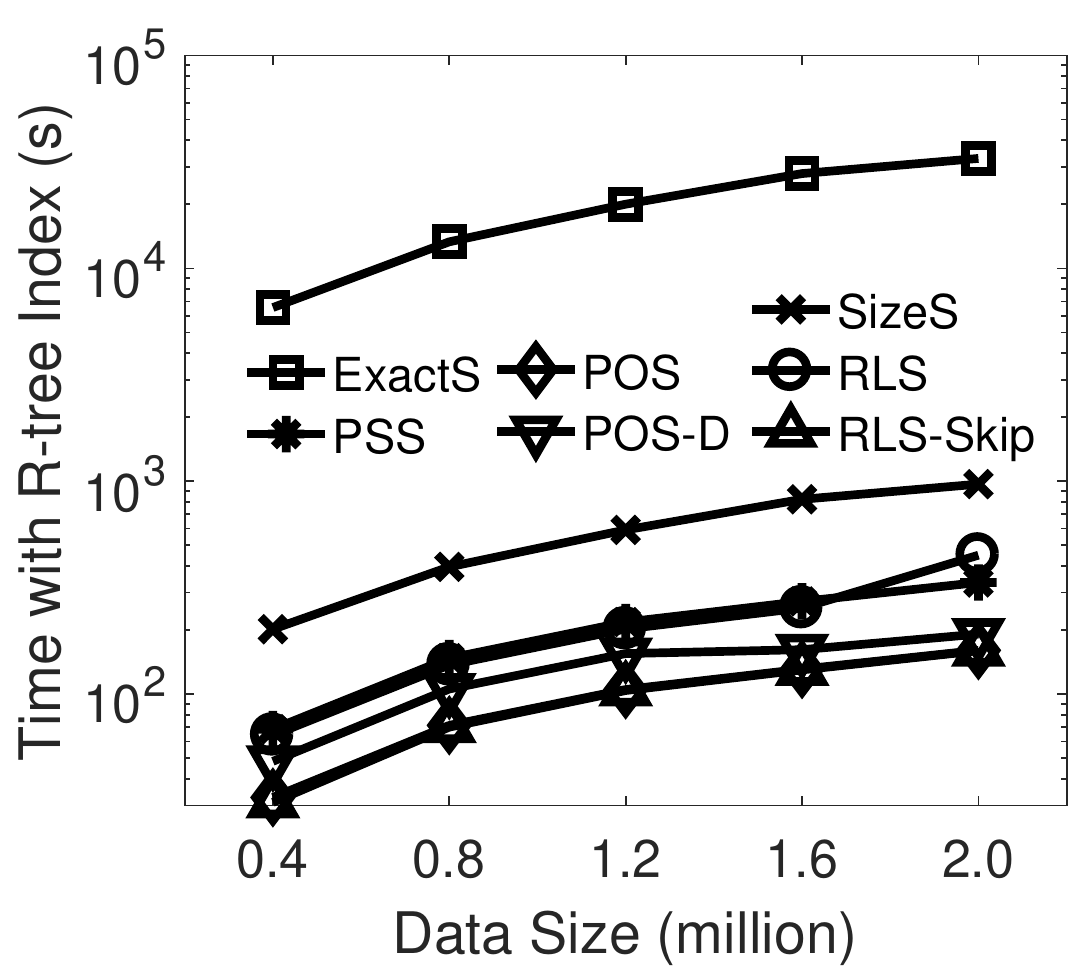}
		\end{minipage}
		&
		\begin{minipage}{2.6cm}
			\includegraphics[width=2.85cm]{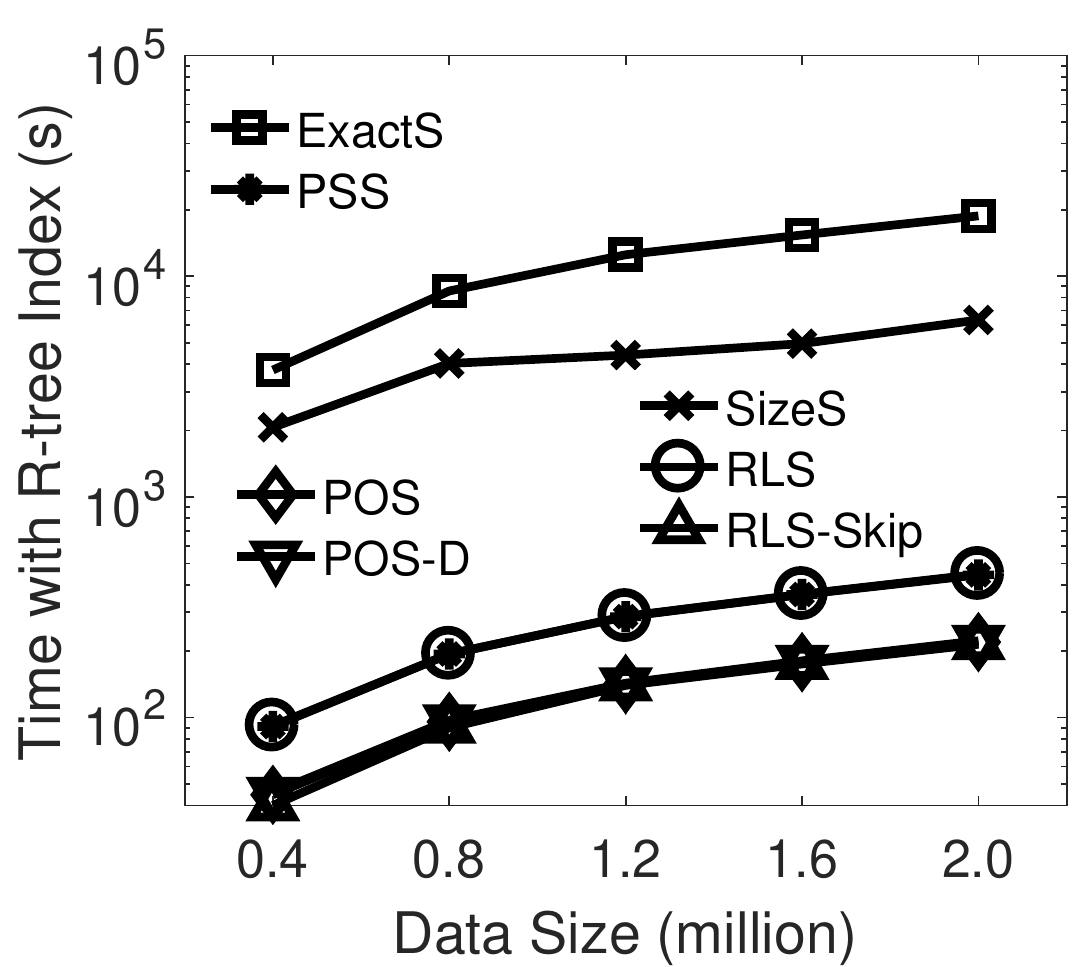}
		\end{minipage}
		\\
		\scriptsize (g) Harbin (t2vec)
		&
		\scriptsize (h) Harbin (DTW)
		&
		\scriptsize (i) Harbin (Frechet)
        &
		\scriptsize (j) Sports (t2vec)
		&
		\scriptsize (k) Sports (DTW)
		&
		\scriptsize (l) Sports (Frechet)

	\end{tabular}
	\caption{Efficiency without index (a)-(f) and with R-tree index (g)-(l) on Harbin and Sports.}
	\label{fig:efficiencyRLS_hs}
\end{figure*}
\begin{figure*}[!ht]
	\hspace*{-.5cm}
	\centering
	\begin{tabular}{c c c c c c}
		\begin{minipage}{2.6cm}
			\includegraphics[width=2.85cm]{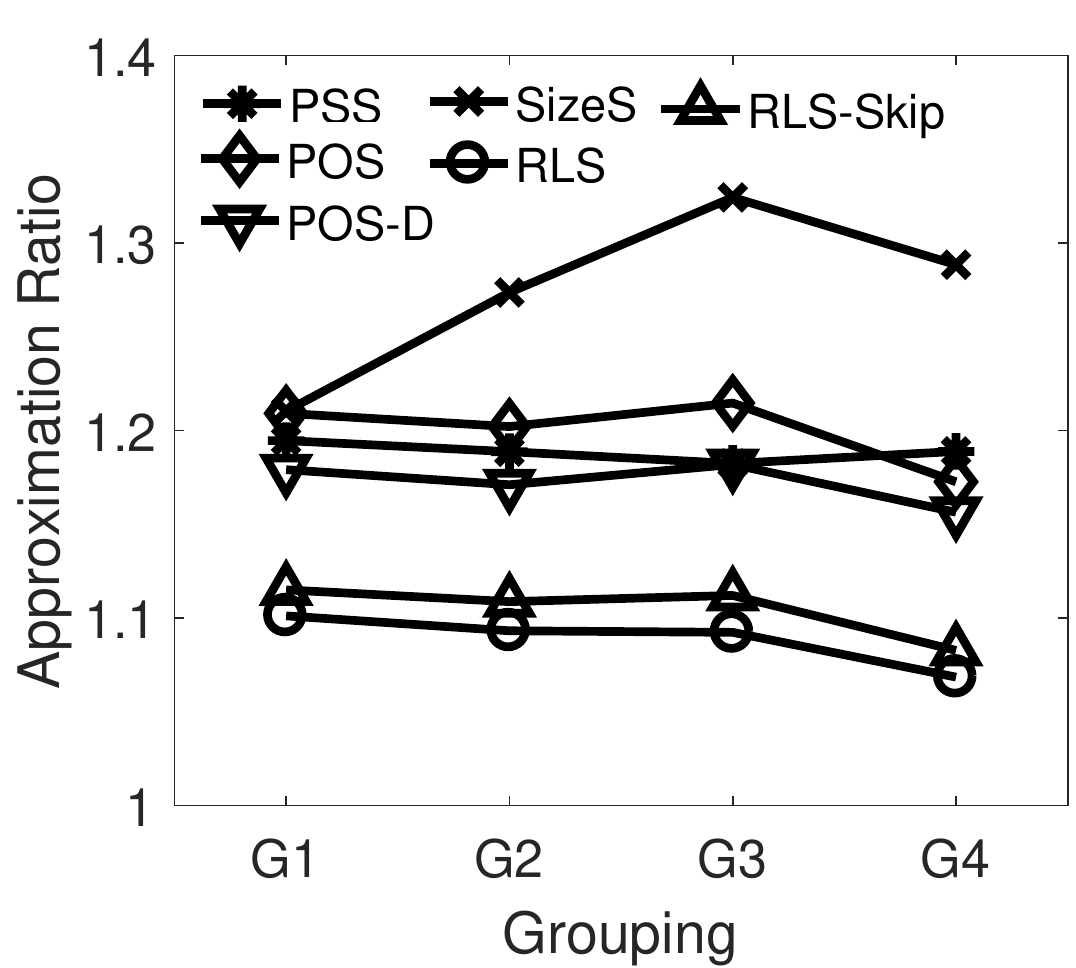}
		\end{minipage}
		&
		\begin{minipage}{2.6cm}
			\includegraphics[width=2.85cm]{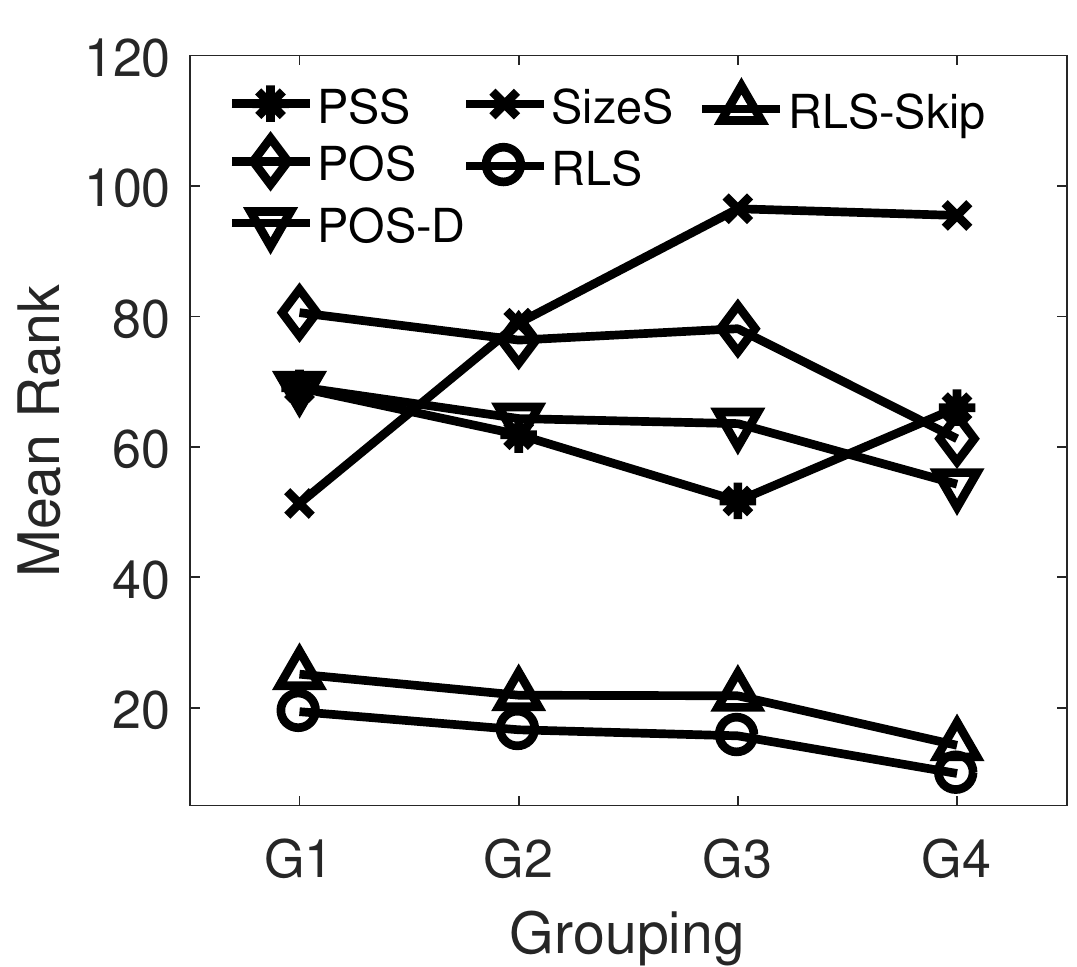}
		\end{minipage}
		&
		\begin{minipage}{2.6cm}
			\includegraphics[width=2.85cm]{t2v_rr_p}
		\end{minipage}
		&
		\begin{minipage}{2.6cm}
			\includegraphics[width=2.85cm]{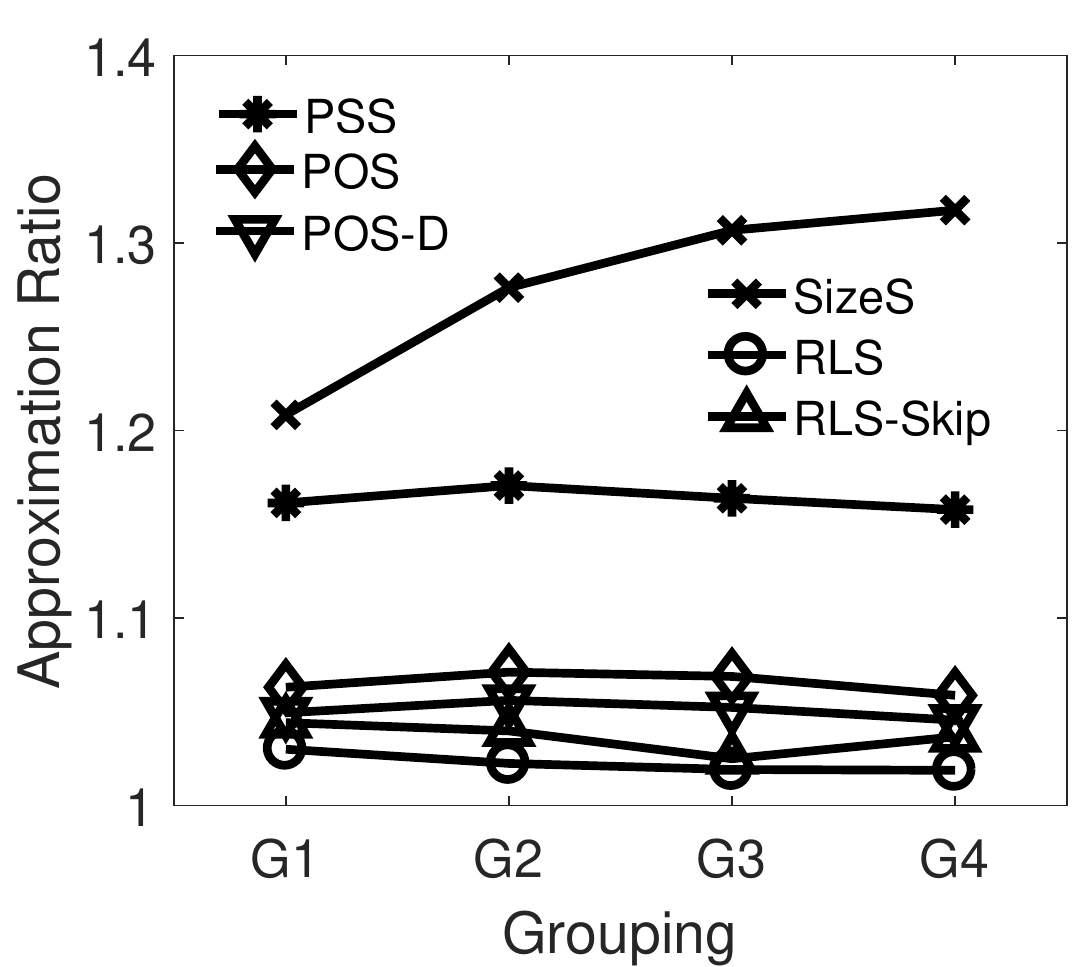}
		\end{minipage}
		&
		\begin{minipage}{2.6cm}
			\includegraphics[width=2.85cm]{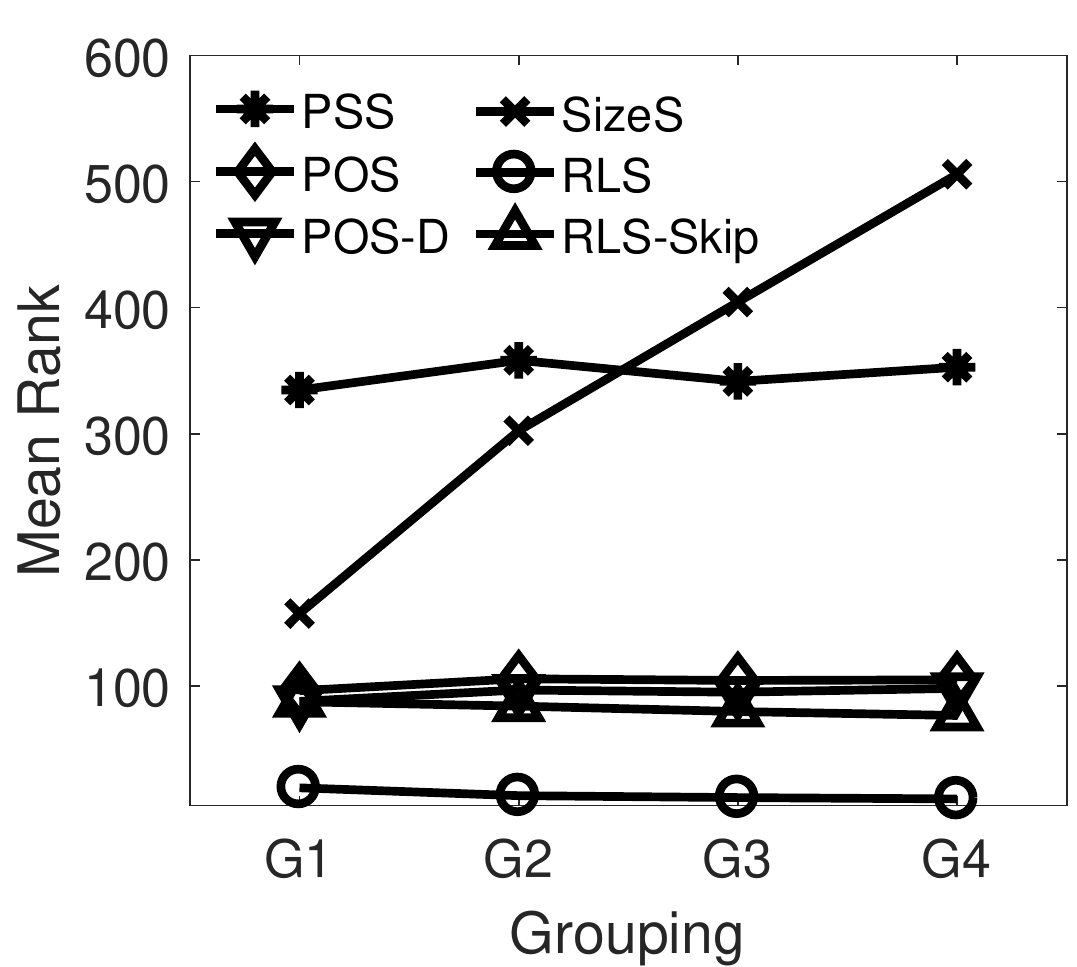}
		\end{minipage}
		&
		\begin{minipage}{2.6cm}
			\includegraphics[width=2.85cm]{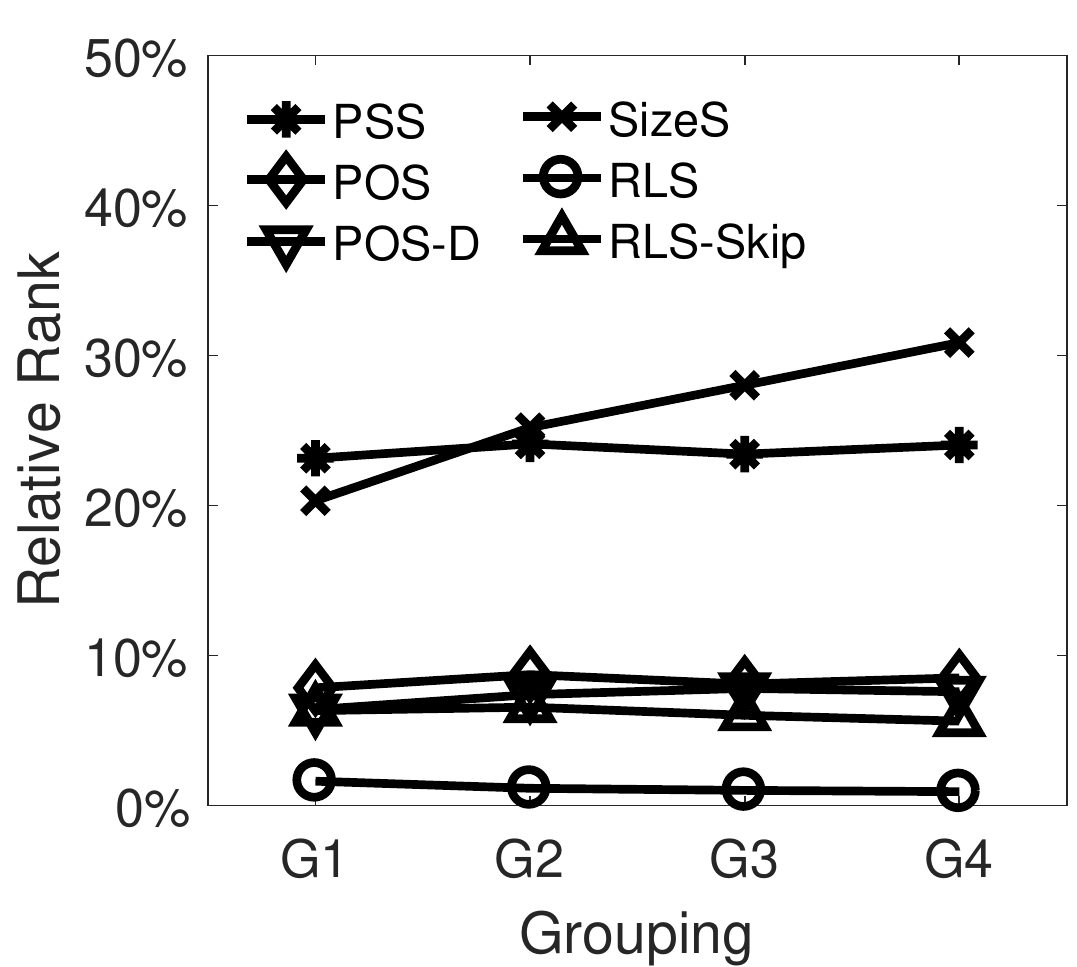}
		\end{minipage}
		\\
		\scriptsize (a) AR Porto
		&
		\scriptsize (b) MR Porto
		&
		\scriptsize (c) RR Porto
		&
		\scriptsize (d) AR Harbin
		&
		\scriptsize (e) MR Harbin
		&
		\scriptsize (f) RR Harbin
		\\
		
		\begin{minipage}{2.6cm}
			\includegraphics[width=2.85cm]{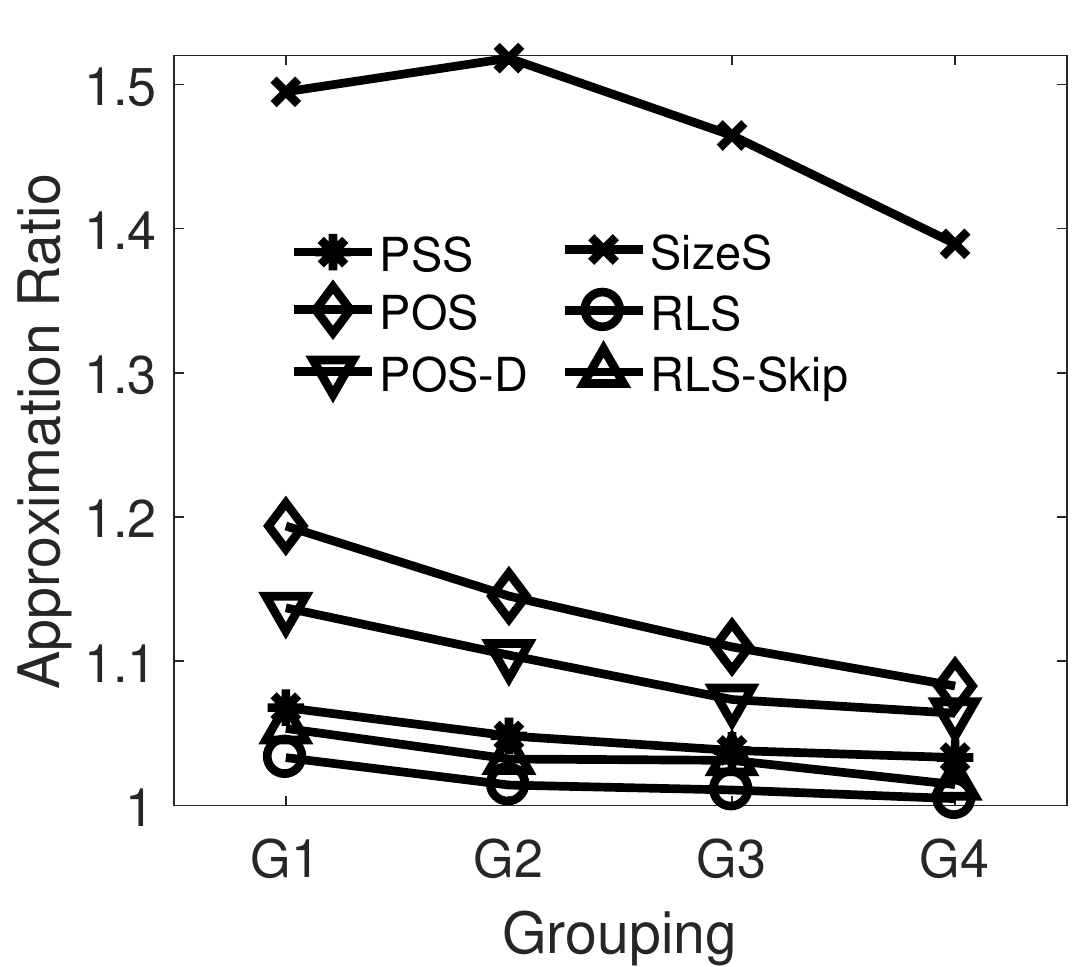}
		\end{minipage}
		&
		\begin{minipage}{2.6cm}
			\includegraphics[width=2.85cm]{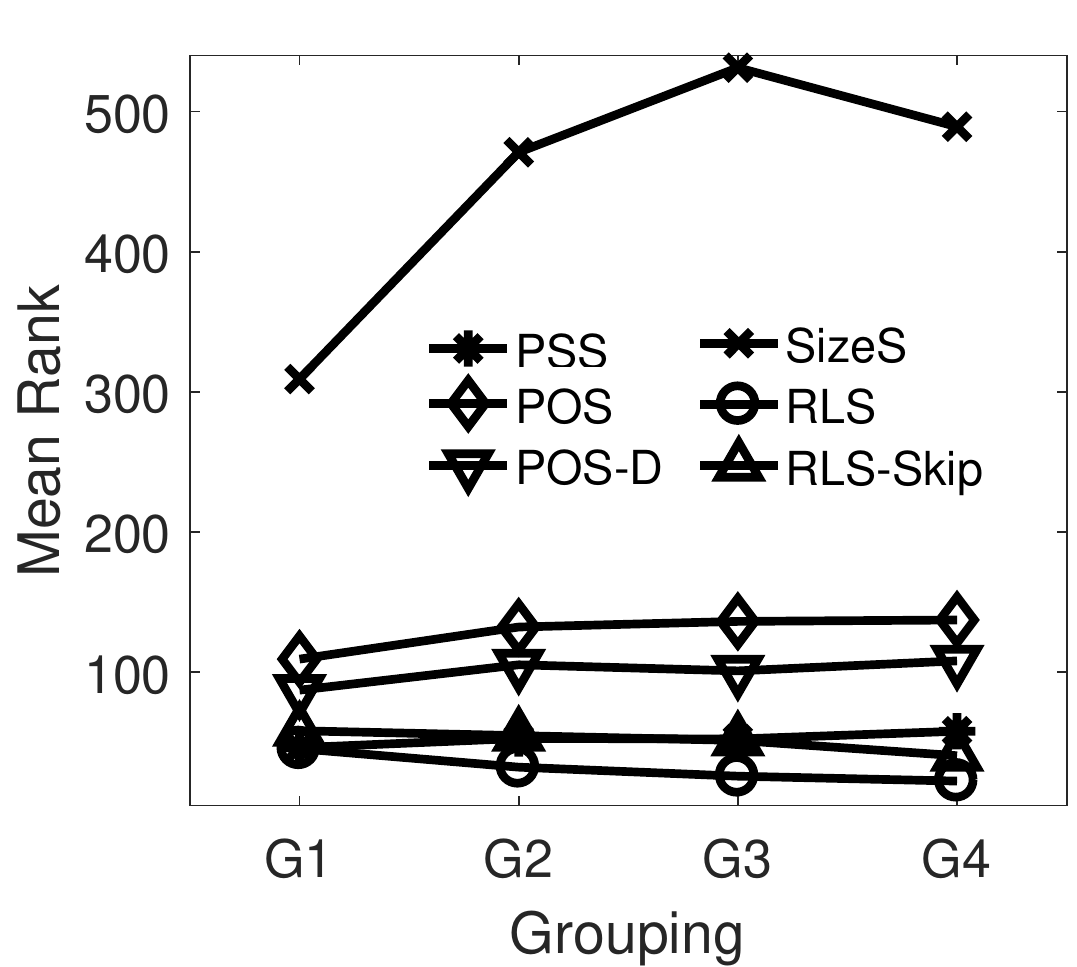}
		\end{minipage}
		&
		\begin{minipage}{2.6cm}
			\includegraphics[width=2.85cm]{dtw_rr_p}
		\end{minipage}
		&
		\begin{minipage}{2.6cm}
			\includegraphics[width=2.85cm]{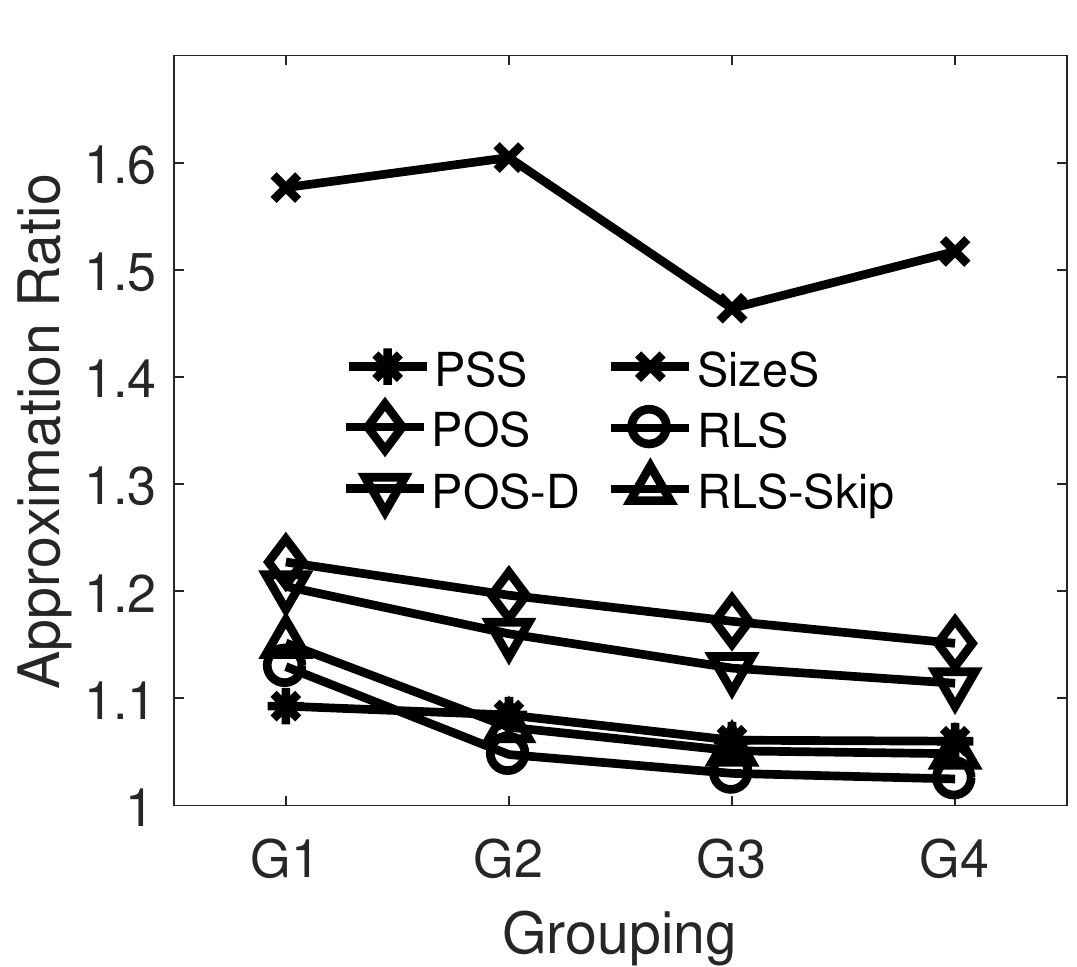}
		\end{minipage}
		&
		\begin{minipage}{2.6cm}
			\includegraphics[width=2.85cm]{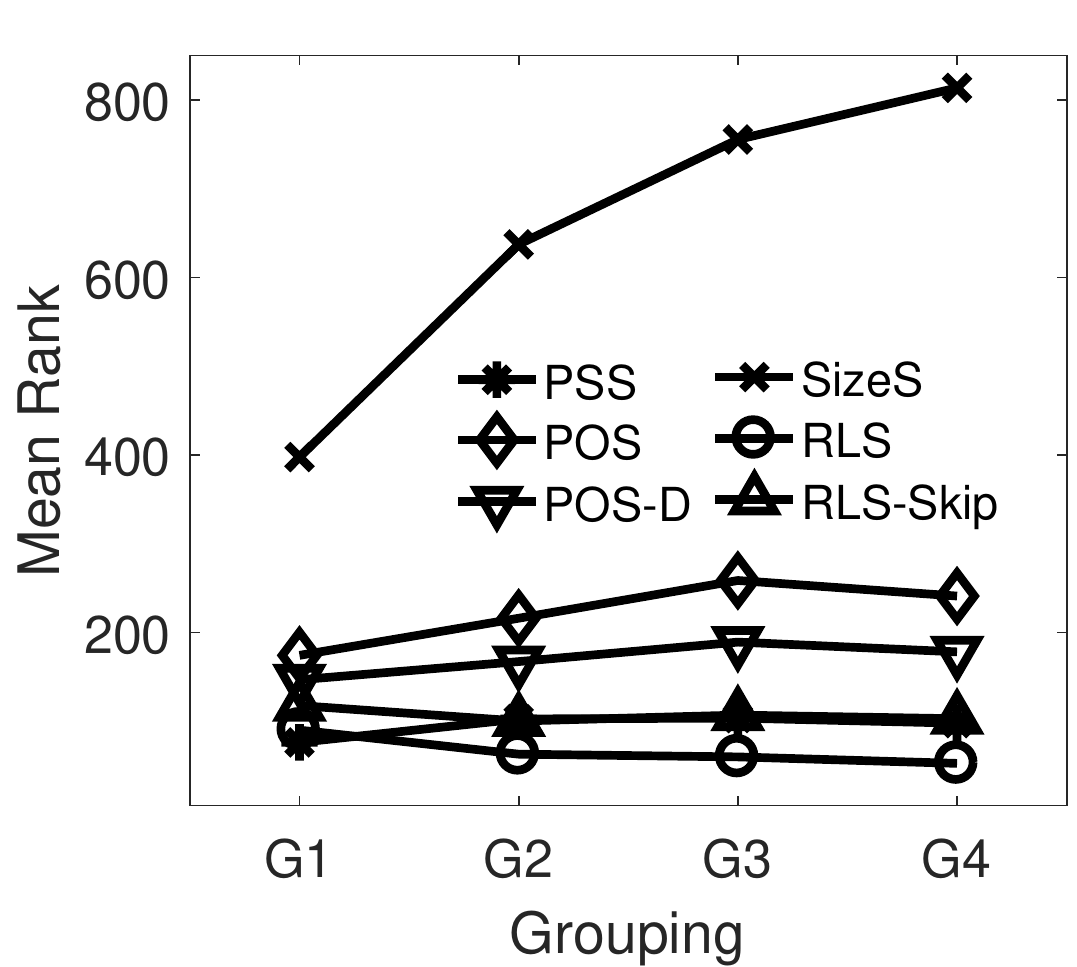}
		\end{minipage}
		&
		\begin{minipage}{2.6cm}
			\includegraphics[width=2.85cm]{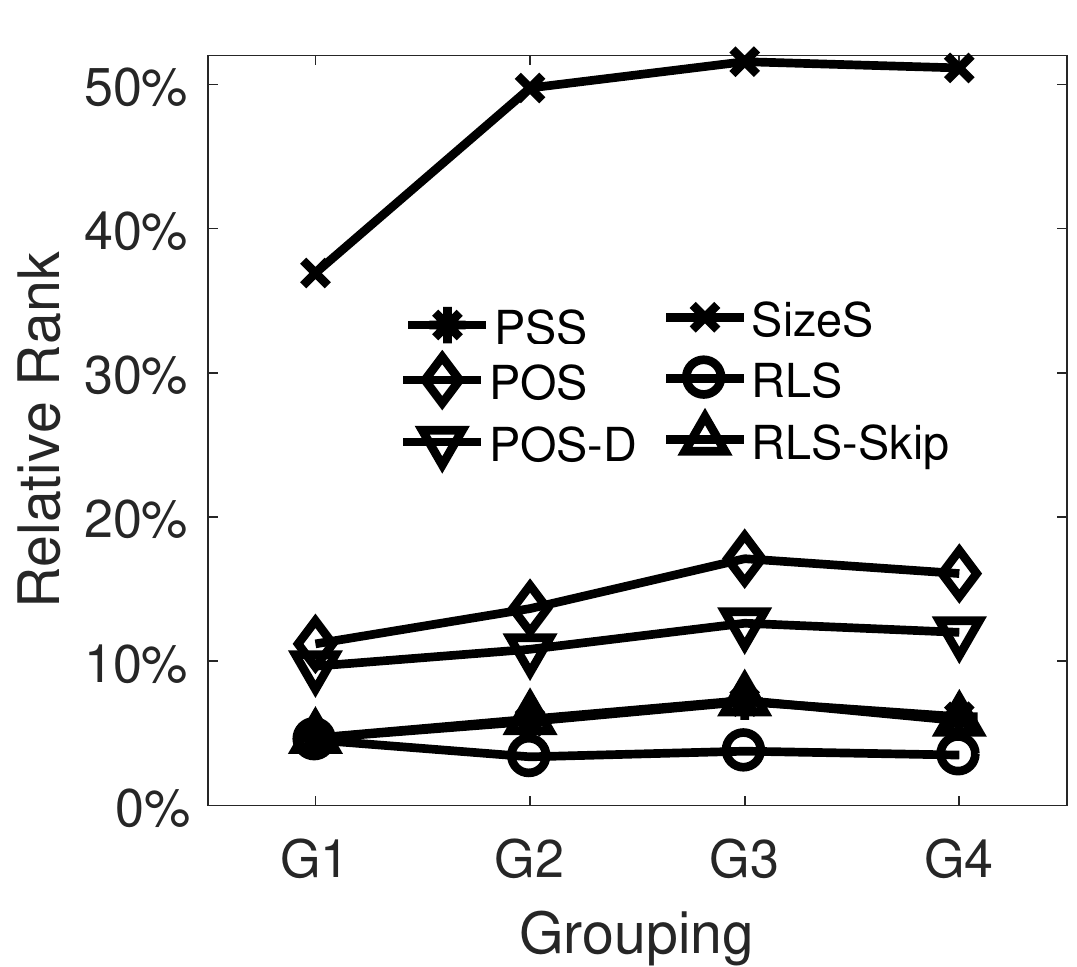}
		\end{minipage}
		\\
		\scriptsize (g) AR Porto
		&
		\scriptsize (h) MR Porto
		&
		\scriptsize (i) RR Porto
		&
		\scriptsize (j) AR Harbin
		&
		\scriptsize (k) MR Harbin
		&
		\scriptsize (l) RR Harbin
		\\
		\begin{minipage}{2.6cm}
			\includegraphics[width=2.85cm]{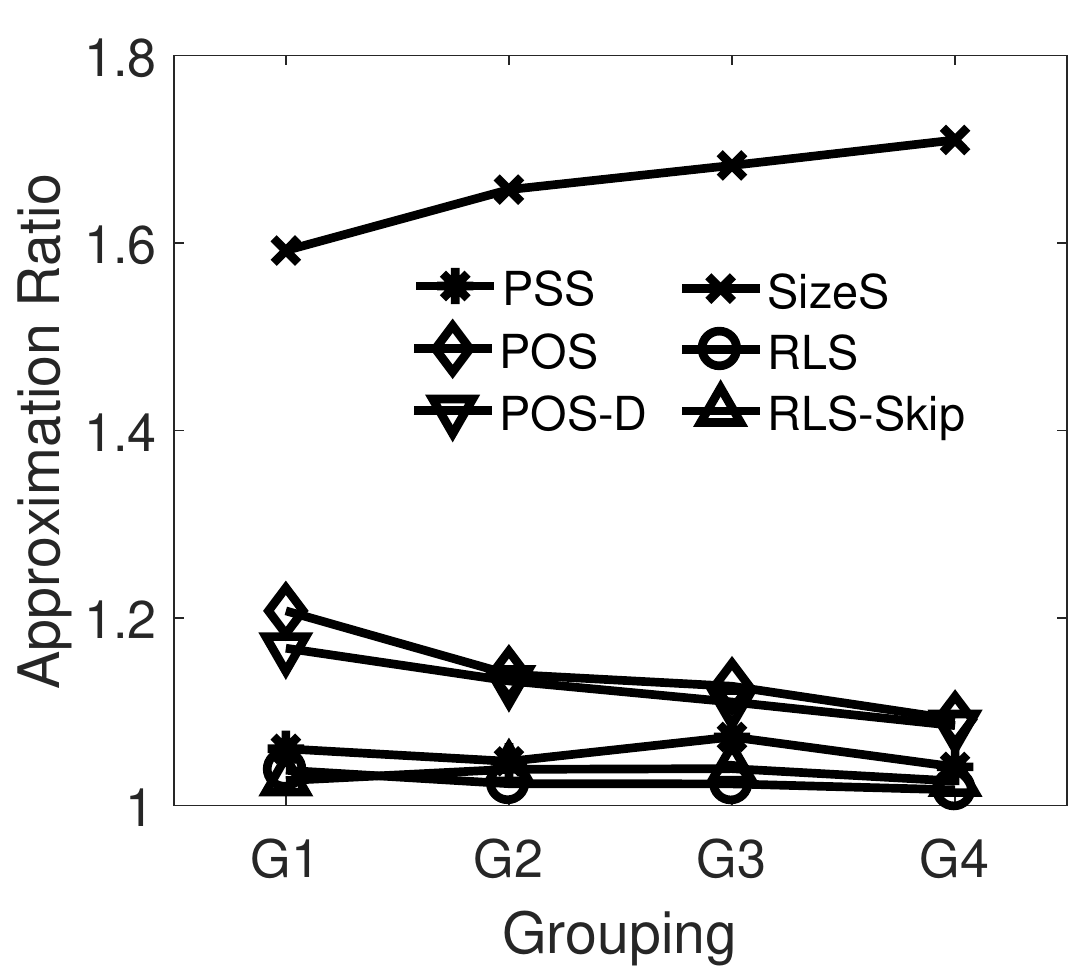}
		\end{minipage}
		&
		\begin{minipage}{2.6cm}
			\includegraphics[width=2.85cm]{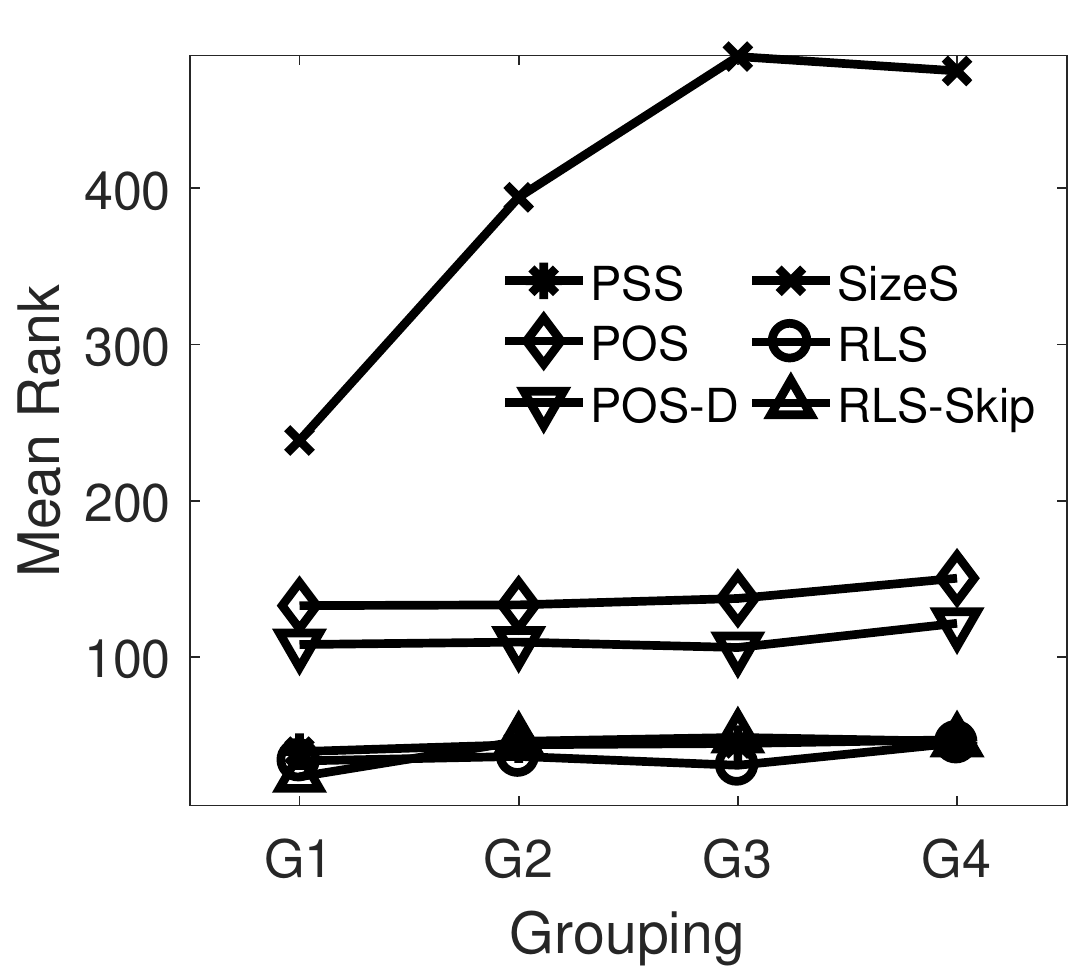}
		\end{minipage}
		&
		\begin{minipage}{2.6cm}
			\includegraphics[width=2.85cm]{fre_rr_p}
		\end{minipage}
		&
		\begin{minipage}{2.6cm}
			\includegraphics[width=2.85cm]{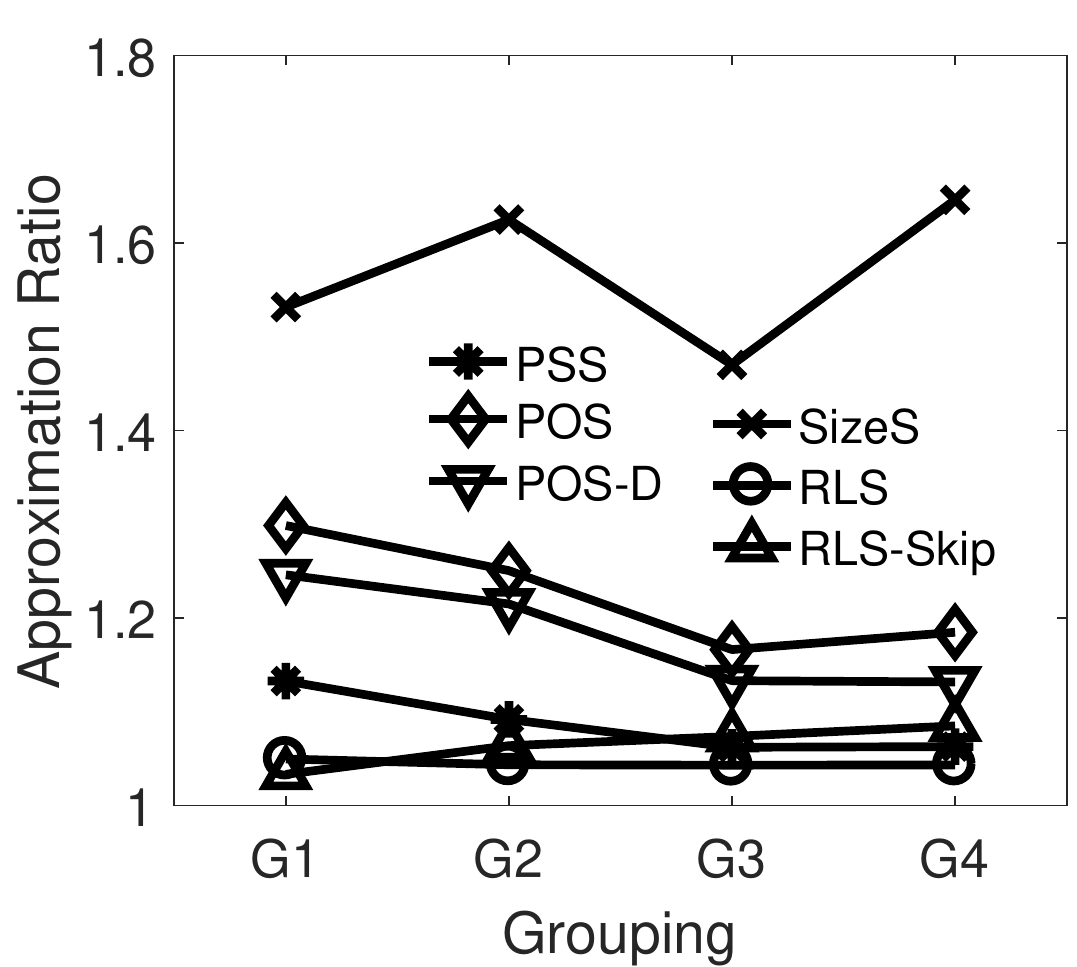}
		\end{minipage}
		&
		\begin{minipage}{2.6cm}
			\includegraphics[width=2.85cm]{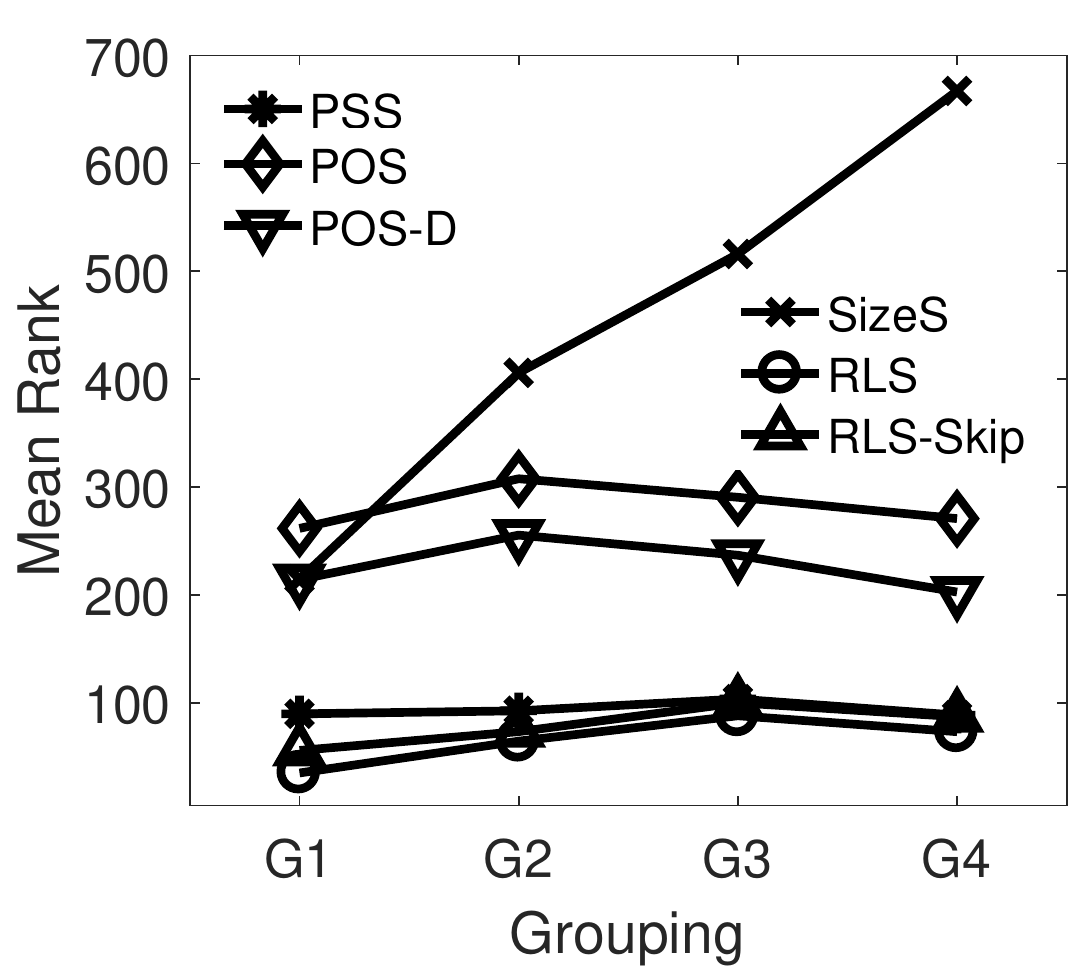}
		\end{minipage}
		&
		\begin{minipage}{2.6cm}
			\includegraphics[width=2.85cm]{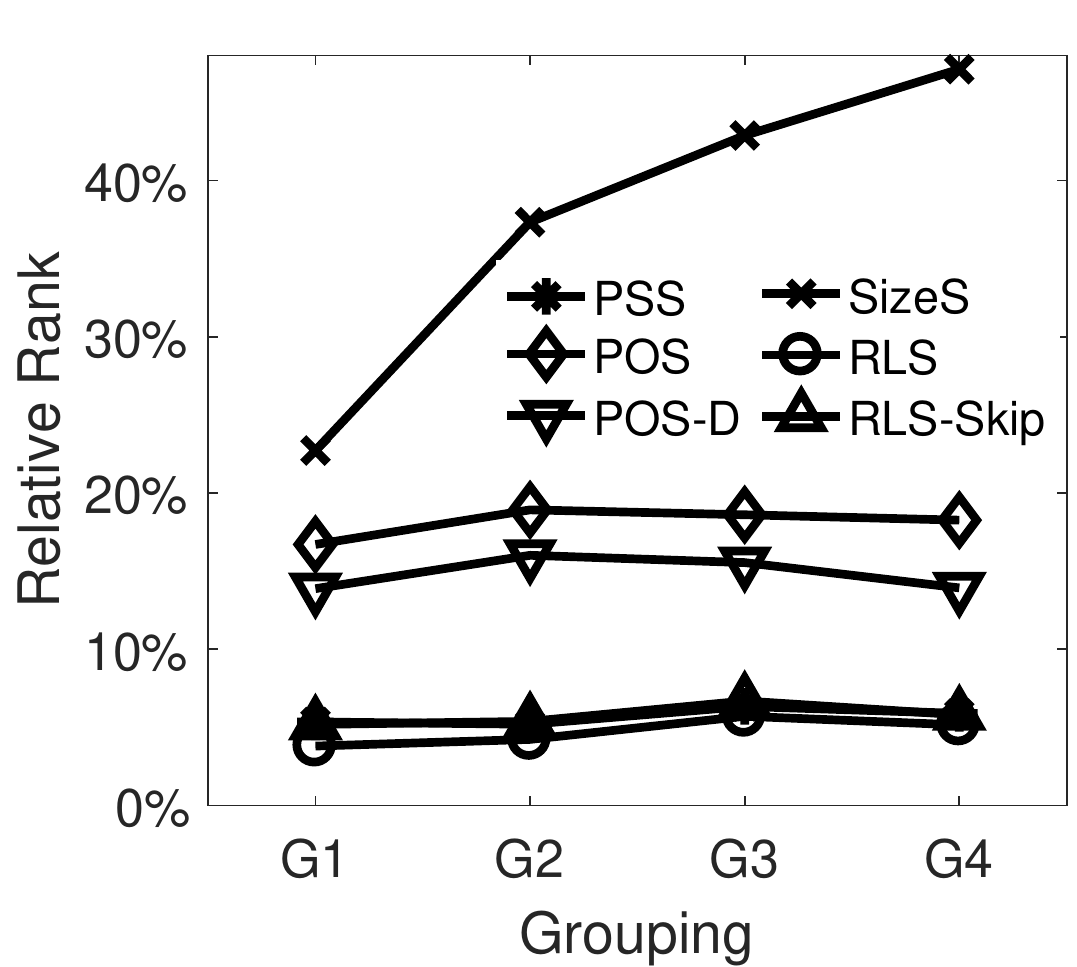}
		\end{minipage}
		\\
		\scriptsize (m) AR Porto
		&
		\scriptsize (n) MR Porto
		&
		\scriptsize (o) RR Porto
		&
		\scriptsize (p) AR Harbin
		&
		\scriptsize (q) MR Harbin
		&
		\scriptsize (r) RR Harbin
	\end{tabular}
	\caption{Results of grouping evaluation for t2vec (a)-(f), DTW (g)-(l) and Frechet (m)-(r).}
	\label{fig:groupingRLS_ph}
\end{figure*}

\begin{figure*}
	\hspace*{-.5cm}
	\centering
	\begin{tabular}{c c c c}
		\begin{minipage}{4cm}
			\includegraphics[width=4.15cm]{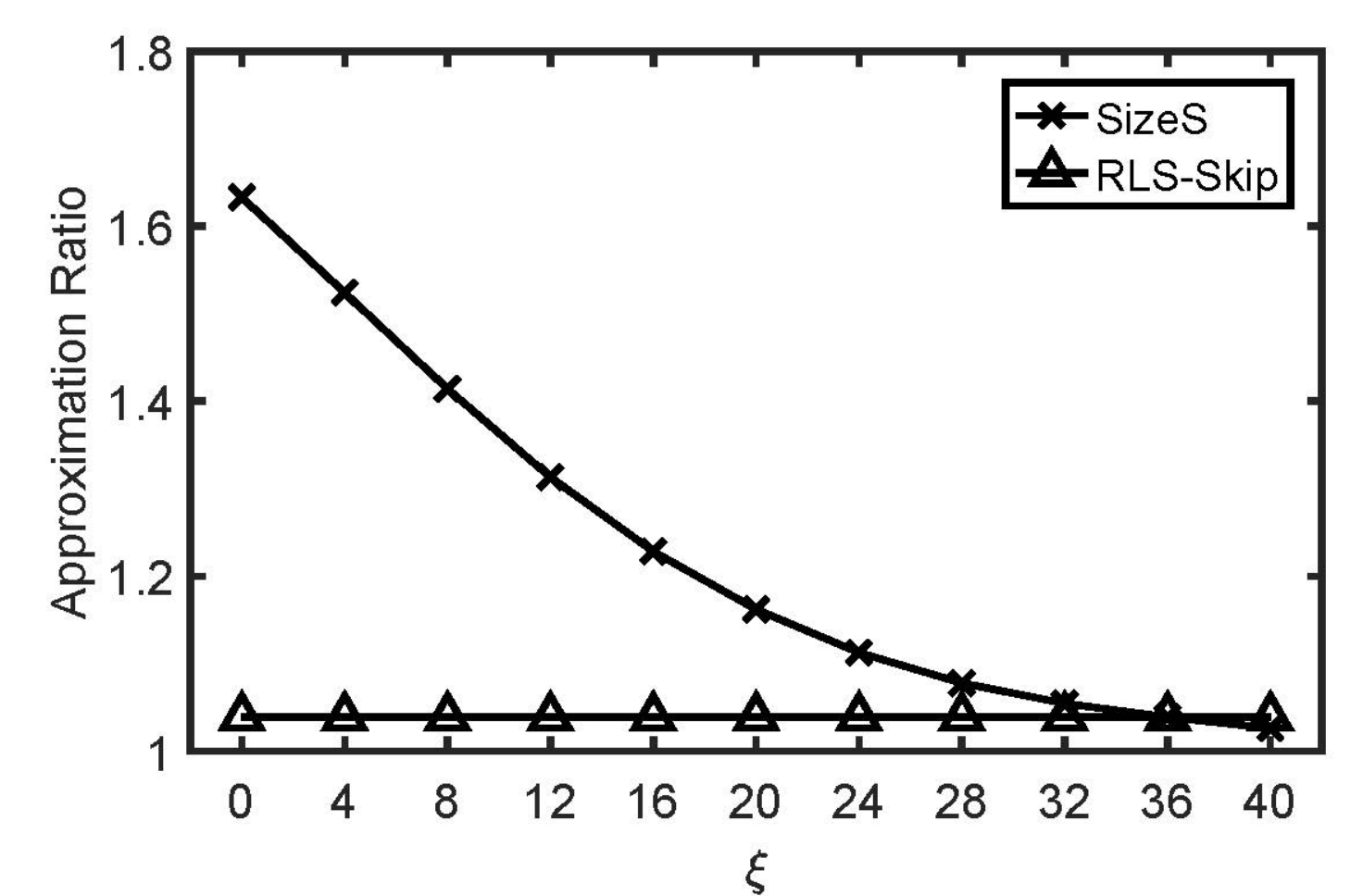}
		\end{minipage}
		&
		\begin{minipage}{4cm}
			\includegraphics[width=4.15cm]{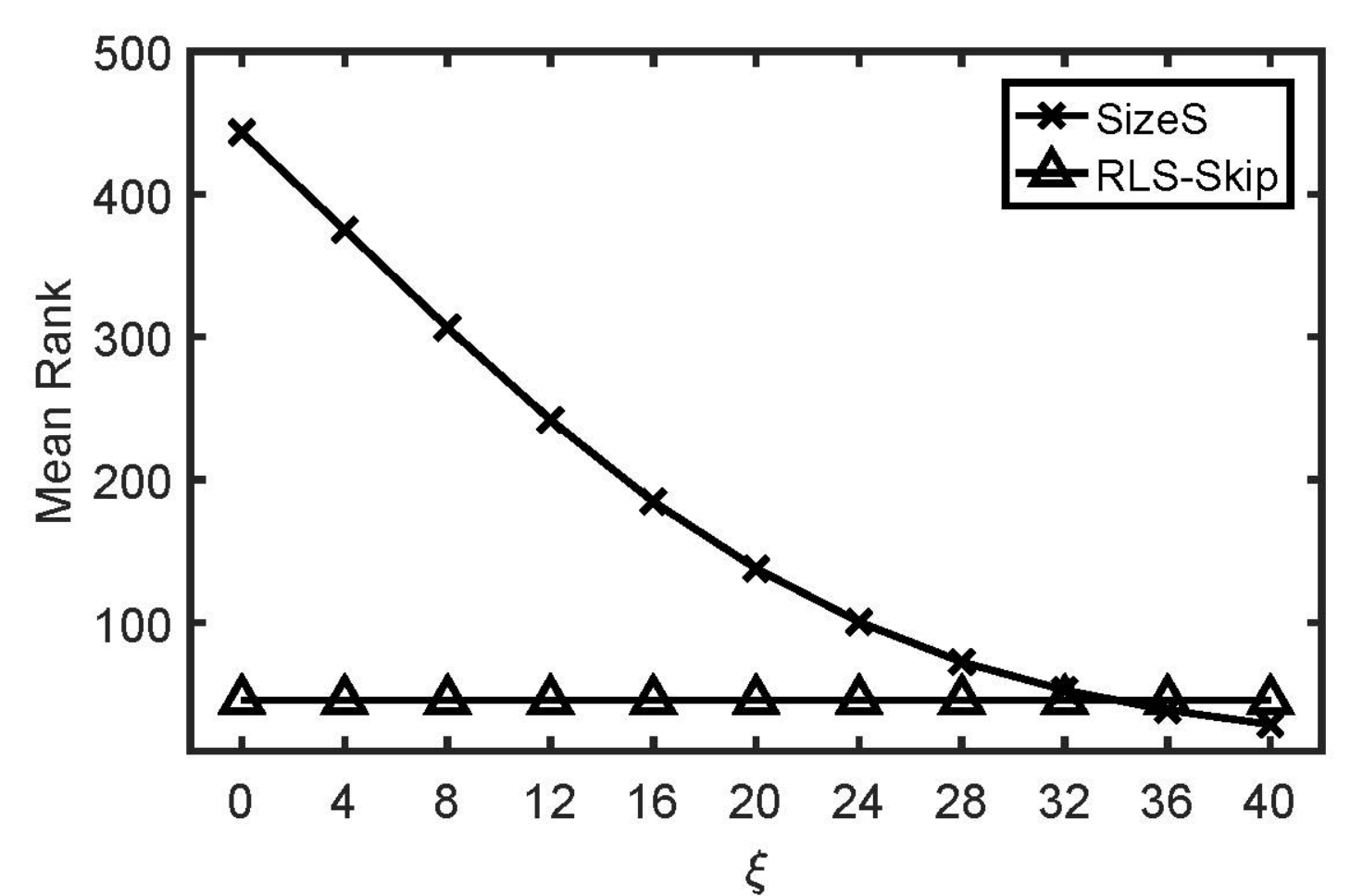}
		\end{minipage}
		&
		\begin{minipage}{4cm}
			\includegraphics[width=4.15cm]{dtw_rr}
		\end{minipage}
		&
		\begin{minipage}{4cm}
			\includegraphics[width=4.15cm]{dtw_time}
		\end{minipage}
		\\
		\small (a) Approximation Ratio (DTW)
		&
		\small (b) Mean Rank (DTW)
		&
		\small (c) Relative Rank (DTW)
		&
		\small (d) Time Cost (DTW)
		\\
	\end{tabular}
	\caption{The effect of soft margin $\xi$ for SizeS.}
	\label{para_fls_sizes}
\end{figure*}

\begin{figure*}[!th]
\centering
\begin{tabular}{c c c}
  \begin{minipage}{4.5cm}
	\includegraphics[width=4.5cm]{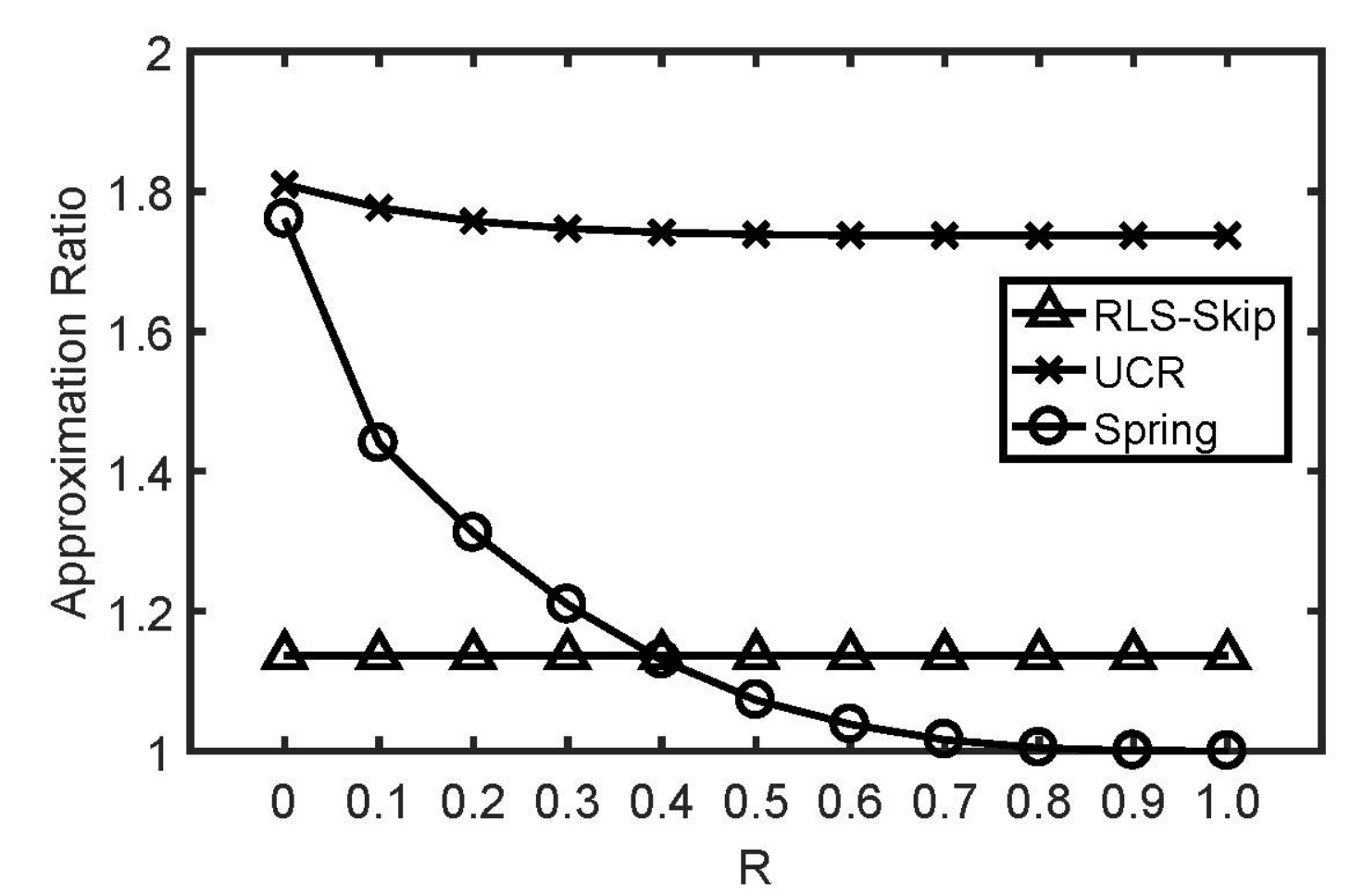}
  \end{minipage}
  &
  \begin{minipage}{4.5cm}
	\includegraphics[width=4.5cm]{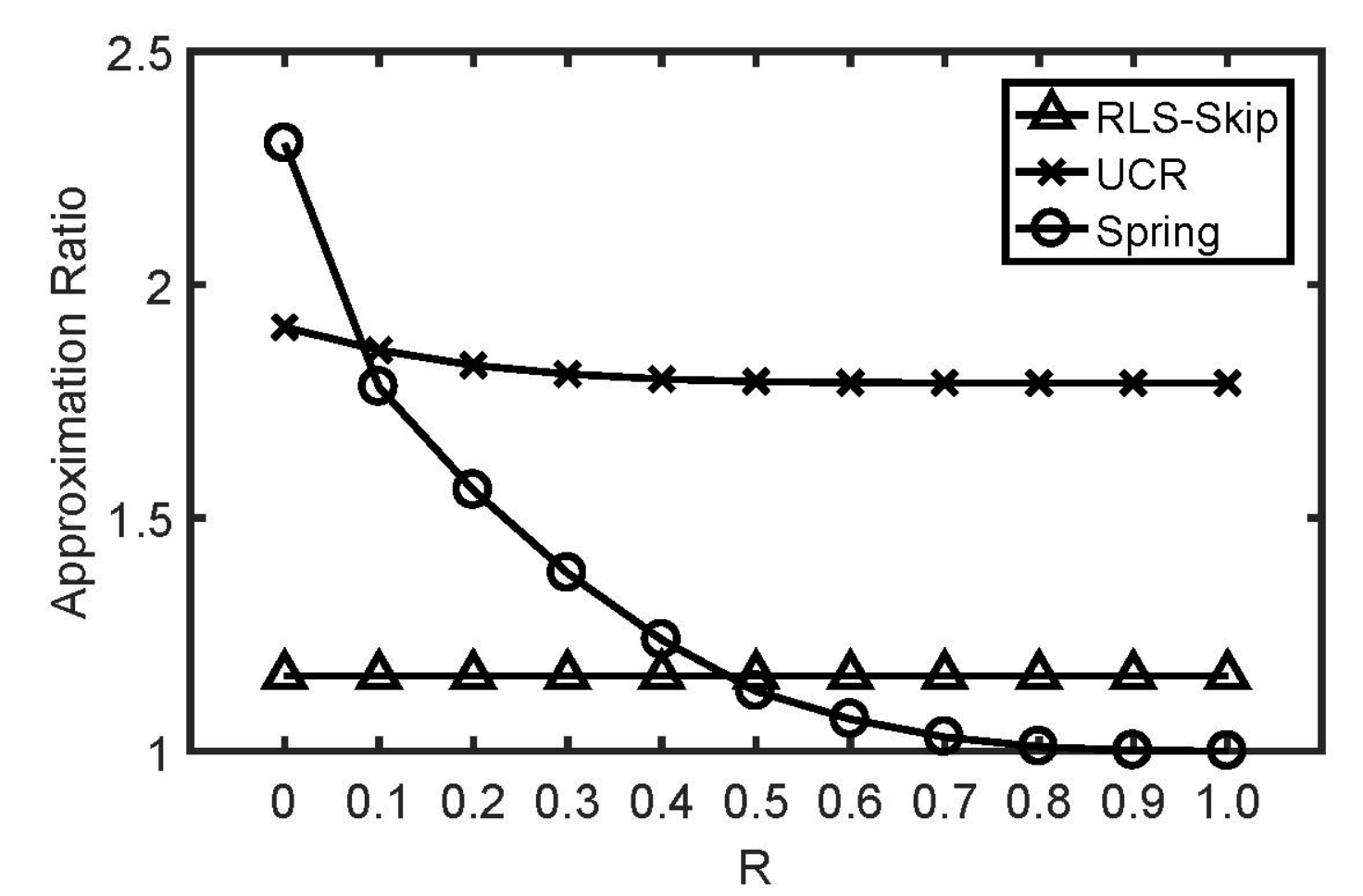}
  \end{minipage}
  &
  \begin{minipage}{4.5cm}
	\includegraphics[width=4.5cm]{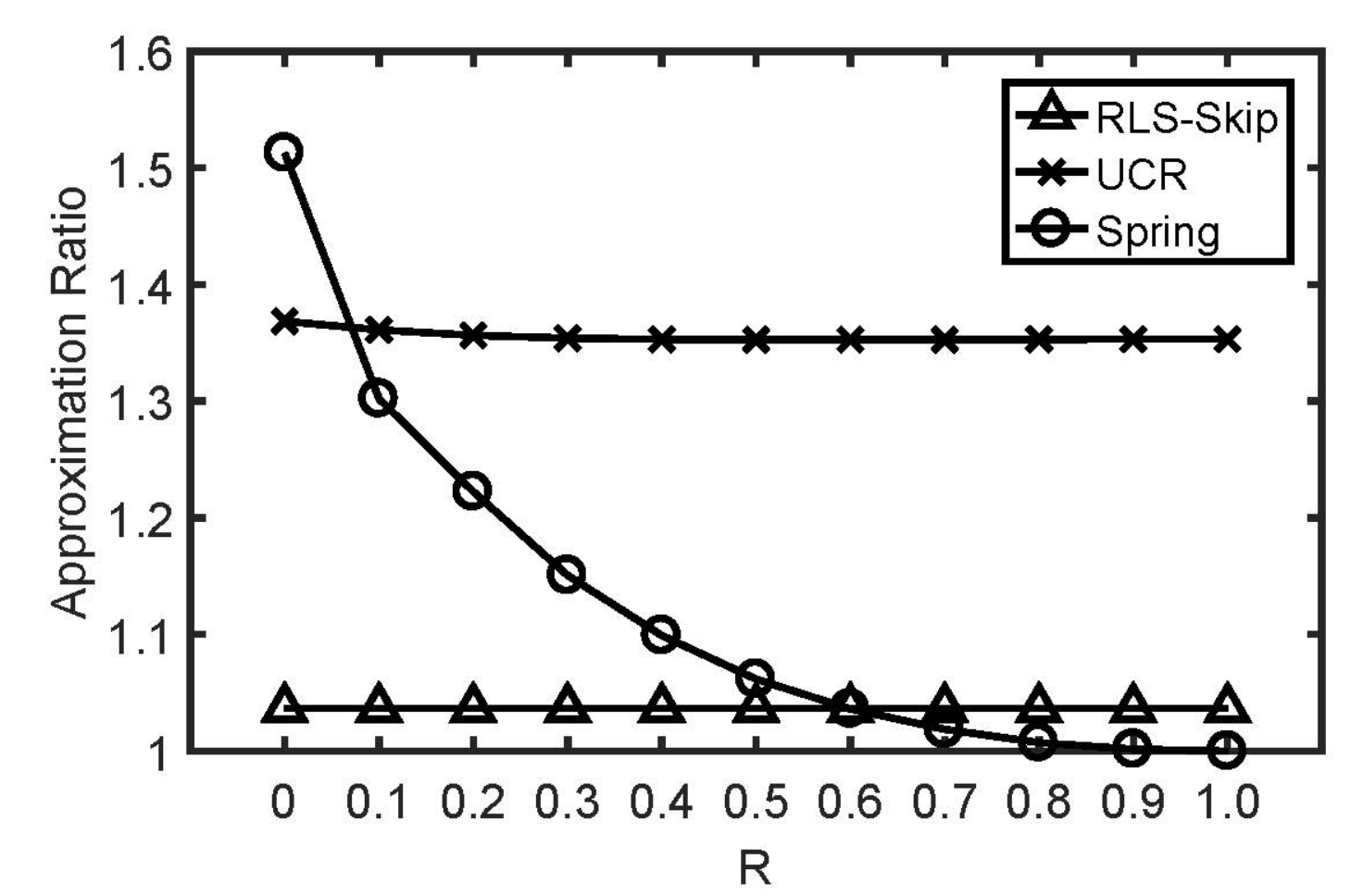}
  \end{minipage}
  \\
  \scriptsize{(a) Approximation Ratio (Porto)}
  &
  \scriptsize{(b) Approximation Ratio (Harbin)}
  &
  \scriptsize{(c) Approximation Ratio (Sports)}
  \\
  \begin{minipage}{4.5cm}
	\includegraphics[width=4.5cm]{spring_ucr_time_dtw_p}
  \end{minipage}
  &
  \begin{minipage}{4.5cm}
	\includegraphics[width=4.5cm]{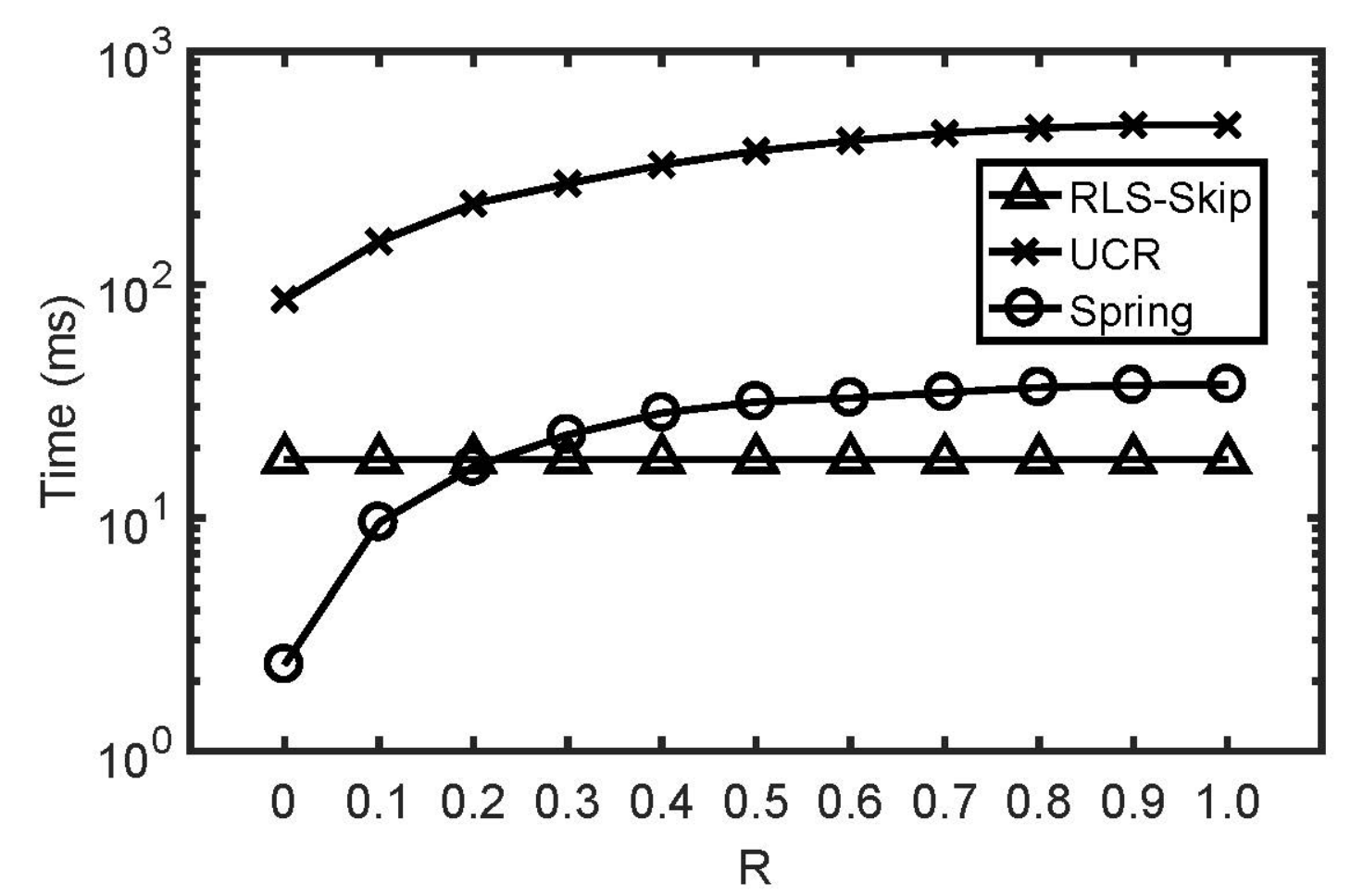}
  \end{minipage}
  &
  \begin{minipage}{4.5cm}
	\includegraphics[width=4.5cm]{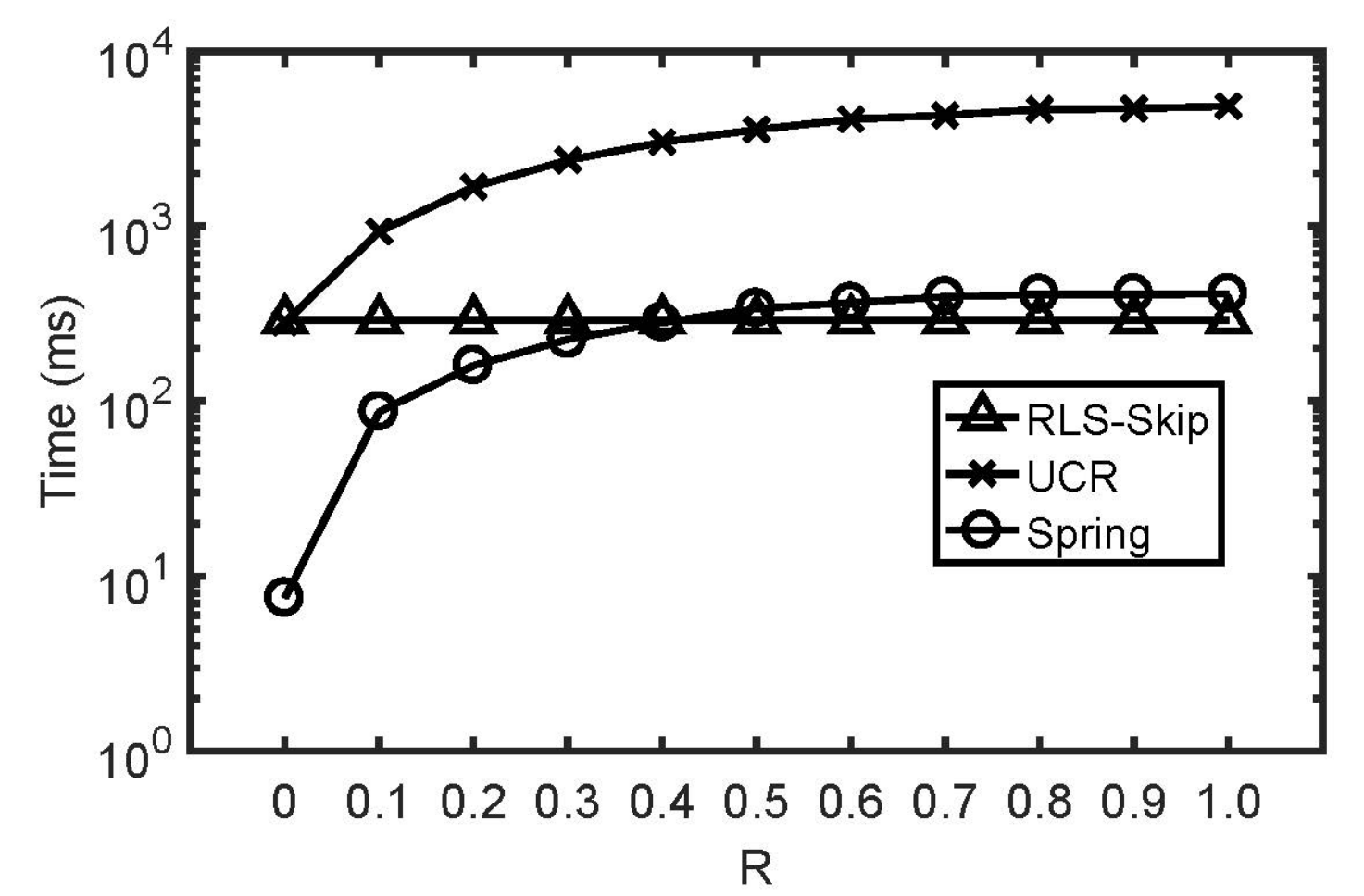}
  \end{minipage}
  \\
  \scriptsize{(d) Time Cost (Porto)}
  &
  \scriptsize{(e) Time Cost (Harbin)}
  &
  \scriptsize{(f) Time Cost (Sports)}
  \end{tabular}
\caption{Comparison with UCR and Spring.}
\label{ucr_result_full}
\end{figure*}

\begin{figure*}
	\hspace*{-.5cm}
	\centering
	\begin{tabular}{c c c c}
		\begin{minipage}{4cm}
			\includegraphics[width=4.15cm]{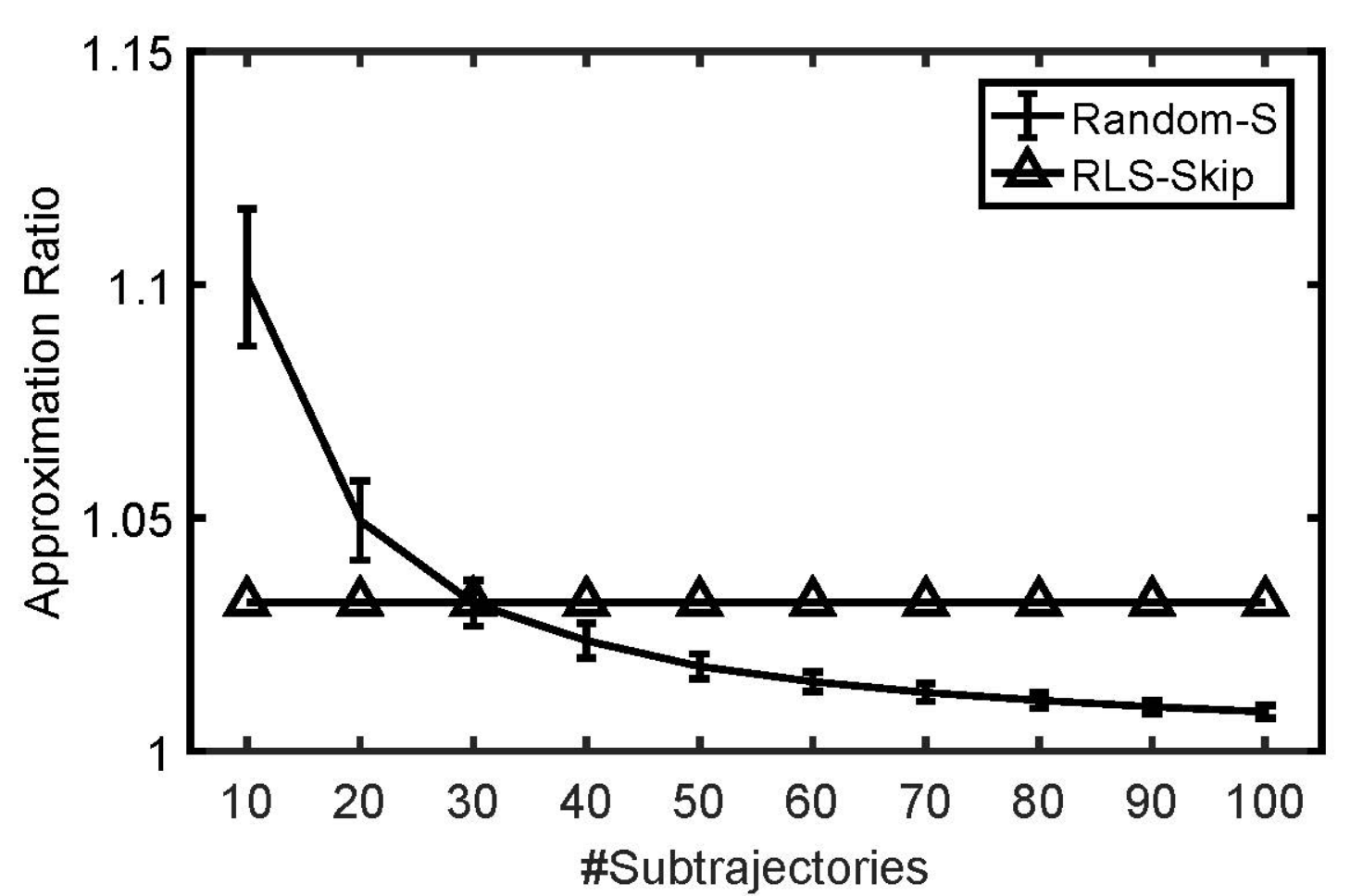}
		\end{minipage}
		&
		\begin{minipage}{4cm}
			\includegraphics[width=4.15cm]{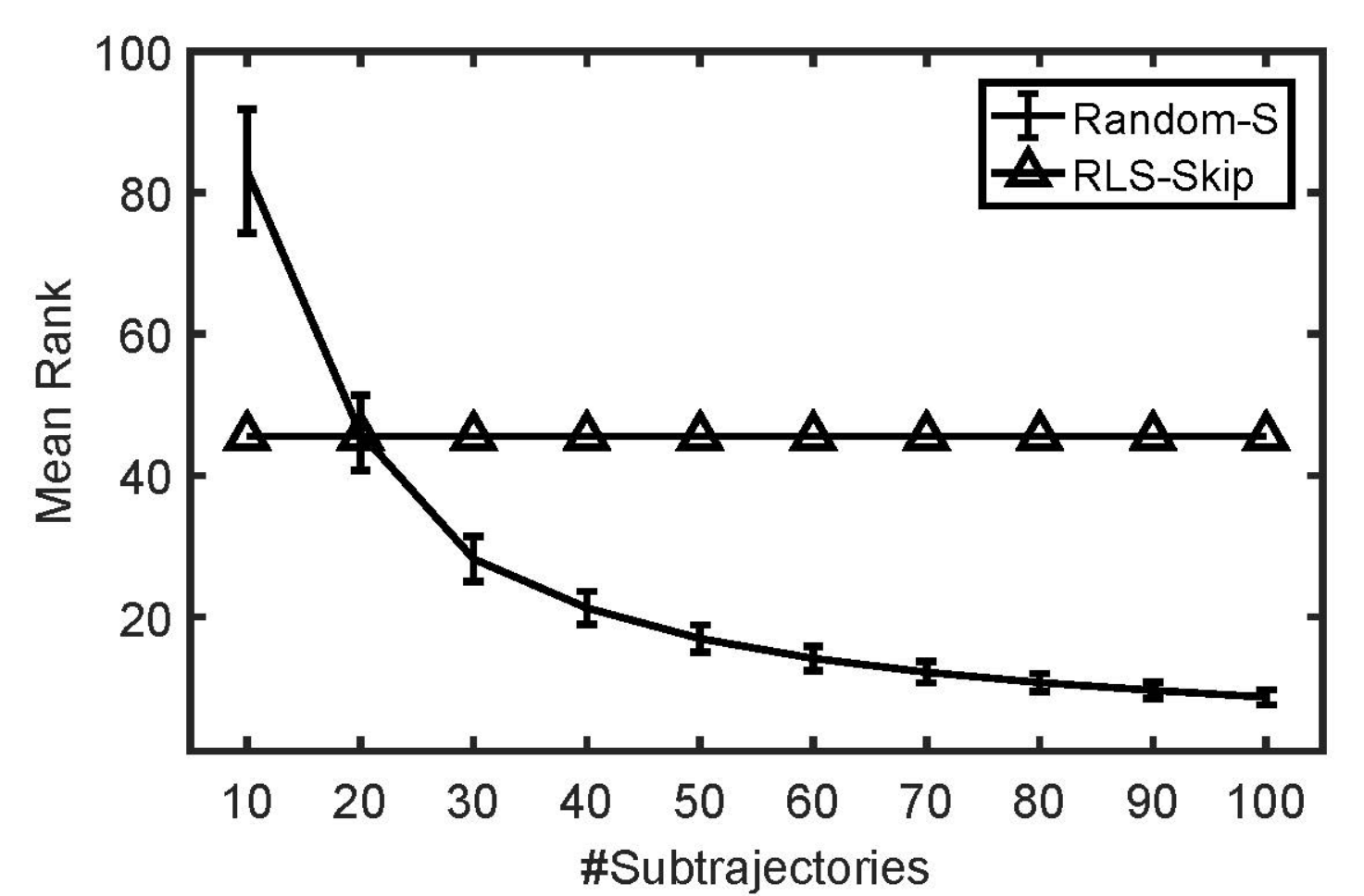}
		\end{minipage}
		&
		\begin{minipage}{4cm}
			\includegraphics[width=4.15cm]{sampling_rr_dtw}
		\end{minipage}
		&
		\begin{minipage}{4cm}
			\includegraphics[width=4.15cm]{sampling_time_dtw}
		\end{minipage}
		\\
		\small (a) Approximation Ratio (DTW)
		&
		\small (b) Mean Rank (DTW)
		&
		\small (c) Relative Rank (DTW)
		&
		\small (d) Time Cost (DTW)
	\end{tabular}
	\caption{Comparison with Random-S.}
	\label{sampling-full}
\end{figure*}

%% file: main.bbl
\begin{thebibliography}{10}

\bibitem{agarwal2018subtrajectory}
P.~K. Agarwal, K.~Fox, K.~Munagala, A.~Nath, J.~Pan, and E.~Taylor.
\newblock Subtrajectory clustering: Models and algorithms.
\newblock In {\em Proceedings of the 37th ACM SIGMOD-SIGACT-SIGAI Symposium on
  Principles of Database Systems}, pages 75--87. ACM, 2018.

\bibitem{alt1995computing}
H.~Alt and M.~Godau.
\newblock Computing the fr{\'e}chet distance between two polygonal curves.
\newblock {\em International Journal of Computational Geometry \&
  Applications}, 5(01n02):75--91, 1995.

\bibitem{athitsos2008approximate}
V.~Athitsos, P.~Papapetrou, M.~Potamias, G.~Kollios, and D.~Gunopulos.
\newblock Approximate embedding-based subsequence matching of time series.
\newblock In {\em Proceedings of the 2008 ACM SIGMOD international conference
  on Management of data}, pages 365--378. ACM, 2008.

\bibitem{brafman2002r}
R.~I. Brafman and M.~Tennenholtz.
\newblock R-max-a general polynomial time algorithm for near-optimal
  reinforcement learning.
\newblock {\em Journal of Machine Learning Research}, 3(Oct):213--231, 2002.

\bibitem{buchin2011detecting}
K.~Buchin, M.~Buchin, J.~Gudmundsson, M.~L{\"o}ffler, and J.~Luo.
\newblock Detecting commuting patterns by clustering subtrajectories.
\newblock {\em International Journal of Computational Geometry \&
  Applications}, 21(03):253--282, 2011.

\bibitem{chen2004marriage}
L.~Chen and R.~Ng.
\newblock On the marriage of lp-norms and edit distance.
\newblock In {\em Proceedings of the Thirtieth international conference on Very
  large data bases-Volume 30}, pages 792--803. VLDB Endowment, 2004.

\bibitem{chen2005robust}
L.~Chen, M.~T. {\"O}zsu, and V.~Oria.
\newblock Robust and fast similarity search for moving object trajectories.
\newblock In {\em Proceedings of the 2005 ACM SIGMOD international conference
  on Management of data}, pages 491--502. ACM, 2005.

\bibitem{cho2014learning}
K.~Cho, B.~Van~Merri{\"e}nboer, C.~Gulcehre, D.~Bahdanau, F.~Bougares,
  H.~Schwenk, and Y.~Bengio.
\newblock Learning phrase representations using rnn encoder-decoder for
  statistical machine translation.
\newblock {\em arXiv preprint arXiv:1406.1078}, 2014.

\bibitem{faloutsos1994fast}
C.~Faloutsos, M.~Ranganathan, and Y.~Manolopoulos.
\newblock {\em Fast subsequence matching in time-series databases}, volume~23.
\newblock ACM, 1994.

\bibitem{gong2019fast}
X.~Gong, S.~Fong, and Y.-W. Si.
\newblock Fast fuzzy subsequence matching algorithms on time-series.
\newblock {\em Expert Systems with Applications}, 116:275--284, 2019.

\bibitem{han2007ranked}
W.-S. Han, J.~Lee, Y.-S. Moon, and H.~Jiang.
\newblock Ranked subsequence matching in time-series databases.
\newblock In {\em Proceedings of the 33rd international conference on Very
  large data bases}, pages 423--434. VLDB Endowment, 2007.

\bibitem{kearns2002near}
M.~Kearns and S.~Singh.
\newblock Near-optimal reinforcement learning in polynomial time.
\newblock {\em Machine learning}, 49(2-3):209--232, 2002.

\bibitem{keogh2005exact}
E.~Keogh and C.~A. Ratanamahatana.
\newblock Exact indexing of dynamic time warping.
\newblock {\em Knowledge and information systems}, 7(3):358--386, 2005.

\bibitem{kim2001index}
S.-W. Kim, S.~Park, and W.~W. Chu.
\newblock An index-based approach for similarity search supporting time warping
  in large sequence databases.
\newblock In {\em Proceedings 17th International Conference on Data
  Engineering}, pages 607--614. IEEE, 2001.

\bibitem{kim2013efficient}
Y.~Kim and K.~Shim.
\newblock Efficient top-k algorithms for approximate substring matching.
\newblock In {\em Proceedings of the 2013 ACM SIGMOD International Conference
  on Management of Data}, pages 385--396. ACM, 2013.

\bibitem{lee2007trajectory}
J.-G. Lee, J.~Han, and K.-Y. Whang.
\newblock Trajectory clustering: a partition-and-group framework.
\newblock In {\em Proceedings of the 2007 ACM SIGMOD international conference
  on Management of data}, pages 593--604. ACM, 2007.

\bibitem{li2019qtune}
G.~Li, X.~Zhou, S.~Li, and B.~Gao.
\newblock Qtune: A query-aware database tuning system with deep reinforcement
  learning.
\newblock {\em Proceedings of the VLDB Endowment}, 12(12):2118--2130, 2019.

\bibitem{li2018deep}
X.~Li, K.~Zhao, G.~Cong, C.~S. Jensen, and W.~Wei.
\newblock Deep representation learning for trajectory similarity computation.
\newblock In {\em 2018 IEEE 34th International Conference on Data Engineering
  (ICDE)}, pages 617--628. IEEE, 2018.

\bibitem{long2013direction}
C.~Long, R.~C.-W. Wong, and H.~Jagadish.
\newblock Direction-preserving trajectory simplification.
\newblock {\em Proceedings of the VLDB Endowment}, 6(10):949--960, 2013.

\bibitem{ma2012ksq}
C.~Ma, H.~Lu, L.~Shou, and G.~Chen.
\newblock Ksq: Top-k similarity query on uncertain trajectories.
\newblock {\em IEEE Transactions on Knowledge and Data Engineering},
  25(9):2049--2062, 2012.

\bibitem{mnih2013playing}
V.~Mnih, K.~Kavukcuoglu, D.~Silver, A.~Graves, I.~Antonoglou, D.~Wierstra, and
  M.~Riedmiller.
\newblock Playing atari with deep reinforcement learning.
\newblock {\em arXiv preprint arXiv:1312.5602}, 2013.

\bibitem{mnih2015human}
V.~Mnih, K.~Kavukcuoglu, D.~Silver, A.~A. Rusu, J.~Veness, M.~G. Bellemare,
  A.~Graves, M.~Riedmiller, A.~K. Fidjeland, G.~Ostrovski, et~al.
\newblock Human-level control through deep reinforcement learning.
\newblock {\em Nature}, 518(7540):529, 2015.

\bibitem{moon2001duality}
Y.-S. Moon, K.-Y. Whang, and W.-K. Loh.
\newblock Duality-based subsequence matching in time-series databases.
\newblock In {\em Proceedings 17th International Conference on Data
  Engineering}, pages 263--272. IEEE, 2001.

\bibitem{mueen2016extracting}
A.~Mueen and E.~Keogh.
\newblock Extracting optimal performance from dynamic time warping.
\newblock In {\em Proceedings of the 22nd ACM SIGKDD International Conference
  on Knowledge Discovery and Data Mining}, pages 2129--2130, 2016.

\bibitem{park2000efficient}
S.~Park, W.~W. Chu, J.~Yoon, and C.~Hsu.
\newblock Efficient searches for similar subsequences of different lengths in
  sequence databases.
\newblock In {\em Proceedings of 16th International Conference on Data
  Engineering (Cat. No. 00CB37073)}, pages 23--32. IEEE, 2000.

\bibitem{puterman2014markov}
M.~L. Puterman.
\newblock {\em Markov Decision Processes.: Discrete Stochastic Dynamic
  Programming}.
\newblock John Wiley \& Sons, 2014.

\bibitem{rakthanmanon2012searching}
T.~Rakthanmanon, B.~Campana, A.~Mueen, G.~Batista, B.~Westover, Q.~Zhu,
  J.~Zakaria, and E.~Keogh.
\newblock Searching and mining trillions of time series subsequences under
  dynamic time warping.
\newblock In {\em Proceedings of the 18th ACM SIGKDD international conference
  on Knowledge discovery and data mining}, pages 262--270. ACM, 2012.

\bibitem{rakthanmanon2013addressing}
T.~Rakthanmanon, B.~Campana, A.~Mueen, G.~Batista, B.~Westover, Q.~Zhu,
  J.~Zakaria, and E.~Keogh.
\newblock Addressing big data time series: Mining trillions of time series
  subsequences under dynamic time warping.
\newblock {\em ACM Transactions on Knowledge Discovery from Data (TKDD)},
  7(3):10, 2013.

\bibitem{ranu2015indexing}
S.~Ranu, P.~Deepak, A.~D. Telang, P.~Deshpande, and S.~Raghavan.
\newblock Indexing and matching trajectories under inconsistent sampling rates.
\newblock In {\em 2015 IEEE 31st International Conference on Data Engineering},
  pages 999--1010. IEEE, 2015.

\bibitem{rumelhart1988learning}
D.~E. Rumelhart, G.~E. Hinton, R.~J. Williams, et~al.
\newblock Learning representations by back-propagating errors.
\newblock {\em Cognitive modeling}, 5(3):1, 1988.

\bibitem{sakurai2007stream}
Y.~Sakurai, C.~Faloutsos, and M.~Yamamuro.
\newblock Stream monitoring under the time warping distance.
\newblock In {\em 2007 IEEE 23rd International Conference on Data Engineering},
  pages 1046--1055. IEEE, 2007.

\bibitem{sha2016chalkboarding}
L.~Sha, P.~Lucey, Y.~Yue, P.~Carr, C.~Rohlf, and I.~Matthews.
\newblock Chalkboarding: A new spatiotemporal query paradigm for sports play
  retrieval.
\newblock In {\em Proceedings of the 21st International Conference on
  Intelligent User Interfaces}, pages 336--347, 2016.

\bibitem{sutton2018reinforcement}
R.~S. Sutton and A.~G. Barto.
\newblock {\em Reinforcement learning: An introduction}.
\newblock MIT press, 2018.

\bibitem{tampakis2019distributed}
P.~Tampakis, C.~Doulkeridis, N.~Pelekis, and Y.~Theodoridis.
\newblock Distributed subtrajectory join on massive datasets.
\newblock {\em arXiv preprint arXiv:1903.07748}, 2019.

\bibitem{tampakis2019scalable}
P.~Tampakis, N.~Pelekis, C.~Doulkeridis, and Y.~Theodoridis.
\newblock Scalable distributed subtrajectory clustering.
\newblock {\em arXiv preprint arXiv:1906.06956}, 2019.

\bibitem{trummer2018skinnerdb}
I.~Trummer, S.~Moseley, D.~Maram, S.~Jo, and J.~Antonakakis.
\newblock Skinnerdb: regret-bounded query evaluation via reinforcement
  learning.
\newblock {\em Proceedings of the VLDB Endowment}, 11(12):2074--2077, 2018.

\bibitem{vlachos2002discovering}
M.~Vlachos, G.~Kollios, and D.~Gunopulos.
\newblock Discovering similar multidimensional trajectories.
\newblock In {\em Proceedings 18th international conference on data
  engineering}, pages 673--684. IEEE, 2002.

\bibitem{wang2019fast}
S.~Wang, Z.~Bao, J.~S. Culpepper, T.~Sellis, and X.~Qin.
\newblock Fast large-scale trajectory clustering.
\newblock {\em Proceedings of the VLDB Endowment}, 13(1):29--42, 2019.

\bibitem{wang2018torch}
S.~Wang, Z.~Bao, J.~S. Culpepper, Z.~Xie, Q.~Liu, and X.~Qin.
\newblock Torch: A search engine for trajectory data.
\newblock In {\em The 41st International ACM SIGIR Conference on Research \&
  Development in Information Retrieval}, pages 535--544. ACM, 2018.

\bibitem{wang2019adaptive}
Y.~Wang, Y.~Tong, C.~Long, P.~Xu, K.~Xu, and W.~Lv.
\newblock Adaptive dynamic bipartite graph matching: A reinforcement learning
  approach.
\newblock In {\em 2019 IEEE 35th International Conference on Data Engineering
  (ICDE)}, pages 1478--1489. IEEE, 2019.

\bibitem{wang2019effective}
Z.~Wang, C.~Long, G.~Cong, and C.~Ju.
\newblock Effective and efficient sports play retrieval with deep
  representation learning.
\newblock In {\em Proceedings of the 25th ACM SIGKDD International Conference
  on Knowledge Discovery \& Data Mining}, pages 499--509, 2019.

\bibitem{TR}
Z.~Wang, C.~Long, G.~Cong, and Y.~Liu.
\newblock Efficient and effective similar subtrajectory search with deep
  reinforcement learning (technical report).
\newblock \url{https://www.ntu.edu.sg/home/wang_zheng/paper/TR-subtraj.pdf}.

\bibitem{watkins1992q}
C.~J. Watkins and P.~Dayan.
\newblock Q-learning.
\newblock {\em Machine learning}, 8(3-4):279--292, 1992.

\bibitem{xie2014eds}
M.~Xie.
\newblock Eds: a segment-based distance measure for sub-trajectory similarity
  search.
\newblock In {\em Proceedings of the 2014 ACM SIGMOD International Conference
  on Management of Data}, pages 1609--1610. ACM, 2014.

\bibitem{yao2019computing}
D.~Yao, G.~Cong, C.~Zhang, and J.~Bi.
\newblock Computing trajectory similarity in linear time: A generic seed-guided
  neural metric learning approach.
\newblock In {\em 2019 IEEE 35th International Conference on Data Engineering
  (ICDE)}, pages 1358--1369. IEEE, 2019.

\bibitem{yi1998efficient}
B.-K. Yi, H.~V. Jagadish, and C.~Faloutsos.
\newblock Efficient retrieval of similar time sequences under time warping.
\newblock In {\em Proceedings 14th International Conference on Data
  Engineering}, pages 201--208. IEEE, 1998.

\bibitem{yuan2019distributed}
H.~Yuan and G.~Li.
\newblock Distributed in-memory trajectory similarity search and join on road
  network.
\newblock In {\em 2019 IEEE 35th international conference on data engineering
  (ICDE)}, pages 1262--1273. IEEE, 2019.

\bibitem{zhang2019end}
J.~Zhang, Y.~Liu, K.~Zhou, G.~Li, Z.~Xiao, B.~Cheng, J.~Xing, Y.~Wang,
  T.~Cheng, L.~Liu, et~al.
\newblock An end-to-end automatic cloud database tuning system using deep
  reinforcement learning.
\newblock In {\em Proceedings of the 2019 International Conference on
  Management of Data}, pages 415--432. ACM, 2019.

\bibitem{zhang2010bed}
Z.~Zhang, M.~Hadjieleftheriou, B.~C. Ooi, and D.~Srivastava.
\newblock Bed-tree: an all-purpose index structure for string similarity search
  based on edit distance.
\newblock In {\em Proceedings of the 2010 ACM SIGMOD International Conference
  on Management of data}, pages 915--926. ACM, 2010.

\end{thebibliography}
